\documentclass[aps,amsmath,twocolumn,amssymb,floatfix,showpacs,superscriptaddress,nofootinbib,longbibliography]{revtex4-1}
\usepackage{MnSymbol}
\usepackage{braket}
\usepackage[dvipsnames]{xcolor}
\usepackage{float}
\usepackage{subfigure}
\usepackage{tikz}
\usepackage[colorlinks=true,linktoc=page,linkcolor=magenta,citecolor=blue,urlcolor=purple]{hyperref}
\usepackage{bm}

\usepackage{multirow}

\newcommand{\vect}[1]{\boldsymbol{\mathrm{#1}}}

\mathchardef\mhyphen="2D 

\newcommand{\ie}{{i.e.,\,\,}}
\newcommand{\eg}{{e.g.,~}}
\newcommand{\ua}{{\uparrow }}
\newcommand{\da}{{\downarrow }}

\newcommand\bea{\begin{eqnarray}}
	\newcommand\eea{\end{eqnarray}}
\newcommand\beq{\begin{equation}}  
	\newcommand\eeq{\end{equation}}

\newcommand{\non}{\nonumber}

\usepackage[normalem]{ulem}
\definecolor{lime}{HTML}{A6CE39}
\usepackage{sidecap,tikz}
\DeclareRobustCommand{\orcidicon}{\hspace{-1.0mm}
	\begin{tikzpicture}
		\draw[lime, fill=lime] (0.0,0.0) 
		circle [radius=0.15] 
		node[white] {{\fontfamily{qag}\selectfont \tiny \,ID}};
		\draw[white, fill=white] (-0.0525,0.095) 
		circle [radius=0.007];
	\end{tikzpicture}
	\hspace{-3.0mm}
}
\foreach \x in {A, ..., Z}{\expandafter\xdef\csname orcid\x\endcsname{\noexpand\href{https://orcid.org/\csname orcidauthor\x\endcsname}{\noexpand\orcidicon}}
}

\AtBeginDocument{%
	\newwrite\bibnotes
	\def\bibnotesext{Notes.bib}
	\immediate\openout\bibnotes=\jobname\bibnotesext
	\immediate\write\bibnotes{@CONTROL{REVTEX41Control}}
	\immediate\write\bibnotes{@CONTROL{%
			apsrev41Control,author="08",editor="1",pages="1",title="1",year="1"}}
	\if@filesw
	\immediate\write\@auxout{\string\citation{apsrev41Control}}%
	\fi
}%

\allowdisplaybreaks

\begin{document}

\title{Floquet first- and higher-order topological superconductors: A review}

\author{Debashish Mondal\orcidD{}}
\affiliation{Institute of Physics, Sachivalaya Marg, Bhubaneswar-751005, India}
\affiliation{Homi Bhabha National Institute, Training School Complex, Anushakti Nagar, Mumbai 400094, India}
\affiliation{Department of Physics,  Saarland University, Campus,
66123 Saarbr\"ucken, Germany}
\affiliation{Center for Quantum Technology (QuTe),  Saarland University, Campus,
66123 Saarbr\"ucken, Germany}

\author{Arnob Kumar Ghosh\orcidA{}}
\affiliation{Department of Physics and Astronomy, Uppsala University, Box 516, 75120 Uppsala, Sweden}

\author{Arijit Saha\orcidC{}}
\email{arijit@iopb.res.in}
\affiliation{Institute of Physics, Sachivalaya Marg, Bhubaneswar-751005, India}
\affiliation{Homi Bhabha National Institute, Training School Complex, Anushakti Nagar, Mumbai 400094, India}

\author{Tanay Nag\orcidB{}}
\email[Corresponding author: ]{tanay.nag@hyderabad.bits-pilani.ac.in}
\affiliation{Department of Physics, BITS Pilani-Hyderabad Campus, Telangana 500078, India}

\begin{abstract}
The current trend in quantum technologies is deeply rooted in quantum computation, with Majorana fermions and their topological protection believed to play a central role. In this context, topological superconductors (TSCs) offer a promising arena for investigating Majorana modes in the context of topological quantum computation. In dimension $d>1$, the spatial localization profile of the topological Majorana modes can be tuned in higher-order TSC extending beyond the realm of conventional first-order topological superconductor (FOTSC). In this review, we systematically survey the zoo of topological superconductivity, ranging from FOTSCs to third-order topological superconductors (TOTSCs), wherein the spatial dimensionality of Majorana zero modes (MZMs) is progressively reduced. We show that periodic driving protocols enable the Floquet engineering of all the above phases, even when the corresponding undriven systems are topologically trivial. We further investigate various driving schemes and the interplay among spin–orbit coupling, magnetic fields, magnetic impurities, and symmetry-breaking mass terms to design Floquet higher-order topological superconducting (FHOTSC) phases, which are characterized by appropriate topological invariants. We extensively discuss the development of invariants that are able to distinguish between regular MZMs and anomalous Majorana $\pi$-modes, appearing at two distinct energies,  where the latter does not exhibit any static analog.  
\end{abstract}

\maketitle

\section{Introduction}


Among the great discoveries of 20$^{{\rm{th}}}$ century, the integer Quantum Hall effect~(IQHE) is one of the most promising ones in the history of modern condensed matter physics~\cite{KlitzingPRL1980}. With the Nobel prize given in the last decade, the theoretical prediction of Chern insulator phases and their connection to the IQHE has provided a massive research thrust for understanding of the topological phases of matter~\cite{HaldanePRL1988}. This is a characteristically different phase of matter that cannot be explained by the Landau theory of phase transition based on spontaneous symmetry breaking~ \cite{huang2008statistical}.
The major distinctions between these two types of phase transitions are as follows: the superconductor and ferromagnetic phases break the global U(1) phase and O(3) symmetry while generating an ordered state from the symmetry-preserved normal and paramagnetic phases, respectively. On the contrary, the topological phases are protected by certain unitary and anti-unitary symmetries, therefore, leading to a distinct phase of matter \cite{RevModPhys.88.035005}. A suitably defined topological marker characterizes such phase in the relevant parameter space inside which it remains quantized. In the case of the ordered state described by the Landau phase transition, the order parameter varies as a function of the system parameter even in the ordered phase. However, in both types of phases, the topological and ordered phases are separated from the trivial and 
disordered phases, respectively,  by a bulk gap-closing transition. This is a signature of a quantum phase transition at zero temperature, indicating the vast applicability of quantum materials across different areas.

As far as the quantized nature of the topological markers/indices in the non-trivial phase is concerned, the quantized Hall conductance in IQHE is a prime example, where TKNN (Thouless, Kohomoto, Nightingale and den Nijs) invariant characterizes the phase~\cite{TKNNPRL1982}.  Here, cyclic evolution of the local curvature in the geometrical parameter space reduces to the cyclic evolution of the wave functions in the periodic Brillouin zone (BZ). Hence, momentum, playing the role of a cyclic parameter, suffices the emergence of the bulk topological index. The topological quantum materials host boundary modes when the materials are exposed to the vacuum by breaking the translation invariance. At the risk of losing momentum as a good quantum number, one discovers a fascinating phenomenon called, bulk-boundary correspondence~\cite{hasan2010colloquium} which connects the number of boundary modes to the bulk invariant.

Having discussed the foundation of the topological phases of matter, we briefly demonstrate a few pedagogical models starting from one-dimensional (1D) Su–Schrieffer–Heeger (SSH) model to three-dimensional (3D) topological insulator~(TI) harboring  electronic  boundary modes. The unitary chiral symmetry-preserved edge modes are characterized by the bulk winding number in 1D SSH model \cite{asboth2016short}. In  two dimension (2D), 
the time-reversal symmetry broken (protected) quantum anomalous (spin) Hall insulators are the typical example of topological quantum materials where electronic charge (spin) transport can be investigated in the presence of chiral (helical) boundary 
modes~\cite{ChangQAHRMP2023,kane2005quantum,kane2005quantumPRL,BernevigBOOK}. Both the above phases exist in the absence of an external magnetic field, however, for the latter case, the spin-orbit coupling~(SOC) plays a significant role and, interestingly, the quantum spin Hall insulators (QSHIs) can be thought of two copies of quantum anomalous Hall insulators (QAHIs) for up and down spins \cite{nag2025extended}. The Chern number can identify QAHI phase while and spin Chern number or $Z_2$ invariant can designate the QSHI phase~\cite{TKNNPRL1982,FuTimereversal2006}. The 2D version of QSHI is elevated to a strong (weak) TI in three dimensions (3D), hosting odd (even) number of gapless surface Dirac cones, that are experimentally realizable as well in  Bi$_2$X$_3$ (X=Se,Te), Sb$_2$Te$_3$ 
etc.~\cite{hasan2010colloquium,FuKane2007,Zhang2009,Xiaexperiment2009,chen2009experimental,fu2007topological}.

Going beyond the charged electronic mode, the topological superconductors (TSCs) support neutral zero energy Majorana fermions which are their own anti-particle.  Thanks to their connection with non-Abelian exchange statistics, the Majorana zero modes (MZMs) are potent enough for the  applications (qubit formation) in topological quantum computation \cite{Kitaev_2001,qi2011topological,hasan2010colloquium,nayak08,das2012zero,Deng1557}. A simple toy model, known as $p$-wave Kitaev chain, hosts MZMs at its boundaries where unitary chiral and anti-unitary particle hole symmetries are preserved \cite{Kitaev_2001}. This model is represented by Majorana fermions while decomposing each complex fermionic site in terms of two real Majorana fermions designating electron and hole degrees of freedom. This Majorana fermion representation of $p$-wave Kitaev chain closely resembles the 1D bi-partite  SSH chain as the former (latter) anchors two distinct unpaired Majorana (sub-lattice electronic) modes in the topological phase. This initiates the field of topological superconductor 
(TSC)~\cite{sato2017topological} where the spin-triplet $p$-wave superconducting pairing is a key ingredient to support an effective spinless pairing between electrons for spinful Hamiltonian. This model while giving basic understanding behind the emergence of MZMs, is not experimentally realizable due to the natural unavailability of $p$-wave nature of the superconductor.

In order to circumvent the above issue, there exists conventional $s$-wave based Rashba nanowire model that is experimentally 
accessible~\cite{Fu2008,Sau2010,Lutchyn10,Hughes2010,Oreg2010,Mourik1003,Zareapour2012,Finck2013,He294,Liu18}. 
The materials with strong spin-orbit coupling such as topological insulators, semiconductor thin films, and Rashba nanowires, when placed in close proximity to the common $s$-wave superconductor, are proposed to provide an efficient platform for the realization of MZMs within the topological superconducting gap in the bulk spectrum. The MZMs  are usually localized at 2D vortex cores or 1D edges of the 2D heterostructures.  In several transport experiments, based on hybrid superconductor-semiconductor nanowire setups, the zero bias peak is observed that can be interpreted as the indirect signatures of the MZMs~\cite{Mourik2012Science,das2012zero,Rokhinson2012,Finck2013,Albrecht2016,Deng2016}. Interestingly, 
the MZMs have been engineered in an alternative route via helical spin chain~\cite{Yazdani2013,Felix_analytics,Loss2013PRL,Felix2014,PhysRevB.89.115109, Bena2015,Loss2016,Christensen2016,Simon2017,Sticklet2019,Rex2020,Mohanta2021, Loss2022} or magnetic adatoms fabricated on the surface of a bulk $s$-wave superconductor
~\cite{Sau2013,PhysRevB.88.180503,Hui2015,Joel2015,Sharma2016,Theiler2019,Mashkoori2020,Teixeria2020}. The quasiparticles in the superconductor interact with the magnetic impurities leading to the formation of Yu-Shiba-Rusinov~(YSR) states~\cite{Shiba_original,Loss2021,Felix2021,Ortuzar2022,PritamPRB2023} inside the superconducting gap. Importantly, the helical spin texture, mimicking the  magnetic impurities, plays the combined role of SOC and magnetic field~\cite{Bena2015,Loss2022}. Hence, so far we discuss the chiral TSC phase that breaks time-reversal symmetry (TRS). Interestingly, the helical TSC phase has also been proposed in the absence of an external magnetic field or magnetic textures for TRS invariant model where extended $s$-wave proximized 2D electron gas with SOC can host MZMs \cite{FZhang2013}. The spinful $p_x+ip_y$ superconductor is another example supporting MZMs that are essentially Majorana Kramer's 
pairs~\cite{Xiao-Liang2009}. In all the above cases, the bulk boundary correspondence is found as the bulk invariant correctly describes the number of boundary modes.

\begin{figure}[]
	\centering
	\subfigure{\includegraphics[width=0.47\textwidth]{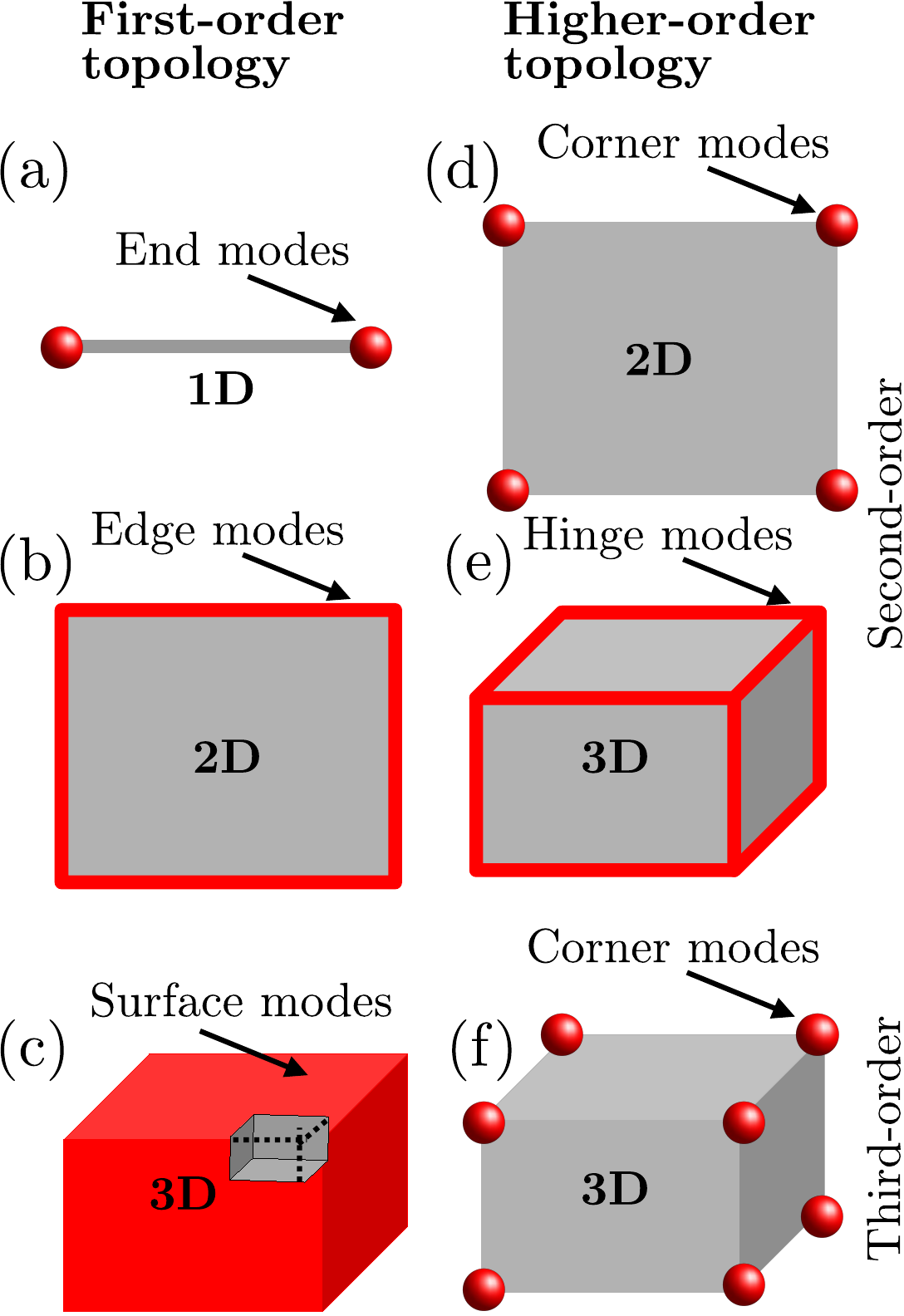}}
	\caption{Schematic representation of $n^{\rm th}$ order TIs/ TSCs are depicted. Panels (a), (b) and (c) represent first oder TIs/TSCs in 1D, 2D and 3D, respectively. For all these cases, systems have gapped-bulk states [gray region] and exhibit 0D end modes~[red spheres], 1D edge modes~[red lines] and, 2D surface modes~[red surface], respectively. Schematics of 2D and 3D SOTI/SOTSC are illustrated in panels (d) and (e), respectively. Both systems have a gapped bulk and $(d-1)$-dimensional boundary states [gray regions] with a  0D~(1D) corner states [represented by red spheres]~(hinge states [represented by red lines]). (f) Schematic of a 3D TOTI/TOTSC is demonstrated. This system hosts a gapped bulk as well as gapped $(d-1)$- and $(d-2)$-dimensional boundary states [gray regions]. However, a TOTI/TOTSC is accompanied by 0D localized corner states [represented by red spheres].	}
	\label{static}
\end{figure}

Till now, we have restricted our discussion to the first-order topological phases namely first-order topological insulators (FOTIs) and first-order topological superconductors (FOTSCs) where the boundary states are localized on ($d-1$)-dimensional space for a $d$-dimensional object~[See Figs.~\ref{static}(a),~\ref{static}(b) and~\ref{static}(c)]. 
Very recently, the conventional bulk-boundary correspondence has been generalized in the context of higher-order topological insulators (HOTI) and higher-order topological superconductors (HOTSC)~\cite{PhysRevB.92.085126,benalcazar2017,benalcazarprb2017,Song2017,Langbehn2017,schindler2018,Khalaf2018,Geier2018,Franca2018,Zhu2018,Liu2018,Yan2018,wang2018higher,PhysRevB.98.165144,Ezawakagome,ezawa2019second,Roy2019,Bomantara2019,Trifunovic2019,Zhang2019,Volpez2019,YanPRB2019,Wu2020,Sumathi2020,jelena2020HOTSC,PhysRevB.101.220506,SongboPRR,SongboPRB,SongboPRR2,GhorashiPRL,GhorashiPRB,ZengPRL2020,PhysRevB.110.064520,PhysRevA.109.063518,PhysRevA.106.063524,PhysRevB.103.L041401,Wu_2022,Tu_2024}. In this case, the dimensionality of the boundary modes, obtained in first-order topological (FOT) phase, is reduced for higher-order topological (HOT) phases. Hence the concept of HOT modes only exists for dimension $d>1$ due to the constraint on the dimensionality of the boundary modes.
To be precise, an $l^{\rm th}$-order topological insulator or superconductor in $p$ dimensions harbors $d_c=(p-l)$-dimensional boundary modes ($l\le p$). For example, 2D {systems} with gapped edge states host 0D corner modes, called second order TIs~(SOTIs) or second order TSCs~(SOTSCs)~[See Fig.~\ref{static}(d)]. On the other hand, a 3D second (third) order TI/TSC with gapped surface~(hinges) hosts gapless propagating (localized) modes on the hinges (corners), characterized  by $d_c=1~(0)$~[See Fig.~\ref{static}(e) and Fig.~\ref{static}(f)].
Consequently, the first-order topological indices fail to characterize the above HOT modes leading to the formulation of
projected  moments of the Berry phases in the real as well as momentum space. The HOT invariants such as the quadrupole, octupole moments, nested polarizations, and  Wannier polarizations can appropriately define the  HOTIs and HOTSCs phase where the FOT invariants vanish~\cite{benalcazar2017,benalcazarprb2017}. 
There exist several experimental demonstrations as well as experimental proposals to realize the HOT modes in solid-state systems~\cite{schindler2018higher,Experiment3DHOTI.VanDerWaals,Aggarwal2021,ShumiyaHOTI2022}, phononic crystals~\cite{serra2018observation}, acoustic systems~\cite{xue2019acoustic,ni2019observation,Experiment3DHOTI.aSonicCrystals,Ni2020}, electric-circuit setups~\cite{imhof2018topolectrical}, photonic lattice~\cite{PhotonicChen,PhotonicXie,mittal2019photonic} etc.


The reduction of the dimensionality of the HOT modes in $d>1$, compared to FOT modes, can be understood by a change in the underlying boundary Hamiltonian. The gapless boundary modes in FOT phases become gapped upon adding a suitable discrete symmetry-breaking term that essentially gaps out the existing boundary modes associated with FOT phase. As a result, under open boundary conditions following a suitable hard wall, one can find zero energy mid-gap states that are localized on a lower dimensional space. The symmetry-breaking term introduces mass gap at the boundaries while its sign changes between two adjacent boundaries. As a consequence of Jackiw-Rebbi theory, exactly at the intersection of these boundaries, one finds zero-energy modes mimicking the HOT phase. The bulk gap does not play a major role in transforming a FOT phase into a HOT phase while the bulk gap closes at the phase transition point~\cite{Arouca20}. Therefore, the bulk gap closes and reopens for a phase transition between the trivial and HOT phases. Therefore, the HOT phases may not always be governed by bulk gap-closing rather they are caused by boundary gap structure. In the literature, boundary gap-mediated HOT phases are referred to as extrinsic while bulk gap-driven are known as intrinsic. 

Having extensively discussed the FOTSC and HOTSC phases in static case (equilibrium), we now turn our attention to the intriguing outcome associated with periodically driven quantum systems among which light-mediated metal-insulator phase transition~\cite{Metalinsulatorlight1998}, light-induced photovoltaic effect~\cite{oka09photovoltaic,Leephotovoltic2008}, photo-thermoelectric effect~\cite{XiaodongPhoto-Thermoelectric2010}, light-induced superconductivity~\cite{FaustiScience}, synchronization transition in an interacting model~\cite{nag17}  are the 
worth mentioning examples. The time-periodic systems are 
efficiently tackled by Floquet theory~\cite{FLoquetpaper1883} where the non-equilibrium steady states, measured stroboscopically after each period, signify the emergence of novel orders that are completely absent in the 
static limit~\cite{nag14,nag15,Nag21_flatband,Tamang21,Nag20_FWSM,nag2021anomalous,ghosh2021systematic,Kundu21,Ghosh2022NHPRBL,gregefalk2023fano,das2023fano}. Unlike the static system, energy of the driven system namely, quasi-energy is defined on a torus rendering its periodicity in the  momentum and drive frequency space.In the context of topological phases, the non-equilibrium generation of Floquet topological states of matter has become another intriguing research direction where drive frequency and amplitude not only lead to new topological phases with zero-energy regular modes, appearing within a given frequency BZ, but also with finite-energy anomalous boundary modes, existing between the intersection of adjacent frequency BZ~\cite{Basov2017,oka09photovoltaic,Kitagawacharacterization2010,lindner11floquet,FloquetGuPRL2011,Rudner2013,Piskunow2014,Usaj2014,Nathan_2015,Eckardt2015,MikamiBW2016,Yan2017, Eckardt2017,oka2019,NHLindner2020,Bao2022}. This indicates the generalization of the bulk boundary correspondence by redefining the static invariant in the quasi-state representation of the Floquet theory where the non-trivial winding of the above state is present in the time/frequency domain in addition to the usual momentum 
space~\cite{Basov2017,oka09photovoltaic,Kitagawacharacterization2010,lindner11floquet,kitagawa11transport,FloquetGuPRL2011,Dora2012,Rudner2013,Thakurathi2013,Piskunow2014,Usaj2014,BenitoPRB2014,Nathan_2015,Eckardt2015,sacramento2015,MikamiBW2016,Yan2017,Eckardt2017,oka2019,NHLindner2020,rxzhangPRL2021,Bao2022}. In this context, FOTSC phases under Floquet drive exhibits interesting transport behavior from which the emergence of Floquet Majorana fermions can be 
inferred~\cite{prnx-47mk,6wnf-b5g8,PhysRevB.111.L121401}.


It can be anticipated that Floquet generation of HOT phases have gained a lot of attention in recent years while the Floquet graphene and Floquet TI had already set the stage~\cite{ghosh2023generation}. The main question that is addressed in this field is how HOT modes or FOT phase are engineered out of trivial system
and can be characterized by appropriate topological invariants. In this context, one can adopt various driving protocols such as discrete time step drive, and continuous time sinusoidal drive, depending on the demand of the final target state. The main aim is to break the symmetry protection of the FOT phase dynamically and at the same time preserve the symmetry for HOT phase such that corner or hinge modes remain protected. In the present context of HOTSC, interestingly, the number and localization profiles of the emergent Floquet MZMs change compared to the static HOTSC scenario. These new Floquet HOTSC (FHOTSC) phases can also host finite-energy anomalous Floquet Majorana modes that require a separate mathematical prescription for characterization. In this review,
we discuss case by case study of the above mentioned issues starting from Floquet FOTSC (FFOTSC) phase in 1D to Floquet third-order topological superconductor (FTOTSC) phase in 3D along with the development of their appropriate topological invariant. 


The remainder of this article is organized as follows. In Sec.~\ref{sec:Floquet_primer}, we briefly discuss about the basics of Floquet theory and periodically driven 1D Kitaev chain as an example of Floquet TSC. Then we introduce proximity induced static TSCs based on {Rashba} nanowire in Sec.~\ref{sec:tsc_static}. In Sec.~\ref{Sec:Floquet_first_order}, we briefly review the emergence of Floquet first order TSCs in 1D platforms, while periodically driven higher order TSCs 
have been discussed in Sec.~\ref{Sec:2D} and Sec.~\ref{Sec:3D} for 2D and 3D, respectively. Sec.~\ref{Sec:discussion_outlook} contains an over all discussion 
and future outlook on the research area. Finally, we summarize and conclude our article in Sec.~\ref{Sec:summary_conclusion}.



\section{Floquet Theory and Floquet FOTSC: A primer} ~\label{sec:Floquet_primer}
In this section, we demonstrate the periodic dynamics of a quantum system using the Floquet theory. The main advantage of this formalism is that it can numerically or analytically yield the quasi-states and quasi-energies, which are the main ingredients to investigate any observable for a time-periodic problem. Another advantage is that Floquet theory can probe the entire frequency domain to characterize the behavior of an observable. It is noteworthy that Floquet theory is also connected to the second-order periodic differential equation, known as the Mathieu equation~\cite{richards2012analysis}. We conclude the section by demonstrating an example of Floquet TSC based on the driven Kitaev model~\cite{kitaev2001unpaired}.

\subsection{Floquet theory}
Floquet theory provides a powerful framework for analyzing the dynamics of a time-periodic Hamiltonian $H(t)=H(t+T)$ where 
$T$ denotes the time period of evolution. It allows the time-dependent Schr\"odinger equation to be reformulated as a time-independent eigenvalue problem for the Floquet Hamiltonian when the system is examined stroboscopically, \ie at the end of each full driving period~\cite{ShirleyPRB1965,GrifoniDrivenquantum1998}.
A discrete time translation operator ${\mathcal T}_{T}$, implementing the shift $t \rightarrow (t+T)$, commutes with the time periodic Hamiltonian $H$ as, $[{\mathcal T}_{T}, H]=0$; such relation is valid only at stroboscopic time scales $t=nT$, where $n$ is an integer. This directly implies the existence of a common set of eigenvectors for ${\mathcal T}_{T}$ and the time-dependent Schr\"odinger 
equation. Hence, the time-translation operator applied to  the state $|\Psi_{\alpha}(t)\rangle$, a solution of time-dependent 
Schr\"odinger equation, 
translates  that state to $t \to t+T$ and simultaneously ${\mathcal T}_{T}$ satisfies an eigenvalue equation with 
$|\Psi_{\alpha}(t)\rangle$
as an eigenstate:
\beq
{\mathcal T}_{T} |\Psi_{\alpha}(t) \rangle=|\Psi_{\alpha}(t+T)\rangle=\lambda_{\alpha} |\Psi_{\alpha} (t)\rangle\ ,
\label{eq_ft1}
\eeq
with $\alpha$ being an appropriate quantum number. One can note from Eq. (\ref{eq_ft1}) that $\lambda_{\alpha}$ is just a phase  factor $\lambda_{\alpha} = \exp(-i \theta_{\alpha})$. Also, since ${\mathcal T}_{T}$ is a unitary operator with unimodular eigenvalues, the Floquet solution of the time-dependent Schr\"odinger equation is given by 
\beq
|\Psi_{\alpha}(t) \rangle = \exp(-i \mu_{\alpha} t) |\Phi_{\alpha}(t)\rangle\ ,
\label{eq_ft2}
\eeq
where, the Floquet modes $|\Phi_{\alpha}(t)\rangle$ carry the same periodicity as that of the Hamiltonian, \ie  $|\Phi_{\alpha}(t+\tau)\rangle=|\Phi_{\alpha}(t)\rangle$ and $\mu_{\alpha}=\theta_{\alpha}/T$ are the corresponding Floquet quasi-energies. One can note that $\mu_{\alpha}$s, defined within a temporal BZ extending from $-\pi/T$ to $\pi/T$, are referred to as quasi-energies. The quasi-energy spectrum is periodic and repeats identically across all higher temporal Brillouin zones. One can see from Eq.~(\ref{eq_ft2}) that the same solution is obtained for each $\theta_{\alpha}\to\theta_{\alpha}+2\pi T$~\cite{ShirleyPRB1965}. It is therefore obvious that the solutions given in Eq.~(\ref{eq_ft2}) satisfy the condition presented in Eq.~(\ref{eq_ft1}). Referring to Eq.~(\ref{eq_ft2}), one can consider the Floquet theory as the temporal counterpart of the Bloch theorem where the wave-function of the Schr\"odinger equation has two parts, one of which is a spatially periodic and the other is simply a plane wave. In the Floquet context, the    Floquet modes $|\Phi_{\alpha}(t)\rangle$ are periodic and the plane wave part $\exp(-i \mu_{\alpha} t)$ involves the Floquet quasi-energy. 

One can define a unitary time evolution operator ${U}(t,0)$ in the basis of the Floquet modes $|\Phi_{\alpha}(t)\rangle$ {up to} an arbitrary time $t$ and given by
\beq
{U}(t,0)=\sum_{\alpha} e^{-i\mu_{\alpha}t} |\Phi_{\alpha}(t)\rangle \langle \Phi_{\alpha}(0)|\ .
\label{eq_ft3}
\eeq
Similarly, an arbitrary state $|\Psi(0)\rangle$ can be written in terms of Floquet basis states as 
$|\Psi(0)\rangle= \sum_{\alpha} r_{\alpha} |\Phi_{\alpha}(0)\rangle$, where $r_{\alpha}= \langle\Phi_{\alpha}(0)|\Psi(0)\rangle$ 
is the overlap of the Floquet modes and the initial 
wave-function. Combining the above two relations, we obtain the time-evolved wave-function at $t\neq 0$ as
\beq
|\Psi (t)\rangle={U}(t,0)|\Psi(0)\rangle=\sum_{\alpha} r_{\alpha} e^{-i\mu_{\alpha}t} |\Phi_{\alpha}(t)\rangle\ ,
\label{eq_ft4}
\eeq
Therefore, unitary time evolution operator $\hat{U}(t,0)$ at time $t=T$ using Eq.~(\ref{eq_ft3}) 
and periodicity of Floquet state \ie $ |\Phi_{\alpha}(0)\rangle= |\Phi_{\alpha}(T)\rangle$, can be written in the following way:
\bea
 {U}_F(T,0)&=&\sum_{\alpha} e^{-i\mu_{\alpha}T} |\Phi_{\alpha}(T)\rangle\langle \Phi_{\alpha}(0)| \non \\
 &=& {\sum_{\alpha} e^{-i\mu_{\alpha}T} |\Phi_{\alpha}(0)\rangle \langle \Phi_{\alpha}(0)|}=e^{-iH_F T}\ , \non
\label{eq_ft5}
 \eea
where $H_F$ is the corresponding Floquet Hamiltonian; ${U}_F(T,0)$ is written in terms of the spectral decomposition of the Floquet Hamiltonian.
We hereafter refer to ${U}_F(T,0)=F(T)$ as the Floquet operator. Similarly, one can easily calculate the stroboscopic
wave-function at $t=nT$ from Eq.~(\ref{eq_ft4}) given by
\bea
|\Psi ({nT})\rangle &=&{ U}_F({nT},0)|\Psi(0)\rangle=  F(T)^n|\Psi(0)\rangle \non \\
&=&\sum_{\alpha} r_{\alpha} e^{-i\mu_{\alpha}{nT}} 
|\Phi_{\alpha}(0)\rangle\ .
\label{eq_ft6}
\eea

Note that, using Eq.~(\ref{eq_ft6}) the time-dependent problem has been reduced to a time-independent problem when viewed stroboscopically. One can further
derive a wave-function for any arbitrary time $t=\delta t+nT$ with $0<\delta t<T$ by suitably choosing the time evolution operator as
${U}(t)={U}(\delta t,0) [F(T,0)]^{n}$. One can explore the behavior of any physical observable in the asymptotic limit of $n \to \infty$. 
The extensive thermodynamic quantities reach a steady state value~\cite{Russomanno12}. This value can be estimated from a stroboscopic decohered
density matrix in large time $\hat{\rho}_{{\rm dec}}=\sum_{\alpha} |r_{\alpha}|^2 |\Phi_{\alpha}(0)\rangle \langle \Phi_{\alpha}(0)|$. For example, stroboscopic
excess energy in large time $W(n\to \infty)$ is given by $ W(n \to \infty)=\sum_{\alpha} |r_{\alpha}|^2 H_{\alpha}$ where  $H_{\alpha}=
\langle \Phi_{\alpha}|H(nT)|\Phi_{\alpha}\rangle$.

Let us revisit the Floquet theory for a generic situation by considering higher order Floquet states with quasi-energy $\mu_{\alpha}+n\Omega$
beyond the range $-\pi/T< \mu_{\alpha}< \pi/T$~\cite{ShirleyPRB1965,GrifoniDrivenquantum1998}. The Floquet Hamiltonian can be identified as $H_F=H(t)-i I~d/dt$, which satisfies an eigenvalue equation with the quasi-states $|\Phi_\alpha\rangle$; $H_F|\Phi_\alpha\rangle= \mu_\alpha |\Phi_\alpha\rangle$, where $\mu_\alpha$ are the quasi-energy of the sub-bands obtained from the Floquet operator. For the Hermitian operator $H_F$,  one can introduce an extended Hilbert space 
${\mathbb L}={\mathbb R}\otimes {\mathbb T}$,  where ${\mathbb R}$ represents the Hilbert space of square integrable
functions $|\alpha(r)\rangle$ on configuration space and ${\mathbb T}$ is spanned by the complete orthonormal set of functions
$\exp(in\Omega t)$ (with $n=0,~\pm 1,~\pm2,~\cdots$) periodic in time  with the period $\Omega=2\pi/T$.  Hence the 
eigenvectors $|\Phi_{\alpha}(r,t)\rangle$ of $H_F$ obey the orthonormality condition in the composite Hilbert space 
${\mathbb R}\otimes {\mathbb T}$ given by
\bea
\langle \Phi_{\alpha'}|\Phi_{\beta'}\rangle&=&{1 \over T}\int_0^T dt \int_{-\infty}^{\infty} dr \Phi_{\alpha'}^*(r,t)\Phi_{\beta'}^*(r,t) \non\\
&=&\delta_{\alpha'\beta'}=\delta_{\alpha,\beta}~\delta_{n,m}\ .
\label{eq_chs}
\eea
 
The wave-function satisfying $|\Psi_{\alpha}\rangle$ given in Eq.~(\ref{eq_ft2})
remains invariant under the transformation $\mu_{\alpha} \to \mu'_{\alpha}=\mu_{\alpha}+n\Omega$ and 
$|\Phi_\alpha(t)\rangle \to |\Phi'_\alpha(t)\rangle=\exp(in\Omega t)|\Phi_\alpha(t)\rangle$. This corresponds to the fact that the Floquet states are physically equivalent if their quasi-energies differ by $n\Omega$. The  states  $|\Psi_{\alpha}(r,t)\rangle$
can be expanded in the Fourier series, 
\beq
|\Psi_{\alpha}(r,t)\rangle=\exp(-i\mu_{\alpha}t)\sum_{n=-\infty}^{\infty} |A_{\alpha}^{(n)}(r)\rangle\exp(-in\Omega t)\ .
\label{eq_fm1}
\eeq

Therefore, the state $|\Psi_{\alpha}(r,t)\rangle$ can be thought of as a superposition of stationary states $|A_{\alpha}^{(n)}(r)\rangle$ 
with energies equal to $\mu_{\alpha}+n\Omega$. One can find the correlation between the quasi-states and the stationary states such that $|\Phi_{\alpha} (t)\rangle \equiv  \sum_{n=-\infty}^{\infty} |A_{\alpha}^{(n)}(r)\rangle\exp(-in\Omega t)$. The stationary states $|A_{\alpha}^{(n)}(r)\rangle$ are associated with $n^{\rm{th}}$-photon sector while quasi-states, associated with a given quasi-energy, are considered as a superposition of stationary states emerging 
from different photon sectors. Furthermore, $|A_{\alpha}^{(n)}(r)\rangle$ can be expanded in the basis
$|\beta(r)\rangle$ associated with the  unperturbed Hamiltonian $H_0$. Note that $|\beta(r)\rangle$ consists of degrees of freedom obtained from $H_0$. 
Using the Schr\"odinger equation and the orthogonality  relation that leads to
$\langle\Psi_{\beta}(r,t)|H|\Psi_{\alpha}(r,t)\rangle=0$, one can obtain the following system of coupled equations
\beq
\sum_n \sum_{\beta}\bigl[ \langle \alpha | H_{m-n}|\beta\rangle -(\mu_{\alpha}+m\Omega)\delta_{m,n}\delta_{\alpha,\beta}\bigr]=0\ ,
\label{eq_fmr1}
\eeq 
where, $ H_{(n)}(r)=(1/T)\int_0^T H(r,t)\exp(in\Omega t)dt$ and
$\mu_{\alpha}\delta_{\alpha \beta}=\langle \alpha |H|\beta\rangle$.
As a result, the matrix element of time-independent
Floquet Hamiltonian $H_{\rm F}$ using Eq. (\ref{eq_fmr1}) is given by
\beq
\langle \alpha n|H_{\rm F}|\beta m\rangle=  H^{\alpha \beta}_{(n-m)}+m\Omega \delta_{m,n}\delta_{\alpha,\beta},
\label{eq_fmr2}
\eeq
where, $|\alpha n \rangle=|\alpha \rangle \otimes |n\rangle$, $|\alpha \rangle$ is the system index and 
$|n\rangle$ is the Fourier modes ($n=0,\pm 1, \pm 2,~\cdots$) \cite{MikamiBW2016,ShirleyPRB1965,SambeFloquet1973}. 
Subsequently, we construct the infinite-dimensional, time-independent Floquet Hamiltonian within the frequency-zone framework as~\cite{Eckardt2015,ShirleyPRB1965,SambeFloquet1973}
\begin{widetext}
	\begin{equation}\label{FloquetHam}
		H_{\rm F}=
		\begin{pmatrix}
			\ddots& & &  \vdots & & & \udots\\
			&H_{-2} & H_{-1} & H_0  -2 \hbar \Omega &  H_{1} &  H_{2} \\
			& & H_{-2} & H_{-1} & H_0 -\hbar \Omega &  H_{1} &  H_{2} &    \\
			& & & H_{-2} & H_{-1} & H_0 & H_{1} & H_{2}\\
			& & &  & H_{-2} & H_{-1} & H_0 + \hbar \Omega &  H_{1} &  H_{2}  \\
			& & & & & H_{-2} & H_{-1} & H_0 +2 \hbar \Omega &  H_{1} &  H_{2} \\
			&&  & &  \udots & & &\vdots  & & \hspace{1cm}\ddots\\
		\end{pmatrix} \ .
	\end{equation}
\end{widetext}
Here, the structure of $H_{\rm F}$ resembles that of a quantum system coupled to a photon-like bath,
where $m$ is the photon number~\cite{Eckardt2015}. The off-diagonal $H_m$'s $(m \ne 0)$ represent an $m$-photon process acting as the coupling between different photon sectors.

Note that, Eq.~\eqref{FloquetHam} represents an infinite dimensional matrix. Thus, to utilize the frequency-zone Hamiltonian, one must adopt an approximation scheme. One of the commonly used limits is the high-frequency limit, in which the periodic drive frequency is much larger than the system bandwidth, i.e., $\Omega \gg$ bandwidth of $H_0$. In this case, one can obtain an analytically tractable form of the Floquet Hamiltonian, employing perturbation theories like Brillouin-Wigner~(BW) perturbation theory~\cite{MikamiBW2016}, van Vleck perturbation theory~\cite{Eckardt2015, MikamiBW2016}, and Floquet-Magnus perturbation theory~\cite{casas2001floquet, blanes2009magnus, Eckardt2015, MikamiBW2016}. However, these formalisms break down for intermediate and low frequencies~\cite{Piskunow2014, Usaj2014}. In such a case, one can use a finite-dimensional  $H_{\rm F}$, by truncating it to a finite number of photon sectors with the number of photon sectors $n$ depending upon the bandwidth of the system's Hamiltonian and the driving frequency~\cite{Piskunow2014, Usaj2014}. Furthermore, there exist other perturbation theories, including high-amplitude perturbation theory or Floquet perturbation theory, which are applicable given the fact that the amplitude of the driving field is much larger than the energy scales of the static Hamiltonian~\cite{Sen2021, Mukherjee2020, Sengupta2022Weyl}.

In the later part of the review, we will discusses three types of periodic drives: step (square) drive, kick drive, and sinusoidal drive. For the step and kick drives, one finds that the Fourier components decay as $H_n \sim 1/n$ for $|n|\ge 1$, whereas for a sinusoidal drive $H_n=0$ for $|n|\ge 2$. In order to have converged results in the low-frequency regime, one has to consider a substantially high number of photon sectors. On the other hand,   in the high-frequency regime one may employ a high-frequency expansions as discussed above, where the contributions from higher photon sectors are suppressed as $H_n \sim 1/(n\Omega)$ for $n\neq 0$. Therefore, in the high-frequency regime, the number of photon sectors can safely be considered less than that of low- or intermediate-frequency regime without loss of generality. 

Note that $0$- and $\pi$-modes emerge from bulk gap closings at quasienergies $0$ and $\Omega/2$, respectively. The corresponding quasienergy gaps depend on the parameters of the driven system and, in general, are independent of each other. Consequently, the $0$ and $\pi$ gaps can possess distinct topological characteristics and must be characterized independently by their respective topological invariants. The above Floquet formalism applies strictly to periodically driven systems and thus cannot be generically adopted for aperiodic/quasiperiodic systems.


\subsection{An example on Floquet dynamics: Driven topological superconductor with $p$-wave pairing
} \label{subsec:KitaevDriven}
\begin{figure*}
\includegraphics[width=6.8in]{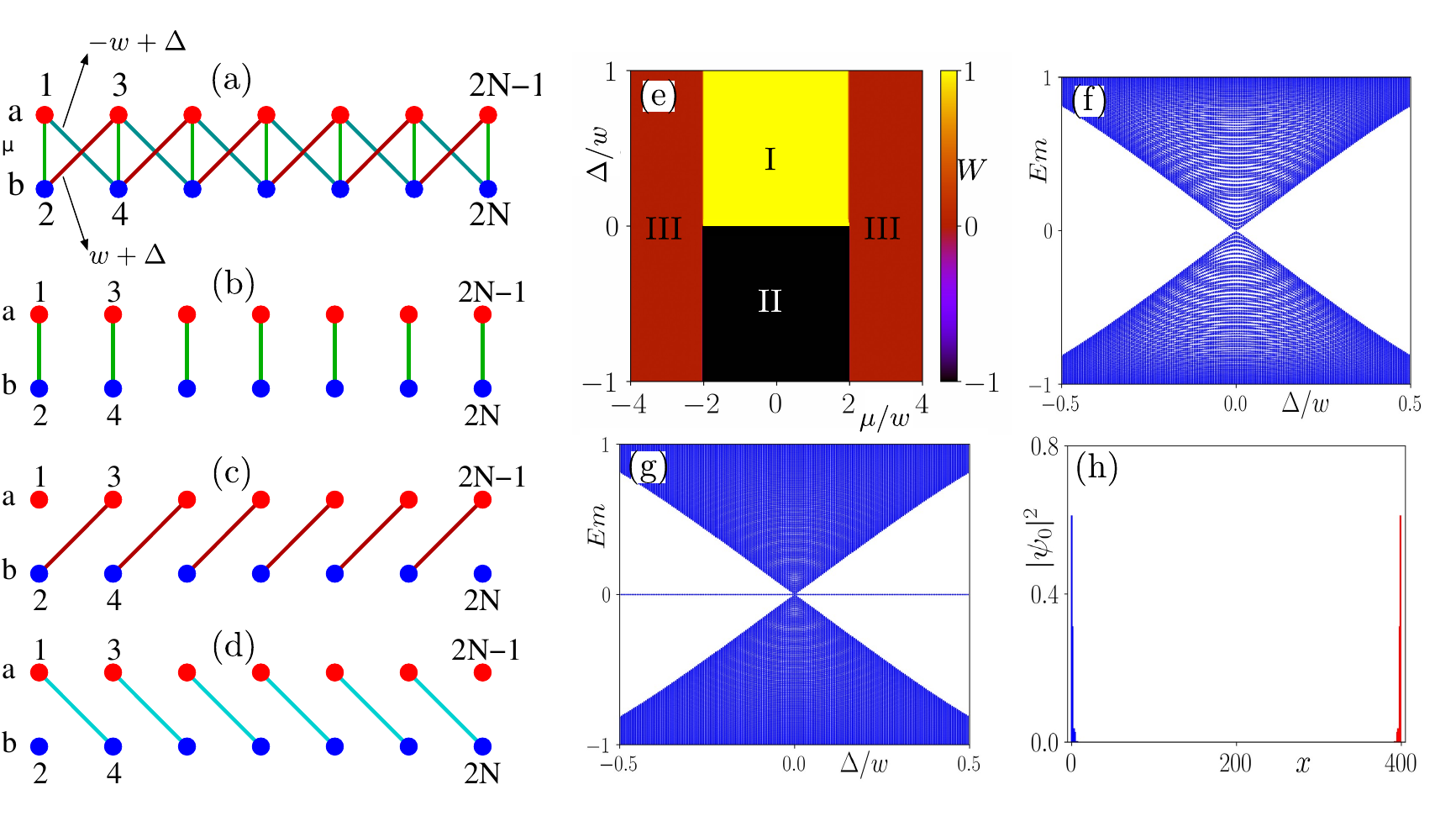}
\caption[]{The model and its spectral properties are shown. (a) Schematic representation of a 1D $p$-wave Kitaev chain is illustrated in the Majorana basis. Panel (b) represents the same with $w=\Delta=0$ and $\mu\neq 0$. No unpaired Majoranas are present at the chain's ends. Panel (c)~[(d)] stands for the same with $\mu=0$ and $w=\Delta$~[$w=-\Delta$] with $a_1$ and $b_N$~[$b_1$ and $a_N$] unpaired Majoranas present at the ends. (e)  Winding number is depicted in $\mu$-$\Delta$ plane. Region I and II denote topological regimes with non-trivial winding $W=1$ and $W=-1$ , respectively, while region III represents a trivial phase with null winding $W=0$. Choosing $\mu=1$~(topological regime), energy eigenvalue spectra of real space Hamiltonian~(Eq.~(\ref{eq:hamr_pwave})) are depicted employing periodic boundary condition~(PBC) and OBC in panels (f) and (g), respectively. Emergence of Majorana zero modes~(MZMs) for $\Delta\neq 0$ is evident in the OBC spectrum. (h) The probability of wave functions corresponding to those MZMs is depicted with respect to the lattice sites $x$. They are localized at the ends of the chain. The remaining model parameters are chosen to be $w = 1.0, \mu=1.0, \Delta = 0.6$.} \label{fig:p-wave} 
\end{figure*}
Having discussed the basics of Floquet theory, we here discuss an example of 1D Floquet
$p$-wave superconductor where the chemical potential is periodically kicked. Before getting into the drive schemes, the Hamiltonian, written in the particle-hole subspace spanned by basis $\psi_k= (c_k, c^{\dagger}_{-k})$, known as Bogoliubov-de-Gennes (BdG) basis, is given by\cite{kitaev2001unpaired} 
\begin{equation}
{\cal H}_{k}=  2 \Delta \sin k \sigma_{y} +(2 w  \cos k +\mu) \sigma_{z}  =\vect{h}_k \cdot \vect{\sigma}\ ,
\label{eq:hamk_pwave}
\end{equation}
where $w$, $\mu$, and  $\Delta$ denote the nearest neighbor hopping amplitude,  chemical potential, and superconducting gap, respectively, with $\vect{h}_k=\{0,h^y_k,h^z_k\}$ and $\vect{\sigma}=\{\sigma_x,\sigma_y,\sigma_z\}$. Note that,
the nearest-neighbor spin-triplet $p$-wave gap function is represented by $\Delta \sin k$ term.  
The energy of the Hamiltonian ${\cal H}_{k}$ is found as  $h^{\pm}_k=\pm \sqrt{(h^y_k)^2 + (h^z_k)^2} $ while the phase boundaries are given by  $\mu =\pm 2 w $ and doesn't depend on $\Delta$. This model preserves chiral symmetry ${\mathcal C}=\sigma_x$, and particle-hole symmetry ${\mathcal P}=\sigma_x {\cal K}$ such that ${\mathcal C} {\cal H}_{k} {\mathcal C}^{-1}= -{\cal H}_{k}$ and  ${\mathcal P} {\cal H}_{k} {\mathcal P}^{-1}= -{\cal H}_{-k}$, respectively, ${\cal K}$ denotes the complex conjugation.  The model breaks TRS as ${\mathcal T} {\cal H}_{k} {\mathcal T}^{-1} \ne {\cal H}_{-k}$ with ${\mathcal T}=i \sigma_y {\cal K}$. 
One can cast the model 
in real space with a fermionic basis as follows
\begin{equation}
{\cal H}=\sum_{n=1}^{N-1} w(c^\dagger_n c_{n+1}+ {\rm h.c}) + \Delta (c_n c_{n+1}+ {\rm h.c}) + \sum_{n=1}^{N}\mu \Big(c^\dagger_n c_{n}-\frac{1}{2} \Big)\ ,
\label{eq:hamr_pwave}
\end{equation}
Here, the fermionic operators satisfy usual anticommutation relations $\{c_n,c_m \}=0$ and $\{c_n,c^{\dagger}_m \}=\delta_{mn}$. This fermionic real-space version of the model closely resembles the transverse $xy$ model, where the h.c terms appear with a minus sign; however, the momentum space version  
of the $p$-wave is an exact look-alike of the earlier model, while $\mu$, $w$, and $\Delta$ denote the transverse field, collective spin-spin interaction terms, and relative difference spin-spin interaction terms known as anisotropy parameter, respectively~\cite{dutta2015quantum}. 

In order to look for the topological edge mode, namely the Majorana zero modes (MZMs), one can adopt the basis of  Majorana operators where one fermionic site is composed of two types of Majorana fermions $a,b$ in the following way $c_n=(a_n + i b_n)/2$ and $c^{\dagger}_n=(a_n - i b_n)/2$. Importantly, $a^{\dagger}_n=a_n$ and $b^{\dagger}_n=b_n$ such that $\{a_n,a_m \}=\{a_n,a_m \}=2 \delta_{mn}$ and $\{a_n,b_m \}=0$.  In terms of the Majorana operators, the Majorana chain then takes the form 
\begin{equation}
{\cal H}=\frac{-i}{2}\left[\sum_{n=1}^{N-1} 
(w+\Delta)a_n b_{n+1} + (w-\Delta)b_n a_{n+1} + \sum_{n=1}^{N}  \mu a_n b_n\right]\ .
\label{eq:hammf_pwave}
\end{equation}

We now exemplify the phases of the model by having simple parameter choices. (i) The topological phase I, hosting unpaired $a_1$ on the left edge and $b_N$ on the right edge, can be understood by considering $\Delta=w$, (ii) the topological phase II, hosting unpaired $b_1$ on the left edge and $a_N$ on the right edge, can be understood by considering $\Delta=-w$, (iii) The trivial phase without any unpaired edge modes, can be understood by considering $\Delta=w=0$ and $\mu\ne 0$.  The topologically (trivially) gapped phases occur for $-2 w <\mu <2 w $ ($|\mu|>|2 w|$) while topological phases I and II reside on $\Delta > 0$ and $\Delta < 0$ regions, respectively, as shown in Fig.~\ref{fig:p-wave}. The bulk boundary correspondence is established when the chiral symmetry protected bulk topological index namely, the winding number $W=\int_{-\pi}^{\pi} (dk/2\pi) d \phi_k/dk$, $\phi_k={\rm arctan}(h^z_k/h^y_k)$ takes $+1$ and $-1$ values for the topological phases I and II, respectively, and they match with the number of Majorana zero modes on a given edge $N_a-N_b$, $N_{a(b)}$ represents the number of unpaired $a$- ($b$)-type of Majorana on that edge. With the above knowledge in mind, the important question is whether topological Majorana zero modes can be engineered by a Floquet drive even when the underlying static limit is trivial.

\begin{figure}[htb]
\includegraphics[width=3.4in]{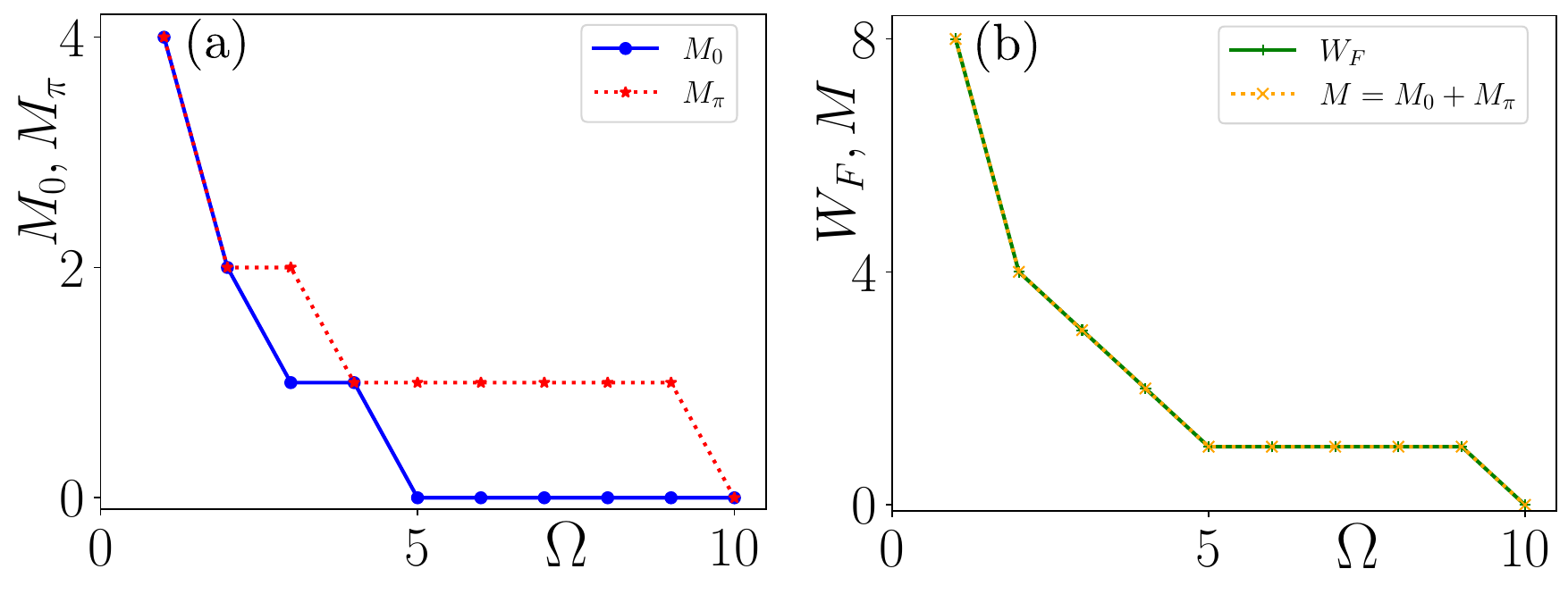}
\caption[]{In panel (a), the number of zero and $\pi$-MEMs $M_{0}$ and $M_{\pi}$ are depicted as a function of driving frequency $\Omega$ (with unit step) for the driven Kitaev chain. Panel (b) represents a comparison between the total number of Floquet Majorana modes $M=M_0+M_{\pi}$ at each end of a 400-site system and the winding number $W_F$ as a function of $\Omega$. All other model parameters are chosen to be $(w,\mu_0,\mu_1,\Delta)$=$(1.0,2.5,0.2,-1.0)$ which represent topologically trivial static system.}
 \label{flo5} \end{figure}


It has been shown that Floquet Majorana modes are generated in a  periodically driven $p$-wave superconductor where the chemical potential follows the following protocol \cite{Thakurathi2013}
\begin{equation}
    \mu(t)= \mu_0 + \mu_1 \sum_{n} \delta(t-nT)\ .
    \label{eq:mut_Floquet}
\end{equation}
Importantly,  $|\mu_0|>|2 w|$, $|\mu_0| \pm |\mu_1|>|2 w|$ such that the underlying static model always resides in the trivial phase.  The full period Floquet  operator consists of two  parts, an evolution with a static Hamiltonian 
${\cal H}(\mu_0)$ having   chemical potential $\mu_0$ followed by a kick-mediated part. The symmetrized form can be explicitly written as  $U(T)= \exp(-i \mu_1 \sigma_z/2) \exp(-i {\cal H}(\mu_0) T) \exp(-i \mu_1 \sigma_z/2)$ where free evolution is sandwiched between kicks at the beginning and the end. The Floquet eigenvalues $\exp(i \theta)$ are found to be $+1$ and $-1$, corresponding to the fact that $\theta=0$ and $\theta=\pi$, respectively.  Therefore, regular  MZMs and anomalous Majorana $\pi$ modes (MPMs)  both appear in the driven system. The MPMs continue surviving for higher values of drive frequency $\Omega=2\pi/T$  than the MZMs, although they both vanish when $\Omega$ increases further, see Fig. \ref{flo5}.

Employing open boundary condition~(OBC), one can diagonalize the Floquet operator $U(T)$ to obtain quasi-energy spectra. From there one can count number of $0$- modes $M_{0}$ and $\pi$-modes $M_{\pi}$. In Fig.~\ref{flo5}(a), these $M_0$ and $M_{\pi}$ are illustrated as a function of driving frequency $\Omega$. Interestingly, one can engineer multiple  Majorana end modes (MEMs), i.e., MZMs located at the ends of the 1D chain, in the low-frequency limit, while a scarcity of these modes is observed in the high-frequency limit. These boundary modes are further confirmed by the bulk-boundary correspondence with a Floquet winding number for which we need to use the bulk Hamiltonian. The general form of  the bulk effective Floquet Hamiltonian is found to be $U_k(T)=\exp(-i \mu_1 \sigma_z/2) \exp(-i {\cal H}_k(\mu_0) T) \exp(-i \mu_1 \sigma_z/2)=\exp(-i {\cal H}_k^{\rm eff} T)$ with ${\cal H}_k^{\rm eff}= \tilde{h}^y_k \sigma_y + \tilde{h}^z_k \sigma_z$ where 
\begin{eqnarray}
    \tilde{h}^y_k &=& \frac{\tilde{h}_0}{\sin (\tilde{h}_0 T)} \bigg( \frac{h^y_k \sin ( h^{+}_k T )}{h^{+}_k} \bigg)\ , \nonumber \\
    \tilde{h}^z_k &=& \frac{\tilde{h}_0}{\sin (\tilde{h}_0 T)} \bigg( \frac{h^z_k \cos \mu_1 \sin ( h^{+}_k T )}{h^{+}_k} + \sin \mu_1 \cos ( h^{+}_k T ) \bigg)\ , \nonumber \\
    \tilde{h}_0 &=& \frac{1}{T} {\rm Arccos}\bigg( \cos \mu_1 \cos ( h^{+}_k T ) - \frac{h^z_k \sin \mu_1 \sin ( h^{+}_k T ) }{h^{+}_k}   \bigg)\ . \nonumber \\
    \label{eq:heff_Floquet}
\end{eqnarray}
Note that the driven system continues satisfying particle-hole and chiral symmetries.
Therefore, the Floquet winding number can be written as $W_F=\int_{-\pi}^{\pi} (dk/2\pi) d \phi^F_k/dk$, $\phi^F_k={\rm arctan}(\tilde{h}^z_k/\tilde{h}^y_k)$. This dynamical winding number $W_{F}$ and total number of Majorana modes $M=M_{0}+M_{\pi}$ are depicted in Fig.~\ref{flo5}(b) in terms of $\Omega$ for comparison.  An exact match between $M$ and $W_F$ thus confirms the bulk-boundary correspondence in the driven Kitaev chain. We note that the symmetrize form of the Floquet operator allows us to obtain two-component  ${\cal H}_k^{\rm eff}$ while the conventional form of the Floquet operator yields three-component  ${\cal H}_k^{\rm eff}$ having $\tilde{h}^x_k$, $\tilde{h}^y_k$, and $\tilde{h}^z_k$. In the latter case, one has to follow the flattened Hamiltonian approach to compute the Floquet winding number~\cite{Mondal23_FNW}. Another interesting point is to mention that $W_F$ provides total number $M=M_{0}+M_{\pi}$ while individual information of $M_0$ and $M_{\pi}$ is missing. To be precise,  $M_0$ and $M_{\pi}$ can be connected with $U_k(T)=\exp(0)=1$  and $U_k(T)=\exp(i \pi)=-1$, respectively.  For more details, one can follow Ref.~\cite{Mondal23_FNW}. In the following part of the review, we will demonstrate the anomalous $\pi$-Majorana mode more extensively.

\begin{figure}[htb]
\includegraphics[width=3.4in]{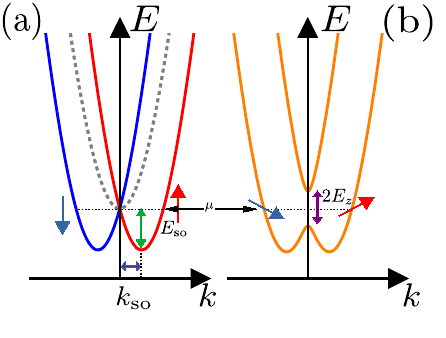}
\caption[]{(a) The bulk energy dispersion of a 1D Rashba nanowire is depicted when placed in closed proximity 
to an $s$-wave superconductor without Zeeman field \cite{Aguado2017Majorana}. Rashba SOC splits spin/ chiral bands towards $\pm$ k. However, spin-degeneracy persists at $k=0$. Panel (b) represents the same with a Zeeman field, resulting in lifted degeneracy. }  \label{fig:p_wave_dispersion} 
\end{figure}

\section{Topological superconductivity in proximity-induced static systems} 
\label{sec:tsc_static}

In this section, we briefly discuss some models for the static FOTSC and HOTSC hosting Majorana boundary modes. We also discuss the topological invariant for the HOTSC.

\subsection{FOTSC}
The spinless $p$-wave TSC~\cite{Kitaev_2001}, as mentioned in Eq.~\eqref{eq:hamk_pwave}, is not readily available in nature and hence has not been found experimentally so far. Therefore, its realization has yet to be examined; however, from a theoretical perspective, it has attracted significant attention for its possible connection to quantum computation and quantum information~\cite{hasan2010colloquium,Alicea_2012,das2012zero,nayak08}. 
Nevertheless, there exist alternative routes to effectively realize the Kitaev model relying on heterostructures where a semiconducting Rashba nanowire with substantial spin-orbit coupling (SOC) is placed in close proximity to a common bulk $s$-wave superconductor, such that there appears a $s$-wave pairing gap in the nanowire via the proximity effect~\cite{Fu2008, Sau2010, Lutchyn10, Hughes2010, Oreg2010}. In the presence of a perpendicular Zeeman field, one can obtain a topological region within which MZMs exist. Note that the spin degrees of freedom associated with the magnetic field and SOC are orthogonal to each other, i.e., $\sigma_\alpha\cdot \sigma_\beta \equiv \delta_{\alpha \beta}$.  The tight-binding hopping yields parabolic dispersions around the origin $k_x=0$ while the band energy reads $\epsilon(k)=-2t \cos k_x $. The Rashba SOC $\vect{g}_k \cdot \vect{\sigma}$ with $\vect{g}_k=2 \lambda(0,-\sin k_x, 0)$, breaks inversion symmetry, and thus lifts the spin-degeneracy at all points $k_x\ne 0$ except at the origin $k_x=0$. Once a perpendicular magnetic field $E_z \sigma_z$ is applied, it further breaks the time-reversal symmetry and the degeneracy at the origin is lifted, and a gap is created (see Fig.~\ref{fig:p_wave_dispersion}). The gap in the bulk energy dispersion is ensured by the fact that $\sigma_y$, associated with SOC, and $\sigma_z$, associated with the magnetic field, do not commute, leading to the topological aspect of the system. The Rashba nanowire system belongs to the class D of the Altland–Zirnbauer~(AZ) 10-fold symmetry class~\cite{AltlandPRB1997}. Thus, the Majorana modes are protected by the particle-hole symmetry $\mathcal{P}$. However, the system also further satisfies a chiral symmetry ${\mathcal C}$, which allows us to construct the winding number~\cite{Mondal23_FNW}.

The proximity-induced superconducting gap $\Delta$ deals with particle-hole degrees of freedom, 
paving the particle-hole flavor to the gap. As a result, under suitable conditions, when the chemical potential $\mu$ is kept inside the bulk gap, such that the topological MZMs appear 
for $E^2_z >\Delta^2 +\mu^2 $. It is important to note that an effective spin-singlet intra-band $p_x$ pairing is generated within the gap~\cite{Sato2009,Sato2010}. This can be easily understood from a dual Hamiltonian ${\cal H}_D(k)=D {\cal H}(k) D^{\dagger}$ written the basis of SOC, where ${\cal H}(k)$ denotes the BdG Hamiltonian. Using the basis $(c_{k\uparrow},c_{k\downarrow},c^{\dagger}_{-k\uparrow},c^{\dagger}_{-k\downarrow})$, ${\cal H}(k)$ can be written as 
\begin{align}
\small
{\cal H}(k)\!=\!\!
\begin{pmatrix}
\epsilon(k)-\mu -E_z \sigma_z + \vect{g}_k\cdot \vect{\sigma}  &\!\!\!\!\!\!\! i \sigma_y \Delta \\
-i\sigma_y \Delta^* &\!\!\!\!\!\!\! \mu-\epsilon(k) + E_z \sigma_z + \vect{g}_k\cdot \vect{\sigma}^*   \\
\nonumber 
\end{pmatrix},\\
\label{eq:bdg_1d}
\end{align}
while the unitary operator $D$ is given by 
\begin{eqnarray}
D=\begin{pmatrix}
1  & i \sigma_y  \\
i\sigma_y  & 1   \\
\end{pmatrix}.
\label{eq:unit_1d}
\end{eqnarray}
This results in the following dual Hamiltonian 
\begin{align}
\small
{\cal H}_D(k)\!\!=\!\!
\begin{pmatrix}
\Delta -E_z \sigma_z  &\!\!\!\!\!\!\!\!\!\!\!\!\!\! -i (\epsilon(k)-\mu)\sigma_y -i (\vect{g}_k\cdot \vect{\sigma}) \sigma_y  \\
i(\epsilon(k)-\mu)\sigma_y + i (\vect{g}_k\cdot  \vect{\sigma}) \sigma_y   & -\Delta + E_z \sigma_z    \\
\nonumber 
\end{pmatrix},\\
\label{eq:bdg_dual_1d}
\end{align}
The $p$-wave nature of the effective gap from the dual Hamiltonian is clearly realizable in 
$-i (\vect{g}_k\cdot \vect{\sigma}) \sigma_y$ and $i (\vect{g}_k\cdot \vect{\sigma}) \sigma_y$
appearing on the off-diagonal part of Eq.~(\ref{eq:bdg_dual_1d}). These parts simplify to $\pm i k_x \sigma_0$, which directly resembles the spin-triplet $p_x$ pairing. However, the superconducting gap amplitude is no longer given by $\Delta$, but rather it becomes the SOC strength $\lambda$.  The diagonal parts cause the gap in the spectrum, while the off-diagonal parts hint towards the non-trivial nature of the gap.   

The above concept of proximity-induced 1D TSC is applicable to 2D also, where an $s$-wave superconductor is placed in close proximity to the surface states of a three-dimensional (3D) strong topological insulator and a magnetic insulator. In this case, the SOC coupling $\vect{g}_k \cdot \vect{\sigma}$ with  $\vect{g}_k=2 \lambda(\sin k_y,-\sin k_x, 0)$ is inherited from the strong topological insulator surface, and an effective chiral $p_x + i p_y$ pairing is generated. This can again be understood from the above argument of the dual Hamiltonian written on the basis of SOC coupling to understand the interplay between the $s$-wave superconductivity, the SOC, and the exchange field $E_z$. Note that the off-diagonal terms in Eq.~(\ref{eq:bdg_dual_1d}) contain $\sin k_y \sigma_z \pm i \sin k_x \sigma_0$, inferring the chiral (spin-polarized) nature of the emergent superconducting pairing, which essentially causes the topological superconductivity hosting Majorana modes at the vortex core~\cite{FuPRL2008}. The physical mechanism behind the formation of a topological gap in the energy spectrum starting from a trivial $s$-wave superconductor remains valid in the case of a parent trivial $d_1 +i d_2$ wave superconductor~\cite{Black-Schaffer2012}.

In general, superconducting pairing can be written in the BdG basis as follows  $\Delta= \Delta_x \sigma_x + \Delta_y \sigma_y+  \Delta_0\sigma_0 $ where the first two terms correspond to the spin-triplet, namely, $p$-wave, and the last term represents the spin-singlet, namely, $s$- and $d$-wave pairings. The combination of  SOC and Zeeman field can turn the spin-singlet sector into an effective spin-triplet sector where the topological superconductivity can appear. The above phases belong to a time-reversal-broken chiral TSC. In the absence of a magnetic field, time-reversal-invariant helical TSC  phases are predicted using extended $s$-wave as well as spinful $p_x + i p_y$ wave superconductivity, while the SOC is only important in the former case, not the latter one~\cite{Xiao-Liang2009,FZhang2013}.

It is, therefore, anticipated that SOC is an essential ingredient in engineering the topological superconductivity from a parent $s$-wave pairing. In recent times, the quest for MZMs has expanded 
to an alternative platform involving helical 
spin chains~\cite{Yazdani2013,Felix_analytics,Loss2013PRL,Felix2014,PhysRevB.89.115109, Bena2015,Loss2016,Christensen2016,Simon2017,Sticklet2019,Rex2020,Mohanta2021, Loss2022} or magnetic adatoms deposited on the surface of a bulk common $s$-wave 
superconductor~\cite{Sau2013,PhysRevB.88.180503,Hui2015,Joel2015,Sharma2016,Theiler2019,Mashkoori2020,Teixeria2020}  where neither the SOC nor the magnetic field is explicitly present. The helical spin texture and the profile  of the magnetic impurities 
give rise to the effective SOC and Zeeman field in the system. The translation symmetry is broken due to the chain of magnetic impurities while it preserves the particle-hole and chiral symmetries as discussed earlier. As a result, the lattice Hamiltonian for a  chain of magnetic atoms implanted on the top of a conventional $s$-wave superconductor is given by~\cite{Yazdani2013} 
\begin{eqnarray}
&H_{\rm lattice}&=\sum_{n,\alpha} t_{h} f_{n,\alpha}^{\dagger} f_{n+1,\alpha} -\sum_{n,\alpha} \mu f_{n,\alpha}^{\dagger} f_{n,\alpha} + \sum_{n} \Delta f_{n,\uparrow}^{\dagger} f_{n,\downarrow}^{\dagger} \nonumber \\
&+& \sum_{n,\alpha,\beta} (\vect{B}_{n}.\vect{\sigma})_{\alpha \beta} f_{n,\alpha}^{\dagger} f_{n,\beta}  + {\rm h.c.}\ \ ,
\label{eq:ham1_lattice}
\end{eqnarray}
where, $n$ denotes the site-index, and $\alpha$ and $\beta$ represent the spin-indices. Here, all the magnetic atoms share an identical magnitude of moment, i.e., $B$. However, their orientations are different at each lattice site \ie $\vect{B}_{n}=B(\sin \theta_{n} \cos \phi_{n} \hat{i}+\sin \theta_{n} \sin \phi_{n} \hat{j}+\cos \theta_{n} \hat{k})$ behaves as classical spins. One can go to a rotated frame on each site such that spin-up and spin-down components mix with each other. The unitary transformation is given by $R_n$ as
\begin{eqnarray}
    R_n= \begin{pmatrix}
\cos(\theta_n/2)  & -\sin(\theta_n/2)e^{-i\phi_n}   \\
\sin(\theta_n/2)e^{i\phi_n}   & \cos(\theta_n/2)  
\end{pmatrix}\ ,
\label{eq:shiba_Urotation}
\end{eqnarray}
which connects the rotated basis in the following way: $(f_{n,\uparrow}, f_{n,\downarrow})^T=R_n (g_{n,\uparrow}, g_{n,\downarrow})^T$.  The rotated Hamiltonian takes the form 
\begin{align}
\tilde{H}_{\rm lattice}=&\sum_{n,\alpha,\beta} t_{h} \Omega_{n}^{\alpha,\beta} g_{n,\alpha}^{\dagger} g_{n+1,\beta} -\sum_{n,\alpha} \mu g_{n,\alpha}^{\dagger} g_{n,\alpha} \nonumber \\ 
&+ B (g_{n,\uparrow}^{\dagger} g_{n,\uparrow} -g_{n,\downarrow}^{\dagger} g_{n,\downarrow})+  \sum_{n} \Delta g_{n,\uparrow}^{\dagger} g_{n,\downarrow}^{\dagger} + {\rm h.c.} \ ,
\label{eq:ham2_lattice}
\end{align}
with $\Omega_{n}=R^{\dagger}_n R_{n+1}$.

Note that this transformation leads to emergent inter-spin hopping terms $\Omega_{n}^{\uparrow \downarrow}$ in addition to intra-spin hopping $\Omega_{n}^{\uparrow \uparrow}$, $\Omega_{n}^{\downarrow \downarrow}$, which already exist. The hopping elements are renormalized  in the rotated Hamiltonian (Eq.~(\ref{eq:ham2_lattice})) as compared to the parent Hamiltonian 
given by Eq.~(\ref{eq:ham1_lattice}). Therefore, the notion of SOC can then be anticipated by the former term. Interestingly, the magnetic moments become diagonal in the spin components while the superconducting gap and chemical potential terms remain unaltered. The helical spin chain, mimicking the composite effect of SOC and perpendicular magnetic field, is shown to host TSC phase for $B^2> \Delta^2 + (\mu+ \tilde{t}_h)^2 $~\cite{Felix_analytics} where $\tilde{t}_h$ represents the renormalized hopping. The scattering between magnetic impurities and the quasiparticles in the superconductor gives rise to Yu–Shiba–Rusinov (YSR) bound states inside the superconducting gap. These bound states can overlap and form an emergent band inside the superconductor, termed the YSR/Shiba band. The MZMs appear within the minigap of the Shiba band that has been experimentally realized by the scanning tunneling microscope measurements~\cite{Yazdani_science,Yazdani2015,HowonSciAdv2018,Schneider2020,Wiesendanger2021,Beck2021,Wang2021PRL,Schneider2022,Crawford2022}.

\subsection{HOTSC}
Going beyond the realm of FOTSC,  we are now in a position to discuss the emergence of Majorana hinge modes (MHMs) and Majorana corner modes (MCMs) as a signature of HOTSC. In order to understand the mechanism behind the emergence of HOTSC, we briefly demonstrate the well-known models of HOTI: In 2D, Benalcazar-Bernevig-Hughes (BBH) model~\cite{benalcazar2017,benalcazarprb2017}, impersonating the 2D stacked Su-Schrieffer-Heeger chains, can host a second-order electronic corner mode while 
a Bernevig-Hughes-Zhang~(BHZ) model~\cite{BHZPRL2006,bernevig2006quantum}, considered in the presence of a four-fold rotation symmetry $C_4$ and TRS $\mathcal{T}$ breaking Wilson-Dirac~(WD) mass term, also harbors corner modes~\cite{schindler2018, Roy2019}. Note that both models have two pseudo-spin degrees of freedom and can therefore be compactly written as a four-band Hamiltonian. However, in the first model, the WD mass term is hidden in the inter- and intra-cell hopping, whereas in the second model, it is solely responsible for obtaining the second-order topological phase. 
Note that, the WD mass term takes the form $(\cos k_x -\cos k_y)$ and $(2\cos k_z -\cos k_x -\cos k_y)$ with appropriate matrix structures in two and three dimensions, respectively which resemble the $d_{x^2-y^2}$- and $d_{3z^2-r^2}$-wave form factors~\cite{Nag19, Nag2021}. Without loss of generality, $(\cos k_x -\cos k_y)$ can be referred to as first WD mass and $(2\cos k_z -\cos k_x -\cos k_y)$ as the second 
WD mass term. These are additional topological mass terms needed to conceive the HOT phase. 
The first-order topological mass term, on the other hand, is solely responsible for band inversion. 
To obtain the SOTI phase, hosting hinge modes, a four-band Hamiltonian is essential, whereas for the TOTI phase, encompassing corner modes, an eight-band Hamiltonian is necessary~\cite{Ghosh_2024}. 
All the constituent matrices are mutually anticommuting, and this constraint is very useful for systematically engineering HOT phases~\cite{Roy2019, Nag2021, Ghosh_2024}.

Now we discuss the construction of the HOTSC models in two and three dimensions. The simplest way to formulate a HOTSC model would be to begin with a $p$-wave FOTSC and add the $d_{x^2-y^2}$- and $d_{x^2-y^2}+id_{3z^2-r^2}$-wave pairing terms, acting like WD mass terms, such that MCMs and MHMs can be generated by gaping out the first-order boundary modes. It has been shown that $p+i d_{x^2-y^2}$-wave pairing in the presence of hopping only can lead to SOTSC in two dimensions~\cite{WangWeak2018}. This SOTSC model is based on the underlying FOTSC, as the $p$-wave pairing ensures Majorana edge modes at the boundaries. The model Hamiltonian is a four-band model that supports first- and second-order phases when only $p$-wave pairing is non-zero, and when both $p$- and $d_{x^2-y^2}$-wave pairings are finite. The SOTSC model can be written in the BdG basis $(c_{k\uparrow}, c_{k\downarrow}, c^{\dagger}_{-k \uparrow},c^{\dagger}_{-k \downarrow})^T$ as follows with $\vect{\sigma}$ and $\vect{\tau}$ act on the spin and particle-hole degrees of freedom~\cite{WangWeak2018}
\begin{align}
H_{p+id}(\vect{k})=&\begin{pmatrix}
\epsilon(\vect{k}) \sigma_0  & i \sigma_y \Delta \\
(i \sigma_y \Delta)^{\dagger}  &  -\epsilon(\vect{k}) \sigma_0   \\
\end{pmatrix}\ ,\nonumber \\
=& (2-\cos k_x -\cos k_y -\mu) \sigma_0 \tau_z \non \\ 
&- \Delta_d(\cos k_x -\cos k_y) \sigma_y \tau_x \non \\
&+\Delta_p (\sin k_x \sigma_z \tau_x - \sin k_y \sigma_0 \tau_y) \ ,
\label{eq:SOTSC_pwave1}
\end{align}    
where, superconducting pairing $\Delta=\Delta_p (\sin k_x \sigma_x + \sin k_y \sigma_y) + i \Delta_d (\cos k_x -\cos k_y) \sigma_0$ and kinetic part $\epsilon(\vect{k})=2-\cos k_x -\cos k_y -\mu$. Note that spin-triplet $p$ ($d$)-channel pairings are diagonal (off-diagonal) in the spin basis while odd (even) in momentum, represented by  $\sigma_z$ and $\sigma_0$ ($\sigma_y$), matrices in the second line of the BdG Hamiltonian. One can find out that the fermionic antisymmetric nature of the pairing is preserved 
$\tilde{\Delta}(\vect{k})=-\tilde{\Delta}^T(-\vect{k})$ with $\tilde{\Delta}(\vect{k})=i \sigma_y\Delta(\vect{k})$ and $T$ denotes the transpose. The FOTSC (SOTSC) phase are obtained for $\Delta_d =0$ ($\Delta_p \ne 0$, $\Delta_d \ne 0$). Note that all the matrices are mutually anti-commuting, and it is enough to have four such matrices to build up the SOTSC in two dimensions.  The Hamiltonian in Eq.~\eqref{eq:SOTSC_pwave1} breaks the time-reversal $\mathcal{T}=i \sigma_y \mathcal{K}$ and four-fold rotational symmetry $C_4$, but preserves the combined $C_4 \mathcal{T}$ symmetry~\cite{WangWeak2018}. Additionally, the Hamiltonian Eq.~\eqref{eq:SOTSC_pwave1} respects the particle-hole symmetry $\mathcal{P}=i \sigma_x \tau_y \mathcal{K}$ owing to the BdG construction~\cite{WangWeak2018}.


To obtain TOTSC in three dimensions, one has to construct at least six anti-commuting matrices, which is only possible if we go beyond the four-band model. We consider an eight-band model where spin $\vect{\sigma}$, sublattice $\vect{s}$, and particle-hole $\vect{\tau}$ degrees of freedom are required to set up the model as follows
\begin{eqnarray}
&&H_{p+id}(\vect{k})=\begin{pmatrix}
\epsilon(\vect{k}) s_0\sigma_0  & \tilde{\Delta}(\vect{k})  \\
\tilde{\Delta}^{\dagger}(\vect{k})  &  -\epsilon(\vect{k}) s_0\sigma_0   \\
\end{pmatrix}\nonumber \\
&=& (2-\cos k_x -\cos k_y -\cos k_z -\mu) s_0 \sigma_0 \tau_z +\Delta_p (\sin k_x s_0 \sigma_z \tau_x \nonumber \\
&-& \sin k_y s_0 \sigma_0 \tau_y + \sin k_z s_0 \sigma_x \tau_x) 
- \Delta^1_d(\cos k_x -\cos k_y) s_x \sigma_y \tau_x \nonumber \\
&-& \Delta^2_d(2\cos k_z-\cos k_x -\cos k_y) s_z \sigma_y \tau_x \ ,
\label{eq:TOTSC_pwave2}
\end{eqnarray}    
with $\tilde{\Delta}(\vect{k})=i \sigma_y\Delta(\vect{k})$ and 
$\Delta= \Delta_p s_0(\sin k_x \sigma_x + \sin k_y \sigma_y + \sin k_z \sigma_z) + i [\Delta^1_d (\cos k_x -\cos k_y) s_x \sigma_0 +\Delta^2_d (2 \cos k_z -\cos k_x -\cos k_y) s_z \sigma_0] $. This ensures that the fermionic anti-symmetric nature of the pairing is preserved $\tilde{\Delta}(\vect{k})=-\tilde{\Delta}^T(-\vect{k})$. The hierarchy of the HOTSC phases can be obtained in a systematic manner: FOTSC with Majorana surface states are observed for  $(\Delta_p,\Delta^1_d,\Delta^2_d)=(\neq 0,0,0)$, SOTSC with MHMs are achieved for $(0,\neq 0,0)$ while TOTSC with MCMs are obtained for $(0,0,\neq 0)$. The HOTSC phases break both TRS and $C_4$ symmetry. 

We now discuss another alternative approach where the diagonal part does not include only the kinetic part but also the SOC, while the superconducting part includes the $d_{x^2-y^2}+ i d_{3z^2-r^2}$ pairing~\cite{Yan2018,RoyPRBL2021,GhoshDynamical2022}. We refer to this class of models as Model I. 
The  generic form of the HOTSC model in $d$-dimension  read as  follows, considering the basis  $(c_{k\uparrow}, c_{k\downarrow}, -c^{\dagger}_{-k \downarrow},c^{\dagger}_{-k \uparrow})^T$, 
\begin{eqnarray}
&H_{dD}(\vect{k})&=\begin{pmatrix}
H^{\rm dD}_{\rm FOTI} (\vect{ k})-\mu & \Delta(\vect{k})  \\
\Delta^\dagger(\vect{k}) & \mu-\mathcal{T}^{-1} H^{\rm dD}_{\rm FOTI} (\vect{-k}) \mathcal{T}\   \\
\end{pmatrix}\ ,\nonumber \\
\label{eq:HOTSC_dwave}
\end{eqnarray}    
that leads to 2D model of SOTSC~\cite{Yan2018}
\begin{eqnarray} H_{2D}(\vect{k})
&=& (m-\cos k_x -\cos k_y) s_z\sigma_0 \tau_z -\mu s_0\sigma_0 \tau_z   \nonumber \\
&+&\sin k_x s_x  \sigma_z \tau_z + \sin k_y s_y \sigma_0 \tau_z \nonumber\\
&+& \Delta^1_d(\cos k_x -\cos k_y) s_0 \sigma_0 \tau_x \ ,
\label{eq:SOTSC_dwave1}
\end{eqnarray}    
and 3D model of TOTSC as~\cite{RoyPRBL2021}
\begin{eqnarray}
 H_{3D}(\vect{k})
&=& (m-\cos k_x -\cos k_y - \cos k_z) s_z\sigma_0 \tau_z -\mu s_0\sigma_0 \tau_z \nonumber \\ &+& \sin k_x s_x  \sigma_x \tau_z  + \sin k_y s_x \sigma_y \tau_z  
+\sin k_z s_x \sigma_z \tau_z \nonumber \\
&+&\Delta^1_d(\cos k_x -\cos k_y) s_0 \sigma_0 \tau_x  \nonumber \\
&+& \Delta^2_d(2 \cos k_z -\cos k_x -\cos k_y) s_0 \sigma_0 \tau_y \ .
\label{eq:TOTSC_dwave2}
\end{eqnarray}    

Here, $H^{\rm dD}_{\rm FOTI}(\vect{k})$ preserves TRS, generated by ${\mathcal T}=i s_0\sigma_y {\mathcal K}$, while $H_{2D}(\vect{k})$ preserves the TRS, but $H_{3D}(\vect{k})$ breaks TRS and ${\mathcal K}$ represents the complex conjugation. Both the SOTSC 
(Eq.~(\ref{eq:SOTSC_dwave1})) and TOTSC (Eq.~(\ref{eq:TOTSC_dwave2})) models are described by an eight-band Hamiltonian with mutually anti-commuting matrices. Note that the $\mu$ term 
does not necessarily anti-commute with the remaining terms, as $\mu$ is an external control not a tight-binding model parameter.   
It is noted that the above models do not carry any FOTSC analog hosting Majorana edge modes and Majorana surface modes in 2D and 3D, respectively. The prescription here is to invoke the topological notion in the $H^{\rm dD}_{\rm FOTI} (\vect{ k})$ part of the BdG Hamiltonian, while the WD mass terms are regarded as the superconducting $d$-wave pairing term. As discussed previously for the first-order case, the dual Hamiltonian can be useful to understand the topological origin. Following a similar line of argument, one can assume that $d$-wave pairing terms appear like a WD higher-order mass terms in the diagonal part of the dual Hamiltonian written in the SOC basis, while the off-diagonal part contains hopping and $p_x+ip_y$-type of pairing. Therefore, the topological gap is caused by effective $p$-wave pairing residing on the off-diagonal part. 

Having discussed the $d$-wave induction of the HOTSC phase, we now focus on the $s$-wave proximity-induced HOTSC phases~\cite{ghosh2020floquet,Ghosh2021PRB}. We refer to this class of models as model II. In this case, the magnetic field is not mandatorily required. One can include the WD mass terms in the kinetic part of the BdG Hamiltonian as
\begin{eqnarray}
H_{\rm dD}(\vect{k})=
\begin{pmatrix}
H^{\rm dD}_{\rm HOTI} (\vect{k})-\mu & \Delta \\
\Delta^* & \mu-\mathcal{T}^{-1} H^{\rm dD}_{\rm HOTI} (-\vect{k}) \mathcal{T}\   \\
\end{pmatrix}\ ,
\end{eqnarray}
that leads to 2D model of SOTSC~\cite{YanPRB2019}
\begin{eqnarray}
&& H_{2D}(\vect{k})= (m-\cos k_x -\cos k_y) s_z \sigma_0 \tau_z + \sin k_x s_x \sigma_x \tau_z \nonumber\\
&+& \sin k_y s_x \sigma_y\tau_z 
+  \Lambda_1(\cos k_x -\cos k_y) s_y \sigma_0\tau_0 + \Delta s_0 \sigma_0 \tau_x\ ,   \nonumber \\
\label{eq:SOTSC_swave1}
\end{eqnarray}
and 3D model of TOTSC~\cite{Ghosh2021PRB}
\begin{eqnarray}
&&H_{3D}(\vect{k})= (m-\cos k_x -\cos k_y -\cos k_z ) \mu_x s_z \sigma_0 \tau_z \nonumber\\
&+& \sin k_x \mu_x s_x \sigma_x \tau_z + \sin k_y \mu_x s_x \sigma_y\tau_z \nonumber\\
&+& \sin k_z \mu_x s_x \sigma_z \tau_z
+ \Lambda_1(\cos k_x -\cos k_y) \mu_x s_y \sigma_0\tau_0 \nonumber \\
&+& \Lambda_2(2 \cos k_z - \cos k_x -\cos k_y) \mu_z s_0 \sigma_0\tau_z+ \Delta \mu_0 s_0 \sigma_0 \tau_x\ , \nonumber \\
\label{eq:TOTSC_swave2}
\end{eqnarray}
where, $\mathcal{T}=i \sigma_y {\mathcal K}$. 
The SOTSC model (Eq.~(\ref{eq:SOTSC_swave1})) is given by an eight-band Hamiltonian, while the TOTSC model (Eq.~(\ref{eq:TOTSC_swave2})) involves a sixteen-band Hamiltonian. All the constituent matrices are not mutually anti-commuting in the HOTSC model $H_{\rm dD}(\vect{k})$; rather, the diagonal parts $H_{\rm HOTI} (\vect{k})$, excluding the superconducting pairing, contain all the anti-commuting matrices. Note that $H_{\rm HOTI} (\vect{k})$ hosts the SOTI and TOTI phases, with broken TRS and $C_4$ symmetry, which are given by  a four- and eight-band Hamiltonian, respectively. The diagonal part supports hinge and corner electronic modes, while $s$-wave superconductivity induces the topological superconductor nature of these electronic modes. One can adopt the dual Hamiltonian to show the emergent $p$-wave nature of the problem. This HOTSC model does not have any FOTSC phases unlike the $p+id$ case and as a result the systematic emergence of the HOTSC phases starting from the FOTSC phases is not possible. The SOC term is incorporated in the diagonal kinetic part of the BdG Hamiltonian for both model I and model II. This choice of the model ensures the topological notion of the diagonal part. In the case of model I, the diagonal part carries FOTI flavor, while for model II, it mimics SOTI and TOTI characters. This is markedly different from HOTSC models with $p+id$-pairing, where the SOC is indirectly included via the $p$-wave pairing.  

In the above HOTSC model conception namely model I and model II, the magnetic field is not a prerequisite. In the case of $s$-wave superconductor proximity-induced FOTSC phases, the magnetic field is inevitable to break the degeneracy at $\vect{k}=0$. Interestingly, the HOTSC phases, obtained from model I and model II, do not rely on the underlying FOTSC nature of the model, rather FOTI/HOTI analog are concealed inside the diagonal part of the BdG. The magnetic field is not an essential requirement to formulate a  FOTI/HOTI model in general. Therefore, the magnetic field does not play a pivotal role in the engineering of the HOTSC phases. However, it has been shown that for model II in 3D, the degeneracy of the MHMs can be tuned by the magnetic field while no such manipulation is possible 
for MCMs~\cite{Ghosh2021PRB}. On the other hand, the MCMs in 2D can be constructed by considering a uniaxial magnetic field without the need for a first WD mass term for the  $s$-wave proximity induced scenario. The emergence of HOTSC phases can be understood from the edge/surface computation of the effective Hamiltonian rather than investigating the bulk band dispersion.  The first-order topological modes have to be gapped out everywhere on the edges (surfaces) except at the corners (hinges)  such that the SOT phase shows up in 2D (3D). The same line of argument is valid for corner modes in 3D where hinge modes are gapped out except for the corners. The qualitative way to understand it is the following:
say, $H^{\rm I}_{\rm edge}=k_x \sigma_x + \tilde{m} \sigma_z$ and $H^{\rm II}_{\rm edge}=k_y \sigma_x - \tilde{m} \sigma_z$  represent edge I and edge II projected Hamiltonian derived from the bulk Hamiltonian under certain conditions. The corner-sharing adjacent edges I and II are positively and negatively gapped, respectively, while there is a sign change of mass terms across the edges. This ensures a zero-energy Jackiw-Rebbi mode to appear at the intersection, which is nothing but the corner modes. Note that $\tilde{m}$ denotes the WD mass term, while the Dirac-like linear gapless dispersion on the edges is guaranteed by the lower-order topological phase in general. For the hinge modes in 3D, the first WD mass term gaps out the gapless surface modes while the second WD mass term gaps out the gapless hinge mode to obtain the corner modes \cite{Ghosh2021PRB}.  


\subsubsection{Topological characterization}

Now, we turn our attention to the appropriate topological invariant to characterize the HOT modes. The lower-order topological invariants are found to vanish once the HOT phase is achieved. As a result, the FOT invariant, such as  Chern number and winding number no longer work, rather, there are several approaches to formulate the HOT invariant based on the complete real space, momentum space, and semi-infinite geometries. The real space polarization is captured by the dipole moment, which renders the FOT invariant. In the case of higher-order real space invariant, one requires higher-moments of the charge to characterize the topology namely, quadrupole moment $Q_{xy}$~\cite{benalcazar2017,benalcazarprb2017}  for 2D  corner modes and  octupole moment $O_{xyz}$~\cite{benalcazar2017,benalcazarprb2017} for 3D corner modes are essential indices. 
On the other hand,  the Berry phase is captured by the polarization, which can be computed by the unitary Wilson loop in the momentum space. The concept of Wilson loop~\cite{benalcazar2017,benalcazarprb2017} is essential to compute the full momentum space HOT invariant namely, the nested polarization.  Apart from these two, there exists another topological invariant, namely, Wannier spectra, where semi-infinite geometry can be exploited to mimic the topological nature of the higher-order modes. In the momentum space, the nested polarization invariant to work, the mirror symmetries have to be respected, while there is no such strict constraint for the other two invariants to characterize the HOT phase.    

Here, we briefly describe the real-space bulk invariant quadrupole moment $Q_{xy}$ that characterizes the SOTSC phase~\cite{benalcazar2017,benalcazarprb2017}. The macroscopic quadrupole moment $Q_{xy}$ is defined for  a crystal obeying periodic boundary condition~(PBC)  as follows ~\cite{wheeler2018many,kang2018many}:

\begin{equation}\label{macroquadint}
	Q_{xy}=\frac{1}{2 \pi} \textrm{Im}\left[ \ln \bra{\Psi_0} e^{2 \pi i \sum_{r}\hat{q}_{xy}(\vect{r})}\ket{\Psi_0}\right]\ ~\text{mod}~ 1,
\end{equation}
where, $\hat{q}_{xy}(\mathbf{r})=\frac{xy}{L^2}\hat{n}(\vect{r})$ denotes the microscopic quadrupole moment at site $\vect{r}$, $L$ is the number of lattice sites considered along one direction, and $\ket{\Psi_0}$ is the many-body ground state constructed from the occupied states~\cite{RestaPRL1998}. In a similar spirit, considering the octupolar moment operator $\hat{o}_{xyz}(\vect{r})=\frac{xyz}{L^3}\hat{n}(\vect{r})$, the macroscopic octupole moment $O_{xyz}$ reads as 
~\cite{wheeler2018many,kang2018many} :

\begin{equation}\label{macrooctint}
	O_{xyz}=\frac{1}{2 \pi} \textrm{Im}\left[ \ln \bra{\Psi_0} e^{2 \pi i \sum_{r}\hat{o}_{xyz}(\vect{r})}\ket{\Psi_0}\right]\ \ ~\text{mod}~ 1.
\end{equation}

The $Q_{xy}$ ($O_{xyz}$) exhibit quantized value of $0.5$ in the SOTSC (TOTSC) phase for 2D (3D). To topologically characterize the MHMs in 3D, one can compute $Q_{xy}$ as a function $k_z$ to show the underlying quadrupole nature in the $xy$ plane.

Having demonstrated the real space HOT invariants, we now briefly discuss the nested polarization, which relies on the momentum-space bulk calculation \cite{benalcazarprb2017}. The position operator $\hat{x}$ is projected on the occupied subspace, yielding the first-order Wilson loop, which reduces to the first-order polarization~\cite{RestaPRL1998}. The prescription is to systematically compute the second-order polarization, obtained from the second-order  Wilson loop, by projecting the position operator along the $y$ direction onto the subspace associated with first-order Wannier bands. The nesting is meant by the intertwined projection of the bands along $y$ as well as $x$. Importantly, the eigenvalues of the Wannier Hamiltonian ${\rm Log}[{\mathcal W}_{x}]$, namely Wannier centers are independent of the initial choice of the momentum in formulating the Wilson loop  ${\mathcal W}_x=F_{x,k_x + (N_x -1) \Delta k_x } \cdots F_{x,k_x + \Delta k_x } F_{x,k_x} $~\cite{benalcazarprb2017}. Here the Wilson loop line element $ \left[F_{x,k_x}\right]_{mn}=\langle \Psi_{n, k_x + \Delta k_x} | \Psi_{m,k_x} \rangle$, is obtained from $m^{\rm{th}}$ occupied band $|\Psi_{m,k_x} \rangle $ and  $\Delta k_x= 2\pi /N_x$ ($N_x$ denotes the number of discrete points along $k_x \in $ BZ). However, the corresponding wave functions do depend on the initial momentum, leading to a complex situation when the projection is taken.  Therefore, one has to perform the momentum average over the initial position of the underlying first-order Wannier bands.  The nested polarization, based on the nested Wilson loop, is found to be $p^{\pm \nu_x}_y=\frac{1}{N_x} \sum_{k_x} \nu^{\pm \nu_x}_y(k_x)=1/2 $ (mod $1$) while first-order polarization $p_x=-\frac{i}{2\pi} \mbox{Log } \mbox{Det} \left[ W_{k+2\pi \leftarrow k}\right]$ is gapped. This identifies the SOTSC phase. Note that the nested Wilson loop captures the complex intertwined projection of the bands along all two directions $x$, and $y$. The initial choice of the momenta $k_{x}$ in formulating the  Wilson loop ${\mathcal W}_{x}$ is to be averaged over to compute $p^{{\pm \nu_x}}_y$. 
In a similar spirit, TOTSC phase is characterized by the third-order Wilson loop-mediated third-order polarization $p^{{\pm \nu_y}^{\pm \nu_x}}_z=\frac{1}{N_x N_y} \sum_{k_x,k_y}\nu_z^{\pm \nu_y^{\pm \nu_{x}}} =1/2 $ (mod $1$) while first-order and second-order polarization are gapped out. Note that the position operator along the $z$ direction is projected onto the subspace associated with second-order Wannier bands. The double nesting is meant by the complex intertwined projection of the bands along all three directions $x,y$ as well as $z$. The initial choice of the momenta $k_{x,y}$ in formulating the  Wilson loop ${\mathcal W}_{x,y}$ are to be averaged over to compute $p^{{\pm \nu_y}^{\pm \nu_x}}_z$.

 We here discuss the Wannier spectra to characterize the SOTSC phase in 2D, where PBC (OBC) is employed along the $x$ ($y$) direction. We need to use the Wilson loop operator along the $k_x$ direction~\cite{benalcazarprb2017} ${\mathcal W}_x$, as stated above. The Wannier Hamiltonian takes the form ${\mathcal H}_{{\mathcal W}_x}= -i \ln {\mathcal W}_x$, whose eigenvalues $2\pi \nu_x$ correspond to the  Wannier spectra (WS). Here, $\nu_x \equiv {\rm mod} (\nu_x,1)$ is the Wannier center, which acquires a quantized value 0.5. The Wannier center shifts half away from the lattice site for the higher-order phases. In 3D, one can consider PBC along two directions $y$ and $z$, and OBC along $x$ direction to construct the Wilson loop operator along $k_y$ direction~\cite{benalcazarprb2017,ghosh2020floquet,ghosh2020floquet2} as follows
\begin{equation}
	{\mathcal W}^{(x)}_y=F^{(x)}_{y,k_y + (N_i -1) \Delta k_y,k_z } \cdots F^{(x)}_{y,k_y + \Delta k_y,k_z } F^{(x)}_{y,k_y,k_z}\ ,
	\label{WS} 
\end{equation} 
with $\left[F^{(x)}_{y,k_y,k_z}\right]_{mn}=\langle \Psi^{(x)}_{n, k_y + \Delta k_y,k_z} | \Psi^{(x)}_{m,k_y,k_z} \rangle$ where $\Delta k_i= 2\pi /N_i$ ($N_i$ denotes the number of discrete points $k_i\in $ BZ) and $|\Psi^{(x)}_{m,k_y,k_z} \rangle $ is the $m^{\rm th}$ occupied state of the Hamiltonian. The Wannier center $\nu^{(x)}_{y,k_z=0}$ is obtained from Wannier Hamiltonian  ${\mathcal H}_{{\mathcal W}^{(x)}_{y}}$, which acquires a half-quantized value characterizing the SOTSC phase in the cubic geometry. On the other hand, the TOTSC phase in 3D manifests itself via the half-quantized value of Wannier center $\nu^{(x,y)}_z$, obtained from  Wilson loop ${\mathcal W}^{(x,y)}_z$ along $k_z$,  by adopting PBC along $z$ direction and OBC along $x$ and $y$ directions. 

\begin{widetext}
To summarize all the static topological superconductor models discussed, we present Table~\ref{tab:static} below.
\begin{table}[h]
\centering

\begin{tabular}{|c|c|c|c|c|c|}
\hline
 Topological order & Dimension  &Boundary modes & Relevant symmetries & Topological indices \\ \hline
  FOTSC [\eg Eqs.~(\ref{eq:hamk_pwave}, \ref{eq:bdg_1d}, \ref{eq:ham1_lattice})]  & 1D     & End modes & $\mathcal{P}$, ${\mathcal C} $& Winding number \\ \hline
   SOTSC [\eg Eqs.~(\ref{eq:SOTSC_pwave1}, \ref{eq:SOTSC_dwave1}, \ref{eq:SOTSC_swave1})]  & 2D   & Corner modes&$\mathcal{P}$, $C_{4} \mathcal{T}$  & Quadrupole moment\\ \hline
   TOTSC [\eg Eqs.~(\ref{eq:TOTSC_pwave2}, \ref{eq:TOTSC_dwave2}, \ref{eq:TOTSC_swave2})] &  3D &  Corner modes  &$\mathcal{P}$, $C_{4} \mathcal{T}$ & Octupole moment \\ \hline
\end{tabular}
\caption{Summary of all the static topological superconductor models discussed, including their Hamiltonians, spatial dimensions, topological order, boundary modes, relevant symmetries, and topological indices.}
\label{tab:static}
\end{table}

\end{widetext}

\section{First-order Floquet TSC in proximity induced 1D systems~\label{Sec:Floquet_first_order}}

In this section, we discuss the emergence of the first-order Floquet superconductor hosting Majorana end modes in different Floquet driving schemes in experimentally realizable 1D systems. Superconductivity is induced in those systems via the proximity effect~\cite{Rudner2013,GhoshDynamical2022}.

\subsection{Driving protocol and Emergence of Floquet Majorana modes in Rashba NW} \label{subsec:FloquetMEMRashba}

In this section, we discuss the emergence of $0$- and $\pi$-Majorana end modes based on a Rashba nanowire~\cite{Fu2008, Sau2010, Lutchyn10, Hughes2010, Oreg2010}. We also discuss the topological characterization of these $0$- and $\pi$-modes using periodized evolution operator.

\subsubsection{Driven Rashba NW}
\begin{figure}[]
	\centering
	\subfigure{\includegraphics[width=0.49\textwidth]{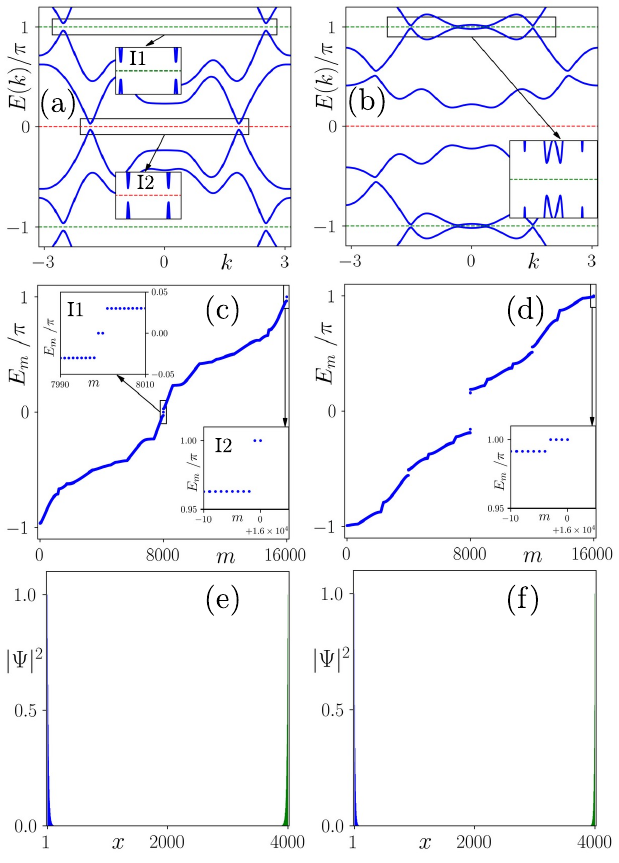}}
	\caption{The bulk quasi-energy spectra for driven Rashba NW are displayed in panels (a) and (b) for $\Omega = 1.5$ and $\Omega = 2.5$, respectively.  	
	We show the bulk gap around quasi-energies $E(k) = \pi$ and $E(k) = 0$, respectively, in the insets I1 and I2 in panel (a). The inset of panel (b) is for $E(k) = \pi$. The quasi-energy spectra $E_m$ for real space under OBC, corresponding to (a) and (b),  are depicted in panels (c) and (d), respectively; the insets offer a magnified view of the quasi-energies for clarity. The emergence of $0$- and $\pi$- MEMs is clearly visible in these spectra. Panels (e) and (f) display the normalized site-resolved probabilities ($\lvert \Psi \rvert^2$) for the $0$- and $\pi$-MEMs, respectively. The remaining model parameters are set as $(c_{0},c_{1},\Delta, B_{x}, t, u)= (1.0, 0.2, 1.0, 1.0, 1.0, 0.5)$.  This figure is adapted from~\cite{Mondal23_FNW} with permission, Copyright (2023) by the American Physical Society.}
	\label{stepeigen}
\end{figure}

Here we consider the underlying Hamiltonian $H_{0}(k) = [(2t-c_0)-2t \cos k] \Gamma_1 + 2u \sin k \Gamma_2 + B_x \Gamma_3 + \Delta \Gamma_4 $~\cite{Mondal23_FNW}. Here, $t$, $c_0$, $u$, $B_x$, and $\Delta$ represent the hopping amplitude, chemical potential, strength of the Rashba SOC, applied magnetic field, and $s$-wave gap, respectively. The $4 \times 4$ $\vect{\Gamma}$ matrices are: $\Gamma_1=\tau_z \sigma_0$, $\Gamma_2=\tau_z \sigma_z$, $\Gamma_3=\tau_0 \sigma_x$, and $\Gamma_4=\tau_x \sigma_0$; with the Pauli matrices $\vect{\tau}$ and $\vect{\sigma}$ acting on the particle-hole and spin subspace. The Hamiltonian $H(k)$ respects the Chiral symmetry, $\mathcal{C}=\tau_y \sigma_z$, particle-hole symmetry $\mathcal{P}=\tau_y \sigma_y \mathcal{K}$, but breaks TRS $\mathcal{T}=i\tau_0 \sigma_y \mathcal{K}$. Here $\mathcal{K}$ represents the complex-conjugation operator. The chiral symmetry ${\mathcal C}$ can be utlized to obtain the topological phase diagram for $H(k)$~\cite{Mondal23_FNW}.
We adopt a three-step periodic driving scheme to generate the FTSC phase starting from the initial non-topological regime  as
\begin{eqnarray}
{H}_{\rm step}(k,t) && = H_{1} \hspace*{2.5 cm} t \in \left[0,T/4\right) \ , \nonumber \\
&& = { H}_{0}(k) \hspace*{2.0 cm} t \in \left[T/4,3T/4\right) \ , \nonumber \\
&& = H_{1} \hspace*{2.5 cm} t \in \left[3T/4,T\right] \
\label{step_drive_protocol},
\end{eqnarray}
where $H_{1}=-c_{1} \Gamma_{1}$ represents an on-site term, modulating the chemical potential only. Here, the time-period~(frequency) of the drive is  represented by $T$~($\Omega$). The Floquet operator is given by 

\begin{eqnarray}
U_{\rm step}(k,T) &=& {\rm TO} \exp \left[-i \int_{0}^{T} dt \  H_{\rm step}(k,t)  \right] \non \\
&=&e^{-i H_{1}T/4 } e^{-i H_{0}(k)T/2} e^{-i H_{1}T/4 } \
\label{step_evolution_operator},
\end{eqnarray}

from which quasi-energy spectrum $E_n(k) \in \left[-\pi,\pi\right]$ and corresponding quasi-states $\ket{\Psi_n(k)}$ can be obtained. The driven system continues to respect the chiral symmetry $\mathcal C$ and particle-hole symmetry $\mathcal{P}$.

We present the bulk quasi-energy bands $E(k)$ with respect to momenta $k$ for two different frequencies, $\Omega=1.5$ and $2.5$ in Fig.~\ref{stepeigen}(a) and (b), respectively.  From these, one can naively anticipate the number of end-modes present within a certain quasi-energy gap. In the insets (I1 and I2) of Fig.~\ref{stepeigen}(a), one can observe two openings of the bulk gap at  $E_k=0$ and $\pi$ associated with the two MEMs. Similarly, the inset of Fig.~\ref{stepeigen}(b) reveals the presence of four energy gaps in the vicinity of $E(k)=\pi$ (highlighted by the green dotted line), which are associated with four topological end modes. The quasi-energy spectra under open boundary conditions (OBC) are presented in Figs.~\ref{stepeigen}(c) and (d) for driving frequencies $\Omega = 1.5$ and $\Omega = 2.5$, respectively. 
In Fig.~\ref{stepeigen}(c), the insets labeled by I1 and I2 exhibit the existence of two $0$-Majorana end modes (0-MEMs), with one localized at each end of the NW, as well as two $\pi$-Majorana end modes ($\pi$-MEMs), again one per end. Conversely, Fig.~\ref{stepeigen}(d) displays only four $\pi$-MEMs (two per end), while no $0$-MEMs are observed, as further illustrated in the corresponding inset.
These observations from finite-size systems are in good agreement with the expectations derived from the bulk quasi-energy band structures, thereby reinforcing the bulk-boundary correspondence. The spatial localization of the in-gap states is shown in Figs.~\ref{stepeigen}(e) and (f), where the normalized site-resolved probability densities ($|\Psi|^2$) for the $0$- and $\pi$-MEMs shown in Fig.~\ref{stepeigen}(e) and (f) respectively. 
It is evident that both types of Floquet modes are strongly localized at the ends of the NW, confirming their topological origin.

In real space, following the BdG basis $\Psi_{j}=\{\psi_{j\ua},\psi_{j\da},\psi_{j\da}^{\dagger},-\psi_{\j\ua}^{\dagger}\}^{\vect{t}}$, at a particular site $j$, the wave function has the form: $\Psi_{j}=u_{j \ua} \psi_{j\ua} + u_{j \da} \psi_{j\da}+v_{j \da} \psi_{j\da}^{\dagger} - v_{j \ua} \psi_{j\ua}^{\dagger}$ with its conjugate, $\Psi_{j}^{\dagger}=u_{j \ua}^{*} \psi_{j\ua}^{\dagger} + u_{j \da}^{*} \psi_{j\da}^{\dagger}+v_{j \da}^{*} \psi_{j\da} - v_{j \ua}^{*} \psi_{j\ua}$. Here, $u_{j \ua}~(v_{j \ua})$ and $u_{j \da}~(v_{j \da})$ are the particle~(hole) components of the wave function at $j^{\rm {th}}$ site from spin-up, and spin-down sectors, respectively. The self-conjugate criteria are given by $|u_{j \ua}|=|v_{j \ua}|$ and $|u_{j \da}|=|v_{j \da}|$ i.e. equal contribution from particle and hole sectors for each spin components. One can check that both the $0$- and $\pi$-MEMs obey the above mentioned self-conjugate criteria, indicating their self-antiparticle nature.

\subsubsection{Topological characterization}
\begin{figure}[]
	\centering
	\subfigure{\includegraphics[width=0.49\textwidth]{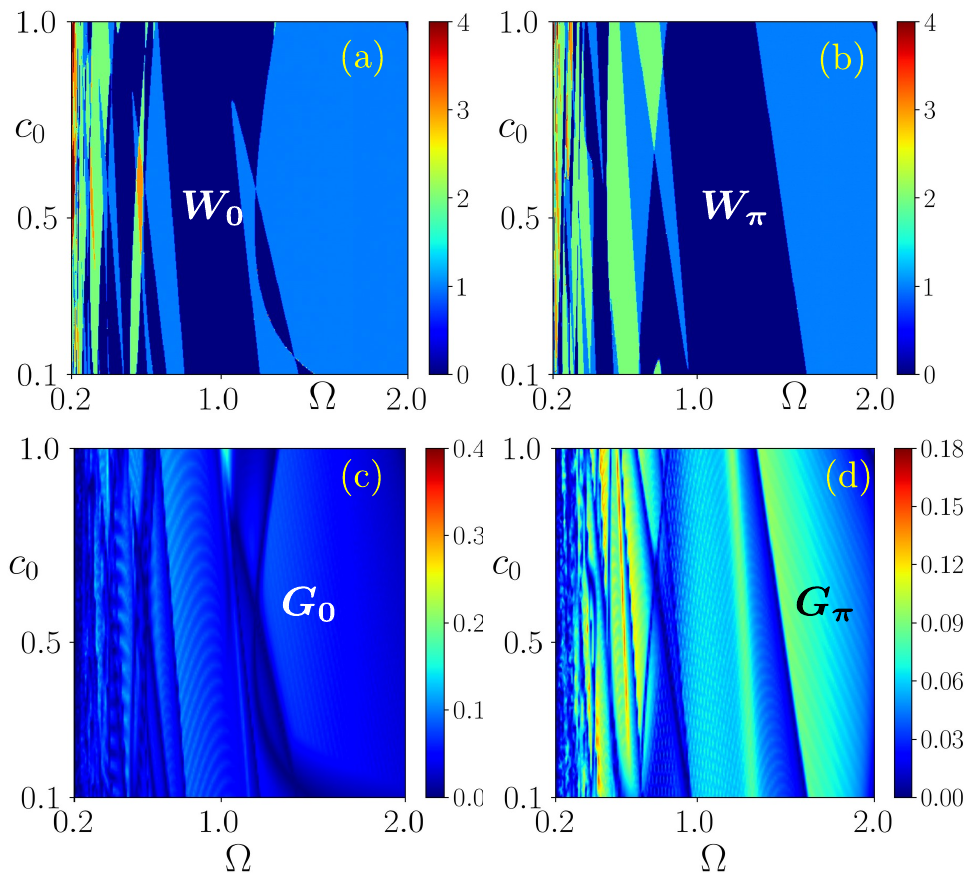}}
	\caption{We present the dynamical winding numbers $W_0$ and $W_\pi$ in the $c_0 \mhyphen \Omega$ plane, shown in panels (a) and (b). Corresponding bulk gap at 0-energy, $G_{0}$, and $\pi$-energy, $G_{\pi}$ are depicted in panels (c) and (d), respectively. All other parameters are set to the same values as those specified in Fig.~\ref{stepeigen}. The color bars represent the winding number~[(a) and (b)] and quasi-energy bulk gap~[(c) and (d)]. This figure is adapted from~\cite{Mondal23_FNW} with permission, Copyright (2023) by the American Physical Society.}
	\label{stepPhase}
\end{figure}

The distinct characterization of the $0$- and $\pi$-MEMs cannot be achieved solely using the Floquet operator, as it has no information about $0$- and $\pi$-MEMs individually~\cite{Rudner2020}. For individual information of these chiral symmetry-protected dynamical $0$- and $\pi$-MEMs, one needs to introduce the notion of gap $\epsilon$ via the periodized evolution operator as~\cite{Rudner2013,GhoshDynamical2022}.
\begin{equation}
U_{\epsilon}(k,t)= U_{\rm step}(k,t) \left[U_{\rm step}(k,T)\right]_{\epsilon}^{-t/T} \ ,
\label{periodise_evolution_operator}
\end{equation}
   where, $U_{\rm step}(k,t)$ represents the usual time-evolution operator and $(-t/T)$$^{\rm th}$ power of the Floquet operator for the $\epsilon$-gap $\left[U_{\rm step}(k,T)\right]_{\epsilon}^{-t/T}$ is given by
\begin{eqnarray}
\left[U_{\rm step}(k,T)\right]_{\epsilon}^{-t/T}&=&  \sum_{n=1}^{N/2} e^{-i(2\pi-2\epsilon-|E_{n}(k)|)t/T} \ket{\psi_n(k)} \bra{\psi_n(k)} \nonumber \\
 &&\hspace{-0.8 cm}+ \sum_{n=N/2 +1}^{N} e^{- \left|E_{n}(k) \right|t/T} \ket{\psi_n(k)} \bra{\psi_n(k)}
 \label{uzero}, \quad 
\end{eqnarray}
where, $n=1 \cdots N/2$ ($N/2+1 \cdots N$) represent valence~(conduction) bands. At the half-period time $t = \frac{T}{2}$, symmetry constraints can be imposed on the evolution operator $U_{\epsilon}\left(k,\frac{T}{2}\right)$ within both the $0$- and $\pi$-quasienergy gaps. These constraints are expressed as follows:
\begin{eqnarray}
	S U_{0}\left(k,\frac{T}{2}\right) S^{-1} &=& -U_{0}\left(k,\frac{T}{2}\right) \ , \nonumber \\
	S U_{\pi}\left(k,\frac{T}{2}\right) S^{-1} &=& U_{\pi}\left(k,\frac{T}{2}\right) \ .
\end{eqnarray}

Using the effective evolution operators $U_{\epsilon}^{\pm}(k)$, one can define the dynamical winding number $W_{\epsilon}$ associated with the quasienergy gap at $\epsilon = 0, \pi$ as~\cite{Michel2016}:
\begin{equation}
W_{\epsilon} = \left| \pm \frac{i}{2\pi} \int_{-\pi}^{\pi} dk \ \mathrm{Tr} \left[ \left\{ U_{\epsilon}^{\pm}(k) \right\}^{-1} \partial_{k} U_{\epsilon}^{\pm}(k) \right] \right|.
\label{dynamical_wind_step}
\end{equation}
The quantity $W_{\epsilon}$ enumerates the number of topologically protected modes localized at each end of the system within the corresponding $\epsilon$-gap. Hence, $W_{\epsilon}$ plays the role of a dynamical counterpart to the static winding number $W$.

Finally, $W_0$ and $W_{\pi}$ are depicted as a function of $\Omega$-$c_0$ in Fig.~\ref{stepPhase} (a) and (b), respectively. They serve as topological phase diagrams. We  illustrate the  corresponding bulk gaps $G_0$ and $G_{\pi}$  in Figs.~\ref{stepPhase}(c) and (d), respectively. One-to-one correspondence between panel  (a) and (c), and between panels (b) and (d) is clearly visible. The change in winding number is always associated with bulk gap closing and the emergence of Floquet MEMs.


\subsection{Emergence of Floquet Majorana modes in driven Helical Shiba chain} \label{subsec:HelicalShiba}

In this section, we discuss another experimentally realistic model to generate the driven first-order TSC. In particular, we discuss the emergence of Majorana modes in a driven helical Shiba chain~\cite{Felix_analytics,Yazdani2013,Mondal23_FSC,PritamPRB2023}. We also discuss the topological characterization of Majorana modes using a real-space topological invariant.

\subsubsection{Floquet generation of MEMs}
Having studied the Floquet Rashba nanowire model, we now exemplify another situation where Floquet MZMs emerge in the helical Shiba chain. We consider a helical spin-chain model constructed from a 1D array of magnetic impurity atoms that emulate an out-of-plane N\'eel-type spin-spiral configuration in $x$-$z$ plane~\cite{Felix_analytics,Yazdani2013,Mondal23_FSC,PritamPRB2023}. The underlying effective 1D Hamiltonian in real space for our system can be 
written as~\cite{Yazdani2013,Mondal23_FSC}
\begin{eqnarray}
H&=& \sum_{j} \Psi_{j}^\dagger \left[-\mu \Gamma_{1} +B \cos(j \theta) \Gamma_{2} +B \sin(j \theta) \Gamma_{3} + \Delta \Gamma_4 \right] \Psi_{j} \nonumber\\
&&+\sum_{j}\Psi_{j,\eta}^\dagger t_{h}  \Gamma_{1} \Psi_{j+1}  +\  {\rm h.c.} 
\label{eq:static_1D_Hamiltonian}\ .
\end{eqnarray}
Note that, here we consider the BdG basis as $\Psi_{j}=\{\psi_{j\ua},\psi_{j\da},\psi_{j\da}^{\dagger},-\psi_{\j\ua}^{\dagger}\}^{\vect{t}}$.
Here, $\mu$, $B$, $\theta$, $\Delta$, and $t_h$ represent the chemical potential, magnetic impurity's strength, angle between two neighboring spins, $s$-wave gap, and the hopping strength, respectively. The $\vect{\Gamma}$ matrices are given as: $\Gamma_1= \tau_z \sigma_0$, $\Gamma_2= \tau_0 \sigma_z$, $\Gamma_3= \tau_0 \sigma_x$, and $\Gamma_4= \tau_x \sigma_0$, with $\vect{\tau}$ and $\vect{\sigma}$ acting on the particle-hole and spin subspace. The Hamiltonian $H$ in Eq.~\eqref{eq:static_1D_Hamiltonian} respects the particle-hole symmetry $\mathcal{P}=\mathbb{I}_N \otimes \tau_y \sigma_y \mathcal{K}$ and chiral symmetry $\mathcal {C}=\mathbb{I}_N \otimes \tau_y \sigma_y$, with $N$ and $\mathcal{K}$ being the number of impurity spins and complex-conjugation operator, respectively.
This model can be experimentally fabricated on the surface of a bulk $s$-wave superconductor~\cite{Yazdani2013}.

Starting from the initial trivial phase of the static Hamiltonian [Eq.~(\ref{eq:static_1D_Hamiltonian})], we employ a periodic sinusoidal modulation of the on-site chemical potential to induce a Floquet TSC phase hosting MEMs. The driving protocol is given by
\begin{eqnarray}
V(t)= \sum_{j} \Psi_{j}^{\dagger}  \left[ V_{0} \cos (\Omega t) \Gamma_{1} \right] \Psi_{j}  \label{eq:drive_propocol}\ ,
\end{eqnarray}
where $V_{0}$ and $\Omega(= 2\pi/T)$ represent the amplitude and frequency of the drive, respectively. Since the spin spiral breaks translation symmetry, 
one has to work in real space. The time-evolution operator can be expressed as a time-ordered (TO) exponential~\cite{AKG_time_evolution}
\begin{align}
U(t,0)&= \text{TO}  \exp \left[ - \int_{0}^{t} dt^{\prime} H(t^{\prime}) \right]  \ , \nonumber \\
&=\prod_{j=0}^{M-1} U(t_{j}+\delta t, t_{j}) \ , \label{eq:evolution_definiton}
\end{align}
where, $U(t_{j}+\delta t, t_{j})=e^{-i H(t_{j}) \delta t}$; with $\delta t = t/M$, $t_{j}=j \delta t$, and $M$ is the total number of Trotterization steps. The driven system continues to satisfy the chiral symmetry $\mathcal{C}$ and particle-hole symmetry $\mathcal{P}$.

\begin{figure}[]
\centering
\subfigure{\includegraphics[width=0.49\textwidth]{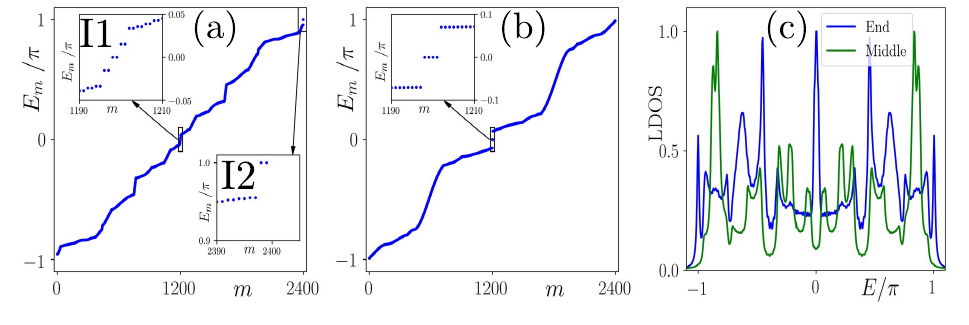}}
\caption{The quasi-energy spectrum obtained from the Floquet operator is presented for $B/\Delta = 2$ and $B/\Delta = 3$ in panels (a) and (b), respectively, employing OBC. The Floquet 0- and $\pi$-MEMs of panel (a) are highlighted in the insets I1 and I2. Panel (c) displays the energy-resolved normalized LDOS evaluated at the end (blue) and the middle (green) of the chain for $B/\Delta=2$. The peaks shown in green represent the emergent Shiba modes embedded in the Floquet quasi-energy spectrum. A 1D chain of 600 lattice sites is considered, with model parameters chosen as ($\mu/\Delta, t_{h}/\Delta, \theta)=(4, 1, 2\pi/3)$ and $\Omega/\Delta=1.5$, $V_{0}/\Delta=5$. This figure is adapted from~\cite{Mondal23_FSC} with permission, Copyright (2023) by the American Physical Society.}
\label{fig:dynamic}
\end{figure}

We present the quasi-energy spectrum as a function of the state index $m$ in Figs.~\ref{fig:dynamic}(a) and \ref{fig:dynamic}(b), corresponding to two different impurity strengths, namely $B/\Delta = 2$ and $B/\Delta = 3$, respectively. These parameters correspond to the non-topological regime of the associated static model. For the case of $B/\Delta = 2$, two MEMs are observed-one localized at each end of the system at quasi-energies $E=0$ and $E=\pi$ [see insets I1 and I2 
in Fig.~\ref{fig:dynamic}(a)]. In contrast, when $B/\Delta = 3$, multiple MEMs emerge at $E = 0$ as illustrated in the inset of Fig.~\ref{fig:dynamic}(b). In Floquet systems, due to the intrinsic particle-hole symmetry of the underlying BdG Hamiltonian, MEMs may also arise at finite quasi-energies, specifically at $E = \pm \pi$. Consequently, the modes located at $E = 0$ and $E = \pi \equiv -\pi$ can act as their own antiparticles~\cite{JiangPRL2011}. To further support these observations, Fig.~\ref{fig:dynamic}(c) illustrates the local density of states (LDOS) resolved in energy, evaluated both at the end (blue curve) and the center (green curve) of the chain. Pronounced peaks at $E = 0$ and $E = \pm \pi$ in the LDOS at the chain's end (blue curve) confirm the presence of Floquet MEMs. In contrast, the LDOS computed at the chain's center (green curve) lacks such features, instead indicating the formation of an emergent bulk Floquet quasi-energy Shiba band within the superconducting gap. The energetic separation between the lowest-lying features of the green curve near $E = 0$ and $E = \pm \pi$ defines the associated effective topological gap in the dynamical scenario.

\subsubsection{Time evolution of the Floquet MEMs}

\begin{figure}[]
\centering
\subfigure{\includegraphics[width=0.47\textwidth]{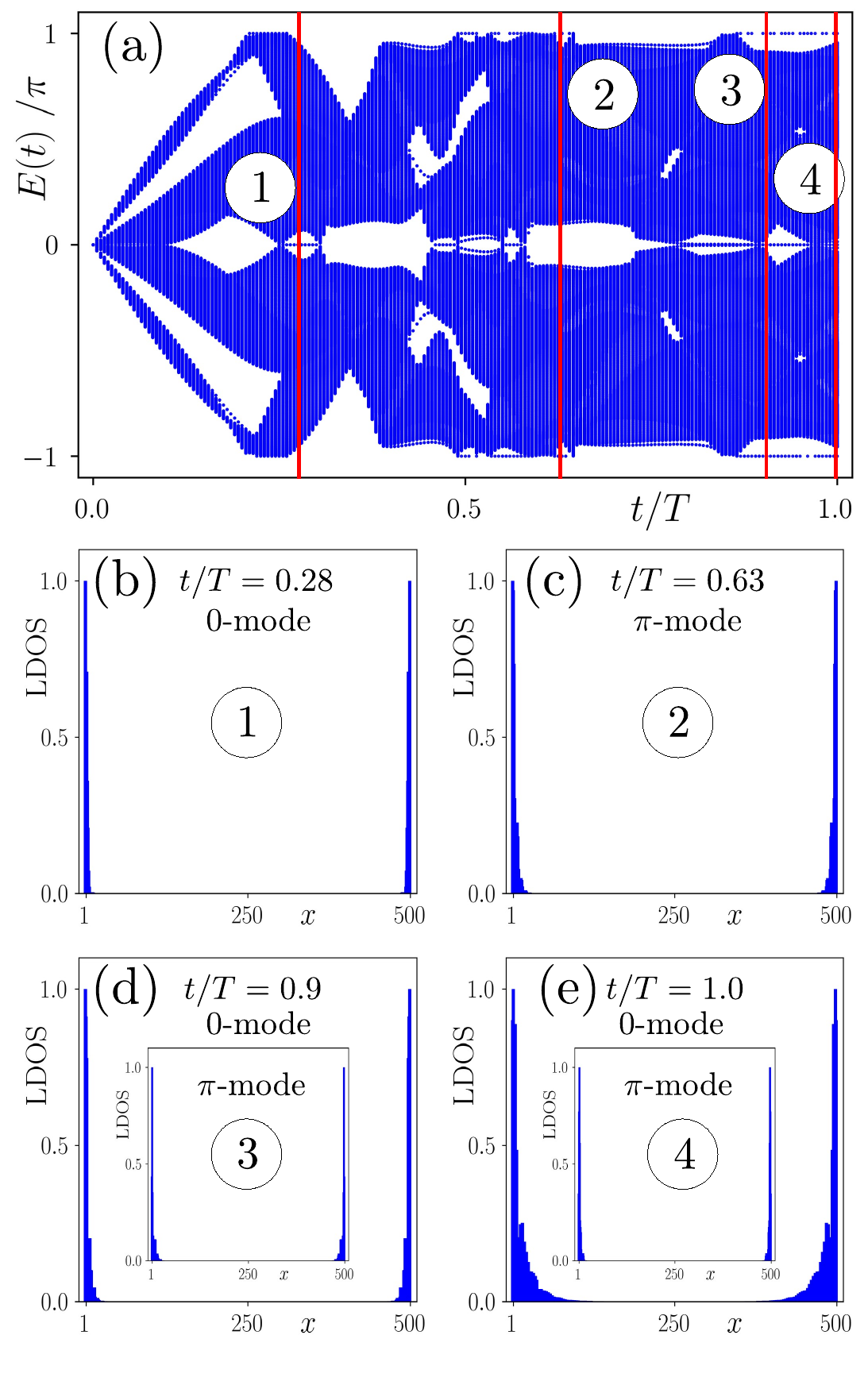}}
\caption{Panel (a) represents the time-dependent eigenvalue spectrum of the usual time-evolution operator \( U(t) \) as a function of time \( t \). Both the \( 0 \)- and \( \pi \)-modes appear and disappear over the time interval \( t \in [0,T] \). We focus on four specific time-instances: \(\textcircled{1}\) (\( t = 0.28T \)) with only \( 0 \)-modes, \(\textcircled{2}\) (\( t = 0.63T \)) with only \( \pi \)-modes, \(\textcircled{3}\) (\( t = 0.9T \)) with both \( 0 \)- and \( \pi \)-modes, and \(\textcircled{4}\) (\( t = T \)) case with both \( 0 \)- and \( \pi \)-modes. The LDOS for the \( 0 \)- and/or \( \pi \)-modes at these time instances are shown in panels (b)–(e). At \( t/T = 0.28 \) (\( t/T = 0.63 \)), only \( 0 \)-modes (\( \pi \)-modes) are present, while at \( t/T = 0.9 \) and \( t/T = 1 \), both \( 0 \)- and \( \pi \)-FMEMs appear. The time-dependent site-resolved LDOS for these FMEMs, evaluated at the specified time instances, are shown in panels (b)–(e), respectively. A 1D chain of 500 lattice sites is considered, with all model parameters set to the values provided in Fig.~\ref{fig:dynamic}. This figure is adapted from~\cite{Mondal23_FSC} with permission, Copyright (2023) by the American Physical Society.}
\label{fig:evolution}
\end{figure}

The explicit real-time dependence of the driving protocol motivates us to investigate the time evolution of the emergent FMEMs~\cite{AKG_time_evolution,Mondal23_FSC}. For this purpose, we diagonalize $U(t,0)$, and depict the instantaneous eigenvalue spectra $E(t)$ as a function of time $t$ in Fig.~\ref{fig:evolution}(a). Initially (at $t=0$), the system by is a trivial superconductor.We further observe that both the $0$- and $\pi$-Floquet MEMs (FMEMs) periodically emerge and disappear within the Floquet TSC phase as a function of time $t$, prior to completing a full driving period $T$. 
This can be identified distinctly from Fig.~\ref{fig:evolution}(a). To have a deeper understanding of the FMEMs at different time-instances, we consider four time-slices: $\textcircled{1}$ at $t=0.28T$ with only $0$-modes, $\textcircled{2}$ at $t=0.63T$ with only $\pi$-modes, $\textcircled{3}$ at $t=0.9T$ with both $0$- and $\pi$-modes, and $\textcircled{4}$ at $t=T$ when both $0$- and $\pi$-modes are present. The LDOS associated with $0$- and/or $\pi$-modes corresponding to the points $\textcircled{1}$-$\textcircled{4}$ are demonstrated with respect to the length of the chain in Figs.~\ref{fig:evolution}(b)-(e), respectively. From these features of LDOS, it is clearly evident that both the $0$- and $\pi$-modes are strongly localized at both ends of the 1D chain at different time instances.

\begin{figure}[]
\centering
\subfigure{\includegraphics[width=0.47\textwidth]{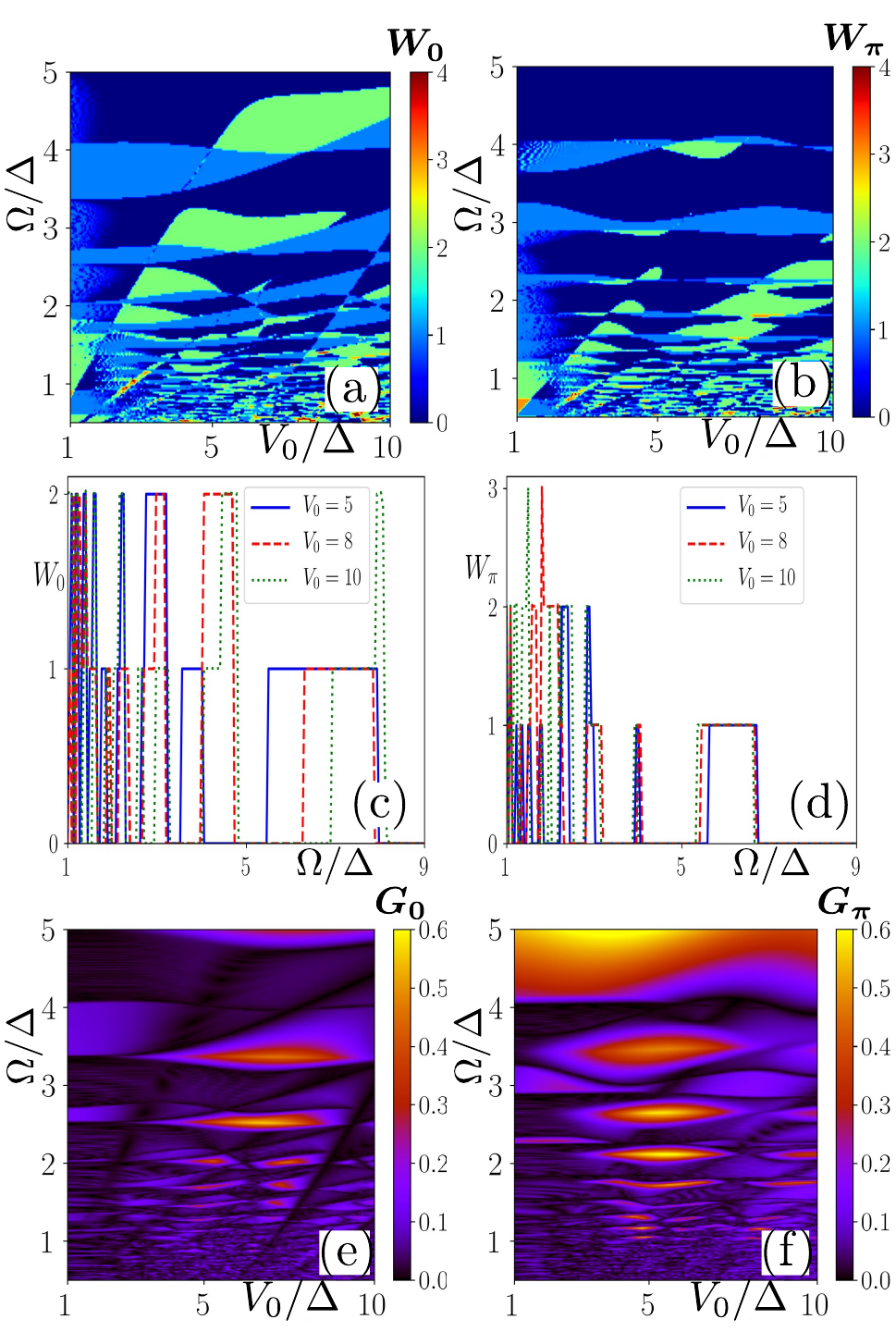}}
\caption{We present the dynamical winding numbers $W_0$ and $W_{\pi}$ in the $V_0$ -$\Omega$ plane in panels (a) and (b), respectively. The winding number $W_{\epsilon}$ characterizes the Floquet MEMs that reside within the quasi-energy gap $\epsilon$. In panels (c) and (d), we show $W_0$ and $W_{\pi}$ as a function of $\Omega$, for three fixed values of $V_0 /\Delta = 5$ (solid blue line), 8 (dashed red line), and 10 (dashed green line), respectively, for better clarity. The quasienergy gaps around quasi-energies 0 and $\pi$ are illustrated in panels (e) and (f), respectively. We set $B/\Delta = 5$ (the topological regime of the static model), while the remaining model parameters remain the same as mentioned in Fig.~\ref{fig:dynamic}. This figure is adapted from~\cite{Mondal23_FSC} with permission, Copyright (2023) by the American Physical Society.}
\label{fig:dynamical_winding}
\end{figure}


\subsubsection{Topological characterization}
Here, we discuss the topological characterization of emergent FMEMs in a driven helical Shiba chain. The helical structure of the spin chain breaks the translation symmetry of the system. Hence, we have to calculate the dynamical winding number $W_{\epsilon}$ for $\epsilon$-gap in real space using the periodized evolution operator and the chiral symmetry of the system as
\begin{equation}
    W_{\epsilon} = \left| \pm \frac{i}{2 \pi} \int_{-\pi}^{\pi} d\eta \, \text{Tr} \left[\left\{U_{\epsilon}^{\pm}(\eta)\right\}^{-1} \partial_{\eta} U_{\epsilon}^{\pm}(\eta)\right] \right| 
    \label{eq:dynamical_wind_HSC},
\end{equation}
where $\eta$ is the phase acquired by a particle to hop from one lattice site to another site in PBC. Note that this kind of PBC is also called as twisted boundary condition~(TBC)~\cite{Demler2010PRB,Titum2016PRX,Sreejith2016} and $\eta$ in real space is equivalent to $k$ in momentum space as far as topological characterization is concerned.

We compute the dynamical winding number $W_{0}$ and $W_{\pi}$ using the Eq.~(\ref{eq:dynamical_wind_HSC}) and depict them in the $V_0 $-$ \Omega$ plane in Figs.~\ref{fig:dynamical_winding}(a) and (b), respectively; the color bar represents the total number of FMEMs. Within specific parameter regimes, one finds that multiple $0$- and/or $\ pi$-modes can emerge. To provide further insight, we additionally illustrate $W_{0}$ and $W_{\pi}$ as a function of $\Omega$ for three different fixed values of $V_{0}$ in Figs.~\ref{fig:dynamical_winding}(c), and (d), respectively. Therefore, the presence of multiple FMEMs is further substantiated by the calculated dynamical winding number.  We further show the bulk gap at quasi-energy $0$ ($G_0$) and $\pi$ ($G_\pi$) choosing the $V_0 $-$ \Omega$ plane in Figs.~\ref{fig:dynamical_winding}(e) and (f), respectively. A closing of the gaps $G_0$ and $G_\pi$ leads to corresponding discontinuities in the winding numbers $W_{0}$ and $W_{\pi}$ respectively, thereby indicating a topological phase transition in the quasi-energy spectrum.
The one-to-one correspondence between the dynamical winding number $W_\epsilon$ and the bulk gap $G_\epsilon$ is evident from Fig.~\ref{fig:dynamical_winding}.

\subsection{Brief Summary}
In this section, we study the Floquet  generation of  FOTSC phases while the initial phase of the matter is trivial. The step drive with on-site and off-site terms in Rashba NW model generates particle-hole and chiral symmetries allowing for Floquet FOTSC phase as discussed in Sec. \ref{subsec:FloquetMEMRashba}. We discuss the effect of  harmonic drive in on-site term for helical Shiba chain in Sec. \ref{subsec:HelicalShiba} where above two {symmetries} are found responsible to protect Floquet FOTSC phase. In both the above instances, a bulk mass term emerges causing the band inversion. This mass term can close the trivial gap and reopen a topological gap within which edge modes reside.

\section{Floquet Second-order TSC in 2D systems} ~\label{Sec:2D}
In this section, we discuss the emergence of Majorana corner modes~(MCMs) in driven 2D SOTSC. First, we discuss the Floquet generation of SOTSC based on a QSHI and $s$-wave superconductor in Sec.~\ref{subsec:2D-TSC1}. We also discuss that, in the high-frequency limit, one can understand the emergence of the corner modes using low-energy edge theory. We further discuss the Wannier spectra and quadrupole moment to characterize the topological nature of the system. The model discussed in Sec.~\ref {subsec:2D-TSC1} represents an \textit{extrinsic} class of SOTSC; the emergence and number of corner modes can be controlled only by closing the edge-gaps, while the bulk gap may remain open throughout the topological transition. In Sec.~\ref{Subsec:DynamicalConstruction}, show the emergence of $0$- and the anomalous $\pi$-MCMs based on QSHI and $d$-wave superconductor. A dynamical prescription is utilized to obtain the topological classification. In this case, the topology can be changed only via the bulk gap closure, thus signifying an \textit{intrinsic} class of SOTSC.


\subsection{Floquet generation of SOTSC} \label{subsec:2D-TSC1}


We extend our discussion of Floquet TSCs to two spatial dimensions in order to find SOTSC. To this end, we consider a static BdG Hamiltonian describing a two-dimensional QSHI in proximity to a conventional bulk $s$-wave superconductor~\cite{Wu2020, ghosh2020floquet}
\begin{equation}
H_0 = \vect{N}(\vect{k}) \cdot \bm{\Gamma} \ ,
\label{ham1}
\end{equation}
where the vector $\vect{N}(\vect{k}) = (N_1, N_2, N_3, N_4)$ is defined through its components as $N_1 = 2 \lambda_x \sin k_x$, $N_2 = 2 \lambda_y \sin k_y$, $N_3 = \xi_{\vect{k}} = m_0 - 2t_x \cos k_x - 2t_y \cos k_y$, and $N_4 = \Delta$. Here, $\lambda_{x,y}$, $m_0$, $t_{x,y}$, and $\Delta$ represent SOC strength, onsite mass, hopping amplitude, and $s$-wave gap. The corresponding $\vect{\Gamma}$ matrices are given by $\Gamma_1 = \sigma_x s_z$, $\Gamma_2 = \sigma_y \tau_z$, $\Gamma_3 = \sigma_z \tau_z$, and $\Gamma_4 = s_y \tau_y$. Here, $\bm{\sigma}$, $\bm{s}$, and $\bm{\tau}$ denote independent sets of Pauli matrices marking the orbital $(a,b)$, spin $(\uparrow,\downarrow)$, and particle-hole $(e,h)$ subspaces, respectively. Here, the Nambu spinor basis is chosen as $C_{\vect{k}} = \left( c_{\vect{k},a\uparrow}, c^\dagger_{-\vect{k},a\uparrow}, c_{\vect{k},a\downarrow}, c^\dagger_{-\vect{k},a\downarrow}, c_{\vect{k},b\uparrow}, c^\dagger_{-\vect{k},b\uparrow}, c_{\vect{k},b\downarrow}, c^\dagger_{-\vect{k},b\downarrow} \right)^{T}$ and our aim is to realize a SOTSC phase characterized by the emergence of localized MCMs, enabled by the interplay of proximity-induced superconductivity and nontrivial topology inherent in the QSHI platform.

Then, in order to explore the emergence of FSOTSC phases, we incorporate a time-periodic driving scheme realized via $\delta$-function kicks, following Ref.~\cite{ghosh2020floquet}. The drive is considered in two distinct cases:

\paragraph*{Case 1:} A time-periodic in-plane Zeeman field applied along the $x$-direction introduces time dependence through
\begin{equation}
V(t) = h_x \Gamma_5 \sum_{r=1}^{\infty} \delta(t - rT) \ , 
\label{kick1}
\end{equation}
where, $T$ is the drive period and $\Gamma_5 = s_x \tau_z$. This driving term explicitly breaks time-reversal symmetry (TRS). In the static limit, the modified Hamiltonian takes the form $H^{\text{I}}_{\text{sta}} = H_0 + h_x \Gamma_5$, which is known to harbor MCMs in the SOTSC phase when the condition $h_x > \Delta$ is satisfied~\cite{Wu2020}. 

\paragraph*{Case 2:} Alternatively, we consider a momentum-dependent perturbation that breaks crystalline symmetry:
\begin{equation}
V(t) = \Lambda(\vect{k}) \Gamma_6 \sum_{r=1}^{\infty} \delta(t - rT) \ ,
\label{kick2}
\end{equation}
with $\Gamma_6 = \sigma_x s_x \tau_z$ and $\Lambda(\vect{k}) = \Lambda(\cos k_x - \cos k_y)$. The latter explicitly violates the four-fold rotational symmetry ($C_4$) of the system. Physically, $\Gamma_6$ corresponds to a hopping process that simultaneously flips both orbital and spin degrees of freedom. In the static regime, the effective Hamiltonian $H^{\text{II}}_{\text{sta}} = H_0 + \Lambda(\vect{k}) \Gamma_6$ supports zero-energy MCMs, protected by unitary chiral symmetry and an anti-unitary particle-hole symmetry~\cite{YanPRB2019}.

Note that, in both cases, the delta-kick driving modifies the quasienergy spectrum, enabling the engineering of Floquet-induced topological superconducting phases that may not be accessible in static systems. See below for the details. In an experimental setup, an ideal $\delta$-kick is an approximation to a short-duration pulse. Specifically, the corresponding physical parameter (e.g. the Zeeman field in Eq.~(\ref{kick1}) is applied for a finite time interval $\Delta t$ satisfying $\Delta t \ll T$. In the limit $\Delta t/T \rightarrow 0$ while keeping the pulse area fixed, the finite pulse reproduces the ideal $\delta$-kick description.  Note that the physical origin of  $\delta$-kick is indeed vast such as, it has been adopted for cooling quantum matter \cite{PhysRevResearch.3.033261,PhysRevLett.78.2088}, handling classical kicked rotor \cite{PhysRevLett.73.2974}, squeezing in quantum optics \cite{PhysRevLett.127.183401}, quantum chaos \cite{PhysRevE.106.034207} ranging quite beyond the conventional solid state systems.  

\subsubsection{Case 1}{\label{sec:III}}
The Floquet operator  for this case (see Eq.~(\ref{kick1})) can ne written as~\cite{ghosh2020floquet}
\begin{equation}
U(T) = \exp(-i H_0 T) \exp(-i h_x \Gamma_5)\ .
\label{fo}
\end{equation}
This operator can be recast into a more compact and insightful form:
\begin{equation}
U(T) = C_T \left(n_0 - i n_5 \Gamma_5\right) - i S_T \sum_{j=1}^{4} \left( m_j \Gamma_j + p_j \Gamma_{j5} \right)\ ,
\end{equation}
where the coefficients are defined as follows:
\begin{align*}
&C_T = \cos\left(\left| \vect{N}(k) \right| T \right), \quad S_T = \sin\left(\left| \vect{N}(k) \right| T \right), \\
&n_0 = \cos h_x, \quad n_5 = \sin h_x, \quad n_j = \frac{N_j(k)}{|\vect{N}(k)|}, \\
&m_j = n_j \cos h_x, \quad p_j = n_j \sin h_x\ ,\\
&\Gamma_{j5} = \frac{1}{2i} [\Gamma_j, \Gamma_5]\quad \text{for } j = 1,2,3,4\ .
\end{align*}

\begin{figure}
\centering
\subfigure{\includegraphics[width=0.48\textwidth]{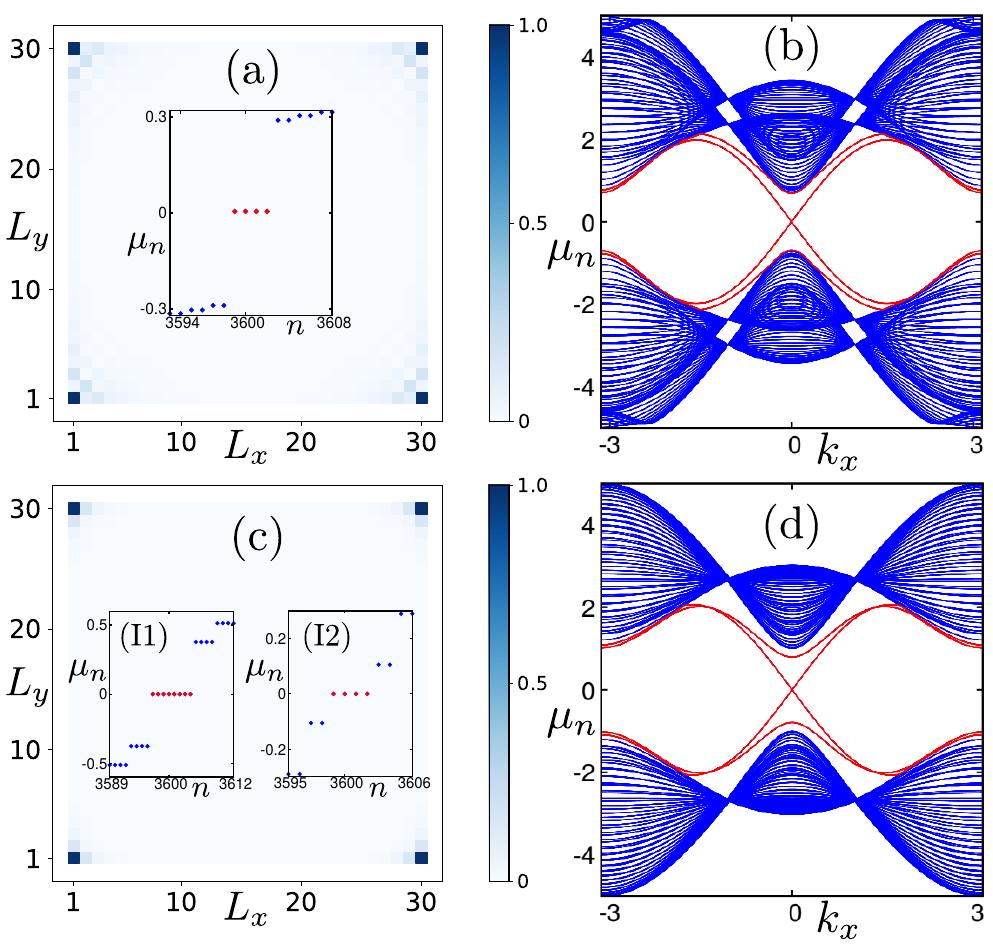}}
\caption{ (a) Local density of states (LDOS) in a finite square geometry as a function of the spatial directions ($L_x = L_y = 30$) is shown for the Floquet protocol defined in Eq.~(\ref{kick1}). The color bar represents the weight of the normalized LDOS. The inset displays the corresponding quasi-energy spectrum. The model parameters used are: $m_0 = t_x = t_y = \lambda_x = \lambda_y = 1.0$, $\Delta = 0.4$, $h_x = 0.4$, and $T = 0.628$. (b) Edge spectrum for the effective Hamiltonian in Eq.~(\ref{heff1}), obtained with OBC along the $y$-direction. The edge gap closes at the critical point $\Delta = h_x / T$. (c) LDOS for the $\Lambda(k)$-kick protocol [Eq.~(\ref{kick2})] is shown with the same system size and parameter values as in panel (a), except $\Lambda = 0.3$. Insets (I1) and (I2) show the quasi-energy spectra for two different regimes: (I1) $|\frac{m \Lambda}{2 t_{x(y)} T}| > \Delta$, where eight zero-energy MCMs appear; (I2) $|\frac{m \Lambda}{2 t_x T}| < \Delta < |\frac{m \Lambda}{2 t_y T}|$, where four MCMs are present. (d) Edge spectrum corresponding to Eq.~(\ref{heff2}) with OBC along the $y$-direction. The closing of the edge gap occurs at $\Delta = |\frac{m \Lambda}{2 t_x T}|$, indicating a topological phase transition. See main text for further details. This figure is adapted from~\cite{ghosh2020floquet} with permission, Copyright (2021) by the American Physical Society.}
\label{LDOS}
\end{figure}

The corresponding effective Hamiltonian, defined through $U(T) = \exp(-i H_{\text{eff}} T)$, takes the form:
\begin{widetext}
\begin{equation}
H^{\text{I}}_{\text{eff}} = \frac{\epsilon_k}{\sin(\epsilon_k T)} \left[
\sin\left(|\vect{N}(k)| T\right) \cos h_x \sum_{j=1}^4 n_j \Gamma_j
+ \cos\left(|\vect{N}(k)| T\right) \sin h_x \Gamma_5
+ \sin\left(|\vect{N}(k)| T\right) \sin h_x \sum_{j=1}^4 n_j \Gamma_{j5}
\right] \ ,
\end{equation}
\end{widetext}
where, the quasienergy $\epsilon_k$ is given by
$\epsilon_k = \frac{1}{T} \cos^{-1}\left[ \cos\left(|\vect{N}(k)| T\right) \cos h_x \right]$. In the high-frequency limit, i.e., $T \to 0$ and $h_x \to 0$, we expand the trigonometric functions and retain leading-order terms. This yields the approximate effective Hamiltonian:
\begin{equation}
H^{\text{I}}_{\text{eff}} \approx \sum_{j=1}^{4} N_j(k) \Gamma_j + \frac{h_x}{T} \Gamma_5 + h_x \sum_{j=1}^{4} N_j(k) \Gamma_{j5}\ .
\label{heff1}
\end{equation}
Importantly, the terms proportional to $\Gamma_{j5}$ in Eq.~(\ref{heff1}) arise purely due to the periodic driving and are absent in the static model. Among these, only $\Gamma_{15}$ does not commute with the static Hamiltonian $H^{\text{I}}_{\text{sta}} = H_0 + h_x \Gamma_5$, implying that the effective Floquet Hamiltonian $H^{\text{I}}_{\text{eff}}$ no longer preserves the chiral symmetry generated by the unitary operator $\mathcal{C}=\sigma_x s_y \tau_z$. Nonetheless, $H^{\text{I}}_{\text{eff}}$ retains particle-hole symmetry $\mathcal{P}=\tau_x \mathcal{K}$, which ensures the protection of zero-quasienergy modes. As a consequence of this dynamical time-reversal symmetry breaking, the system can host FSOTSC phases featuring MCMs. The local density of states (LDOS) associated with the zero (numerically approximated) quasienergy Floquet eigenstates is shown in Fig.~\ref{LDOS}(a), demonstrating that these states are spatially localized at the four corners of the sample 
- a hallmark of the FSOTSC phase.

\begin{figure}
\centering
\subfigure{\includegraphics[width=0.48\textwidth]{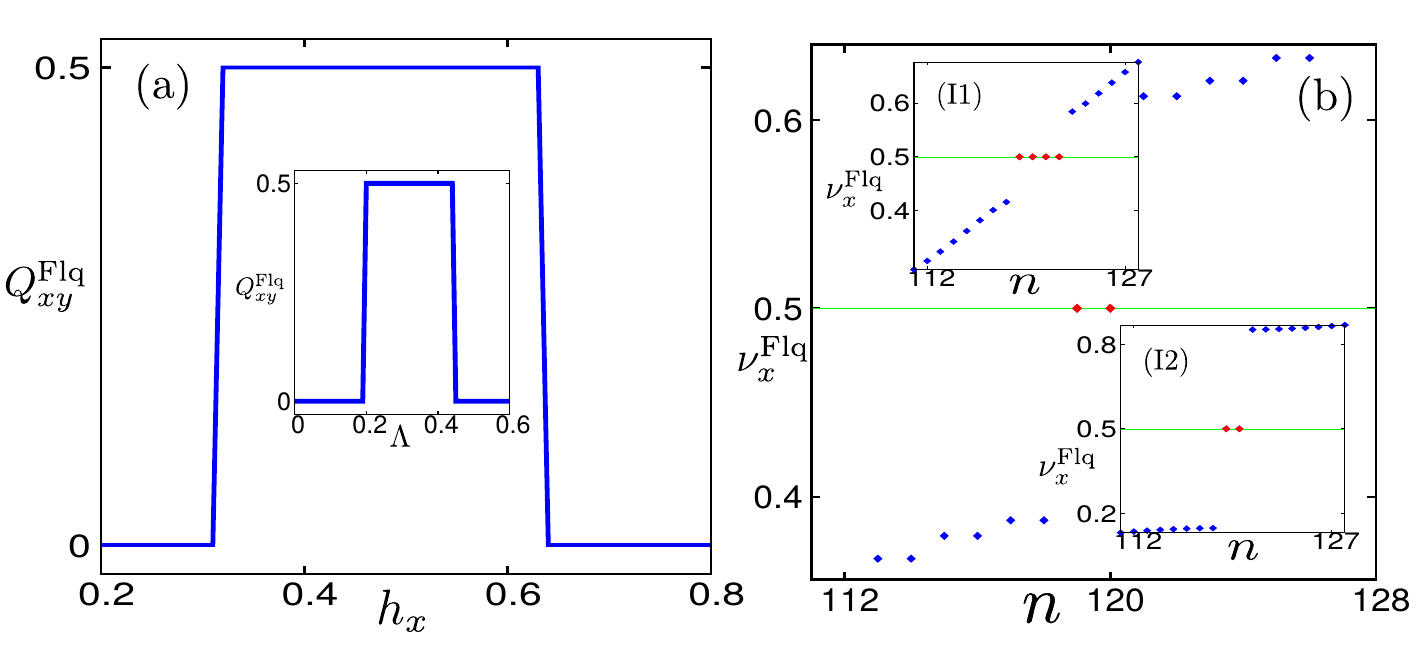}}
\caption{ 
(a) Floquet quadrupole moment (FQM) $Q^{\rm Flq}_{xy}$ is depicted against Zeeman field amplitude $h_x$ for the driving protocol described in Eq.~(\ref{kick1}). The inset shows $Q^{\rm Flq}_{xy}$ as a function of $\Lambda$ for the symmetry-breaking protocol in Eq.~(\ref{kick2}). The model parameters used are $L_x = L_y = 16$, $m_0 = t_x = t_y = \lambda_x = \lambda_y = 1.0$, $\Delta = 0.4$, and $T = 0.628$. 
(b) Floquet Wannier spectrum (FWS), characterized by the eigenvalues $\nu^{\rm Flq}_x$, is displayed as a function of the eigenstate index for protocol Eq.~(\ref{kick1}) (main panel), with $h_x > \Delta T$. Insets (I1) and (I2) correspond to the protocol in Eq.~(\ref{kick2}) under different parameter regimes. Inset (I1) shows the case where $|\frac{m\Lambda}{2t_{x(y)}T}| > \Delta$, indicating the presence of eight Majorana corner modes (MCMs). Inset (I2) corresponds to the regime $|\frac{m\Lambda}{2t_xT}| < \Delta < |\frac{m\Lambda}{2t_yT}|$, where four MCMs are realized. This figure is adapted from~\cite{ghosh2020floquet} with permission, Copyright (2021) by the American Physical Society.}
\label{invariant}
\end{figure}

To examine the topological robustness of the Floquet-generated MCMs in the SOTSC phase~\cite{ghosh2020floquet}, we numerically compute two complementary topological invariants: the Floquet quadrupole moment (FQM), denoted by $Q^{\rm Flq}_{xy}$, and the Floquet Wannier Spectra (FWS), characterized by $\nu^{\rm Flq}_x$. These quantities are shown in Fig.~\ref{invariant}(a) and Fig.~\ref{invariant}(b), respectively. The FQM $Q^{\rm Flq}_{xy}$ is computed using the many-body Floquet ground state $n_F$, constructed by filling all quasi-energy states with eigenvalues $\mu_n$ lying within the interval $[-\omega/2, 0]$, where $\omega = 2\pi/T$ is the driving frequency. Explicitly, $n_F = \sum_{\mu_n \le 0} |\phi_n\rangle \langle \phi_n|$, where $|\phi_n\rangle$ are the eigenstates of the Floquet operator [cf. Eq.~(\ref{fo})] obtained via exact diagonalization~\cite{Nag19,Ghosh2020}. We find that $Q^{\rm Flq}_{xy}$ remains quantized as $Q^{\rm Flq}_{xy} \equiv \text{mod}(Q^{\rm Flq}_{xy}, 1) = 0.5$, over a finite range of the driving amplitude $h_x$ in the protocol defined by Eq.~(\ref{kick1}). This quantization, shown in Fig.~\ref{invariant}(a), serves as strong evidence of the topological protection of the MCMs in the Floquet SOTSC phase. In addition to the FQM, we compute the FWS along the $x$-direction, obtained from the Wannier Hamiltonian ${\mathcal H}^{\rm Flq}_{{\mathcal W}_x}$ constructed using the high-frequency effective Hamiltonian $H^{\text{I}}_{\text{eff}}$ [cf. Eq.~(\ref{heff1})] in the $T \to 0$ limit. 
The spectrum $\nu^{\rm Flq}_x$ exhibits two isolated eigenvalues pinned at $0.5$, as illustrated in Fig.~\ref{invariant}(b), indicating the presence of localized MCMs at the system’s boundaries. A similar structure holds for $\nu^{\rm Flq}_y$ along the $y$-direction, further corroborating the existence of a Floquet SOTSC phase under this driving protocol. Overall, the quantized values of both $Q^{\rm Flq}_{xy}$ and $\nu^{\rm Flq}_x$ signify that the corner modes are topologically protected and robust against moderate variations in the driving amplitude.

We now proceed to derive the low-energy edge theory in the Floquet regime, starting from the effective Hamiltonian $H^{\text{I}}_{\text{eff}}$ [cf. Eq.~(\ref{heff1})] in the high-frequency limit $T \to 0$. Expanding the bulk Hamiltonian around the $\Gamma$ point ($\vect{k} = 0$), the low-energy continuum model takes the form:
\begin{align}
H^{\text{I}}_{\text{eff}, \vect{k}} &\approx \left(m + t_x k_x^2 + t_y k_y^2\right) \Gamma_3 + 2\lambda_x k_x \Gamma_1 + 2\lambda_y k_y \Gamma_2 \nonumber \\
&\quad + \Delta \Gamma_4 + \frac{h_x}{T} \Gamma_5 + 2\lambda_x h_x k_x \Gamma_{15}\ ,
\label{model1lowenergy}
\end{align}
where $m = m_0 - 2t_x - 2t_y$, and we consider $m < 0$ to ensure a topologically nontrivial regime~\cite{FuKane2007}.

As a representative case, we consider Edge I, which is defined by imposing OBC along the $x$-direction and PBC along the $y$-direction. The effective Hamiltonian can be decomposed as:
\begin{equation}
H^{\text{I}}_{\text{eff}} = H_I(-i\partial_x) + H_P(-i\partial_x, k_y, h_x)\ ,
\end{equation}
where we neglect the $k_y^2$ term as a higher-order correction. The components are given by:
\begin{align}
H_I &= \left(m - t_x \partial_x^2\right) \Gamma_3 - 2i\lambda_x \partial_x \Gamma_1\ , \nonumber \\
H_P &= 2\lambda_y k_y \Gamma_2 + \Delta \Gamma_4 + \frac{h_x}{T} \Gamma_5 - 2i \lambda_x h_x \partial_x \Gamma_{15}\ .
\end{align}

Assuming a zero-energy eigenstate of the form $\Psi(x) = \psi(x) \Phi$, where $\Phi$ is a spinor and $\psi(x)$ decays exponentially along the $x$-direction, we identify the relevant low-energy spinor subspace as:
\begin{align*}
\Phi_1 &= \ket{\sigma_y = +1} \otimes \ket{s_z = +1} \otimes \ket{\tau_z = -1}\ , \\
\Phi_2 &= \ket{\sigma_y = -1} \otimes \ket{s_z = +1} \otimes \ket{\tau_z = +1}\ , \\
\Phi_3 &= \ket{\sigma_y = -1} \otimes \ket{s_z = -1} \otimes \ket{\tau_z = -1}\ , \\
\Phi_4 &= \ket{\sigma_y = +1} \otimes \ket{s_z = -1} \otimes \ket{\tau_z = +1}\ .
\end{align*}

A similar construction applies to other edges by appropriately modifying boundary conditions. The resulting low-energy edge Hamiltonian along the $l^{\rm{th}}$ edge takes the general form:
\begin{equation}
H^{\text{I}, \text{Edge}}_l = -i A_l s_z \partial_l - i B_l s_y \tau_z \partial_l - \Delta s_y \tau_y - h_l s_x \tau_z\ ,
\label{Model1edge}
\end{equation}
with parameters defined as:
\begin{align*}
A_l &= \{-2\lambda_y, 2\lambda_x, -2\lambda_y, 2\lambda_x\}\ , \\
B_l &= \{0, -2\lambda_x h_x, 0, -2\lambda_x h_x\}\ , \\
h_l &= \{0, \frac{h_x}{T}, 0, \frac{h_x}{T}\}\ ,
\end{align*}
corresponding to edges I–IV, respectively.

From Eq.~(\ref{Model1edge}), it is evident that the superconducting pairing term $\Delta s_y \tau_y$ opens a gap on all four edges, as it anticommutes with the kinetic term $s_z$ (i.e., $\{s_z, s_y \tau_y\} = 0$). However, the Zeeman term $h_x$ only contributes to edges II and IV, introducing a finite mass gap $h_l$ via $s_x \tau_z$.

To illustrate this behavior, Fig.~\ref{LDOS}(b) displays the quasi-energy spectrum in a semi-infinite slab geometry. A clear gap closing is observed at $\Delta = h_x / T$, indicating a topological phase transition. Closer analysis reveals two decoupled Dirac blocks with mass terms $\Delta \pm \frac{h_x}{T}$ on edge II, whereas both blocks on edge I retain the same mass sign. This difference implies that for $h_x > \Delta T$, edges I and II host Dirac masses of opposite signs, resulting in localized MCMs at their intersection-i.e., at system corners. Notably, in contrast to the static case, the Floquet SOTSC phase allows the emergence of MCMs at significantly smaller values of the in-plane magnetic field $h_x$ as the drive frequency increases ($T \to 0$). This illustrates how periodic driving enables enhanced tunability of topological phases.

\subsubsection{Case 2}{\label{sec:VI}}
We now consider the driving protocol defined by Eq.~(\ref{kick2}) and analyze the resulting Floquet SOTSC phase. Numerical calculations reveal the presence of strongly localized MCMs. In this case, we observe the emergence of eight Floquet MCMs at quasi-energy $\mu_n = 0$ (within numerical precision), as shown in Figs.~\ref{LDOS}(c)-(I1). These MCMs exhibit distinct character from those generated via the in-plane magnetic field kick drive protocol [cf. Eq.~(\ref{kick1})], where only four zero-energy MCMs are observed. Specifically, for $h_x$-kicking, each quasi-zero-energy Floquet state is found to localize at a single corner. 
In contrast, under the $\Lambda(k)$-kicking protocol, each corner hosts at least two quasi-degenerate Floquet states, indicating that multiple MCMs share the same spatial region. In the high-frequency limit $T \rightarrow 0$ and for small kicking amplitude $\Lambda \rightarrow 0$, the effective Floquet Hamiltonian acquires the following approximate form:
\begin{equation}
H^{\rm II}_{\rm eff} \approx \sum_{j=1}^{4} N_j(k) \Gamma_j + \frac{\Lambda(k)}{T} \Gamma_6 + \Lambda(k) \sum_{j=1}^{4} N_j(k) \Gamma_{j6}\  ,
\label{heff2}
\end{equation}
where, $\Gamma_{j6} = \frac{1}{2i} [\Gamma_j, \Gamma_6]$. Similar to the previous case, it also preserves the chiral and  particle-hole symmetries.

Following a procedure analogous to that for the $h_x$-kicking case, we derive the corresponding low-energy edge Hamiltonian for the $l^{\rm{th}}$ edge in coordinate space as
\begin{equation}
H^{\rm II, Edge}_l = -i A_l s_z \partial_l + i B_l s_x \partial_l - \Delta s_y \tau_y + \Lambda_l s_y\ ,
\label{Model2edge}
\end{equation}
with coefficients defined as:
\begin{align*}
A_l &= \{-2\lambda_y, 2\lambda_x, -2\lambda_y, 2\lambda_x\}, \\
B_l &= \left\{\frac{m\lambda_y\Lambda}{t_x}, \frac{m\lambda_x\Lambda}{t_y}, \frac{m\lambda_y\Lambda}{t_x}, \frac{m\lambda_x\Lambda}{t_y} \right\}, \\
\Lambda_l &= \left\{ -\frac{m\Lambda}{2t_x T}, \frac{m\Lambda}{2t_y T}, -\frac{m\Lambda}{2t_x T}, \frac{m\Lambda}{2t_y T} \right\}.
\end{align*}

To gain physical insight, we first consider the case $\Delta = 0$. The position-dependent $\Lambda_l$ changes sign across adjacent edges, inducing Dirac mass domain walls at each corner. These domain walls result in localized zero-energy Jackiw-Rebbi bound states~\cite{jackiw1976solitons,Nag19,Ghosh2020}. With superconducting pairing turned on ($\Delta \ne 0$), the edge Hamiltonian can be block-diagonalized in the particle-hole space:
\begin{equation}
H^{\rm II, Edge}_l = H_{\tau_y=+1} \oplus H_{\tau_y=-1}\ ,
\end{equation}
where
\begin{align}
H_{\tau_y = +1} &= -i A_l s_z \partial_l + i B_l s_x \partial_l + (\Lambda_l - \Delta) s_y\ , \nonumber \\
H_{\tau_y = -1} &= -i A_l s_z \partial_l + i B_l s_x \partial_l + (\Lambda_l + \Delta) s_y\ .
\label{ham2_edge}
\end{align}

The presence of domain walls in both $\tau_y = \pm 1$ sectors occurs when $|\Lambda_l| > \Delta$, i.e., when $\left|\frac{m\Lambda}{2t_{x/y} T}\right| > \Delta$. In this regime, each corner hosts two MCMs, leading to the observed eight-fold degeneracy [see Fig.~\ref{LDOS}(c)-(I1)]. The localized wavefunction at the corner $C$ on edge-I can be written as a superposition:
\begin{equation}
\Psi_C(y) \sim \alpha\, e^{- \frac{M_I - \Delta}{2\lambda_y} y} \Phi_C^1 + \beta\, e^{- \frac{M_I + \Delta}{2\lambda_y} y} \Phi_C^2\ ,
\end{equation}
where, $M_I = |\Lambda_I|$ and $\Phi_C^{1,2}$ are spinors associated with the two decoupled blocks.
This behavior contrasts with the $h_x$-kicked case, where only one of the edge Hamiltonian blocks hosts a domain wall, while the other remains gapped, yielding a single MCM per corner. Interestingly, when the mass inversion condition is satisfied only for one block-e.g., $\left|\frac{m\Lambda}{2t_x T}\right| < \Delta < \left|\frac{m\Lambda}{2t_y T}\right|$-only one MCM appears per corner. This crossover is illustrated in Fig.~\ref{LDOS}(c)-(I2), and the corresponding spectral gap closing at $\Delta = \left| \frac{m\Lambda}{2t_x T} \right|$ is demonstrated in Fig.~\ref{LDOS}(d).

To characterize the topological nature of these phases, we compute the FQM $Q^{\rm Flq}_{xy}$. 
As shown in the inset of Fig.~\ref{invariant}(a), we find $Q^{\rm Flq}_{xy} = 0$ in the phase with eight MCMs [Fig.~\ref{LDOS}(c)-(I1)] and $Q^{\rm Flq}_{xy} = 0.5$ when only four MCMs are present [Fig.~\ref{LDOS}(c)-(I2)]. This indicates that when two MCMs coexist at each corner, they can hybridize into finite-energy fermionic states, leading to a vanishing quadrupole moment. In contrast, the Floquet Wannier spectrum (FWS) remains sensitive to the number of MCMs. Specifically, the spectrum shows four (two) eigenvalues pinned at $\nu^{\rm Flq}_{x(y)} = 0.5$ for the eight-MCM (four-MCM) phase, as shown in Figs.~\ref{invariant}(b)-(I1) and \ref{invariant}(b)-(I2), respectively. For the $h_x$-kick protocol, only two such eigenvalues are observed, consistent with the presence of four MCMs. Thus, the phase with eight MCMs may be interpreted as a weak Floquet SOTSC phase, while the one with four MCMs represents a strong SOTSC phase.


\subsection{Dynamical construction of Quadrupolar topological superconductors} \label{Subsec:DynamicalConstruction}
We now discuss a full hierarchy to generate the dynamical modes, namely, MPMs, following the step-drive protocol.  
For the case of an unconventional $d$-wave superconducting pairing, we adopt a three-step periodic driving scheme designed to realize both the 2D FSOTSC and the 3D FTOTSC~\cite{GhoshDynamical2022}. The driving protocol is given by
\begin{eqnarray}
H_{d \rm D}(\vect{k},t) &=& J_1' h_{1,d \rm D}(\vect{k}) \ ; \quad t \in [0,T/4], \nonumber \\
&=& J_2' h_{2,d \rm D}(\vect{k}) \ ; \quad t \in (T/4,3T/4]\ , \nonumber \\
&=& J_1' h_{3,d \rm D}(\vect{k}) \ ; \quad t \in (3T/4,T]\ .
\label{drive}
\end{eqnarray}
Here, $J_i' h_{i,d \rm D}(\vect{k})$ represents the Hamiltonian in the $i^{\rm th}$ step for a $d$-dimensional system. The constants $J_1'$ and $J_2'$ have dimensions of energy. We further define dimensionless parameters $(J_1,J_2)=(J_1'T, J_2'T)$, where $T$ is the period of the drive. To generate a 2D FSOTSC, we choose 
$h_{1,2 \rm D}(\vect{k})=h_{3,2 \rm D}(\vect{k})=\tau_z\sigma_z$ and  
$h_{2,2 \rm D}(\vect{k})=\epsilon_{2 \rm D}(\vect{k}) \tau_z \sigma_z + \Lambda_{2 \rm D}(\vect{k}) + \Delta_{2 \rm D}(\vect{k})$,  
with  
\begin{eqnarray}
\epsilon_{2 \rm D}(\vect{k}) &=& \cos k_x + \cos k_y\ , \nonumber \\
\Lambda_{2 \rm D}(\vect{k}) &=& \sin k_x \, \tau_z \sigma_x s_z + \sin k_y \, \tau_z \sigma_y\ , \nonumber \\
\Delta_{2 \rm D}(\vect{k}) &=& \Delta \, (\cos k_x - \cos k_y)\,\tau_x\ .
\end{eqnarray}

Here, $\epsilon_{2 \rm D}(\vect{k})$ and $\Lambda_{2 \rm D}(\vect{k})$ account for the hopping and spin–orbit coupling terms in 2D, respectively, while $\Delta_{2 \rm D}(\vect{k})$ corresponds to the $d_{x^2-y^2}$ pairing symmetry~\cite{Yan2018,ghosh2020floquet2}. In the first and last steps of the drive, only the on-site contribution [$h_{1,d \rm D}(\vect{k})$] appears in the Hamiltonian, which simplifies the analytical treatment and facilitates topological characterization~\cite{Rudner2013,Huang2020}. Importantly, each $h_{i,d \rm D}(\vect{k})$ respects particle–hole, chiral, and mirror symmetries, with the mirror symmetry being especially significant for the ensuing analysis.

The quasienergy spectrum is obtained from the eigenvalue equation for the Floquet operator as,
\begin{eqnarray}
U_{2 \rm D}(\vect{k},T)\ket{\Psi} = e^{-iE(\vect{k})}\ket{\Psi}\ ,
\end{eqnarray}
yielding
\begin{eqnarray}\label{eigen}
E(\vect{k}) &=& \pm \arccos \Big[ 
\cos \!\left(\tfrac{\alpha_{2 \rm D}(\vect{k}) J_1}{2}\right) 
\cos \!\left(\tfrac{\beta_{2 \rm D}(\vect{k}) J_2}{2}\right) \nonumber \\
&& - \sin \!\left(\tfrac{\alpha_{2 \rm D}(\vect{k}) J_1}{2}\right) 
\sin \!\left(\tfrac{\beta_{2 \rm D}(\vect{k}) J_2}{2}\right) 
\chi_{2 \rm D}(\vect{k}) \Big]\ ,
\end{eqnarray}
where ,
$\alpha_{2 \rm D}(\vect{k})=\lvert h_{1,2 \rm D}(\vect{k}) \rvert=\lvert h_{3,2 \rm D}(\vect{k}) \rvert$,  
$\beta_{2 \rm D}(\vect{k})=\lvert h_{2,2 \rm D}(\vect{k}) \rvert$, and  
$\chi_{2 \rm D}(\vect{k})=\frac{\epsilon_{2 \rm D}(\vect{k})}{\alpha_{2 \rm D}(\vect{k}) \beta_{2 \rm D}(\vect{k})}$. Topological phase transitions take place when the quasienergy gap closes at $E(\vect{k})=0,\pm \pi$ for momenta $(k_x,k_y)=(0,0)$ or $(\pi,\pi)$. This condition leads to the generalized phase boundary in 2D~\cite{ghosh2021systematic},
\begin{eqnarray}
{|J_2|} = \frac{|J_1|}{2} + n\pi, \qquad n \in \mathbb{Z}\ .
\label{eq_pd}
\end{eqnarray}

\begin{figure}[]
	\centering
	\subfigure{\includegraphics[width=0.49\textwidth]{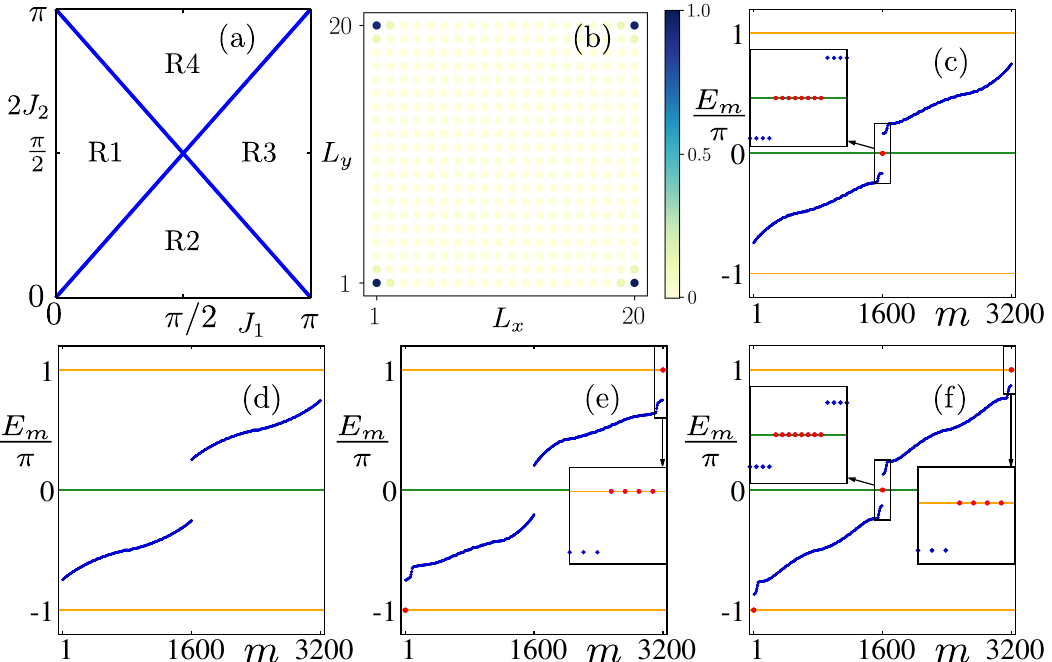}}
\caption{ (a) Phase diagram of the 2D FSOTSC in the $J_1$–$J_2$ plane, obtained from Eq.~(\ref{eq_pd}).  
(b) LDOS for a 2D square lattice of size $L_x \times L_y$ at quasi-energies $E_m=0$ and $E_m=\pm \pi$ is shown. The color bar represents the weight of the normalized LDOS. In panels (c)–(f), quasienergy spectra $E_m$ are plotted as a function of the state index $m$ for the four distinct regions: R1, R2, R3, and R4. The driving parameters are chosen as $(J_1,2J_2) = \left(\tfrac{\pi}{4},\tfrac{\pi}{2}\right)$, $\left(\tfrac{\pi}{2},\tfrac{\pi}{4}\right)$, $\left(\tfrac{3\pi}{4},\tfrac{\pi}{2}\right)$, and $\left(\tfrac{\pi}{2},\tfrac{3\pi}{4}\right)$ for R1, R2, R3, and R4, respectively. We fix the pairing amplitude at $\Delta=1.0$ for all numerical computations. This figure is 
adapted from~\cite{GhoshDynamical2022} with permission, Copyright (2022) by the American Physical Society.}
\label{2DSOTSC}
\end{figure}

The resulting phase diagram in the $J_1$-$J_2$ plane is displayed in Fig.~\ref{2DSOTSC}(a)~\cite{GhoshDynamical2022}. It separates into four regions: R1 with only zero-energy modes, R2 with no modes, R3 supporting only $\pi$-modes, and R4 where both $0$ and $\pi$ MCMs coexist. The corresponding LDOS of the MCMs is shown in Fig.~\ref{2DSOTSC}(b) for a 2D square lattice. The quasienergy spectra ($E_m$) are illustrated in Fig.~\ref{2DSOTSC}(c)–(f): R1 exhibits eight zero-energy MCMs, R2 hosts none, R3 supports four MCMs at $E_m=\pm \pi$, and R4 features both eight zero-energy MCMs and four MCMs each at $E_m=\pm \pi$. The latter has no static analog. At the outset, we mention that the driven system continues to satisfy the mirror symmetries, which allow us to obtain the topological characterization for the system, which we discuss next. The particle-hole symmetry remains respected in the driven system. 

\subsubsection{Topological characterization of SOTSC phase}\label{Sec:III}
In the anomalous Floquet phase (where both $0$ and $\pi$ modes coexist), the key difficulty lies in distinguishing the topological nature of both $0$- and $\pi$-MCMs. To address this subtle issue, we begin by analyzing the Wannier-sector polarization for the 2D FSOTSC and 3D FTOTSC using the nested Wilson loop formalism~\cite{GhoshDynamical2022,benalcazarprb2017,Ni2020}, with the Floquet operator $U_{2 \rm D}(\vect{k},T)$.  For the 2D FSOTSC, the averaged first-order nested Wannier polarization $\langle \nu_{y,\rm Flq,\mu'}^{\pm \nu_x} \rangle$ takes the quantized value $0.5$ within phases R1 and R3. However, this quantity alone cannot determine whether the MCMs reside in the $0$- or $\pi$-gap. In contrast, both the trivial phase (R2) and the anomalous phase supporting coexisting $0$- and $\pi$-modes (R4) yield a vanishing nested polarization. This limitation motivates us to introduce the periodized evolution operator (PEO), which encodes explicit information about the quasienergy gaps where the MCMs appear~\cite{Rudner2013,Huang2020}. The PEO is defined as $U_{d \rm D,\epsilon}(\vect{k},t) = U_{d \rm D}(\vect{k},t)\left[ U_{d \rm D}(\vect{k},T)\right]^{-t/T}_\epsilon$, where $U_{d \rm D}(\vect{k},t)$ denotes the full time-evolution operator. This construction separates into two distinct parts: an anomalous component $U_{d \rm D,\epsilon}(\vect{k},t)$, which captures the nontrivial Floquet dynamics, and a normal quasi-static component $\left[U_{d \rm D}(\vect{k},T)\right]^{t/T}_\epsilon$.

To characterize the anomalous Floquet modes, we employ the notion of a dynamical mean polarization, which captures the relative displacement of a particle between two time instants~\cite{Huang2020,GhoshDynamical2022}. It is defined as $\hat{\bar{x}}(t) = \frac{\hat{x}(t)+\hat{x}(0)}{2}$,
where $\hat{x}(0) \equiv \hat{x}$ corresponds to the static polarization operator~\cite{RestaPRL1998},
$\hat{x} = \sum_{im} \hat{c}_{im}^\dagger \ket{0} \, e^{-i \Delta_x x_i} \bra{0} \hat{c}_{im}$, with $\Delta_i = 2\pi/L_i$ and $\hat{c}$ denoting the quasiparticle creation operators. The time-evolved operator is given by 
$\hat{x}(t) = U^\dagger_{2 \rm D,\epsilon}(\vect{k},t) \, \hat{x} \, U_{2 \rm D,\epsilon}(\vect{k},t)$. The eigenvalues of $\hat{\bar{x}}(t)$ are directly linked to the dynamical Wilson loop operator $W_{x,\epsilon,\vect{k}}(t)$, defined through 
$W_{x,\epsilon,\vect{k}}(t) \ket{\nu_{x,\epsilon,\mu}(\vect{k},t)} = 
e^{-2\pi i \nu_{x,\epsilon,\mu}(k_{j\neq x},t)} \ket{\nu_{x,\epsilon,\mu}(\vect{k},t)}$.
Here, the Wilson loop $W_{x,\epsilon,\vect{k}}(t)$ is constructed as a path-ordered product of the link matrices $Q_{x,\epsilon,\vect{k}}(t) = \frac{\mathbb{I} + U^\dagger_{2 \rm D,\epsilon}(\vect{k}+\Delta_x \vect{e}_x,t) U_{2 \rm D,\epsilon}(\vect{k},t)}{2}$.
From this construction, one obtains the dynamical first-order branches $\nu_{x,\epsilon,\mu}(k_{j\neq x},t)$, which describe the relative motion of a particle along the $x$-direction with respect to the lattice position $x_i$ during the time interval $t \in [0,T]$~\cite{Huang2020,benalcazarprb2017}.

\begin{figure}[H]
	\centering
	\subfigure{\includegraphics[width=0.49\textwidth]{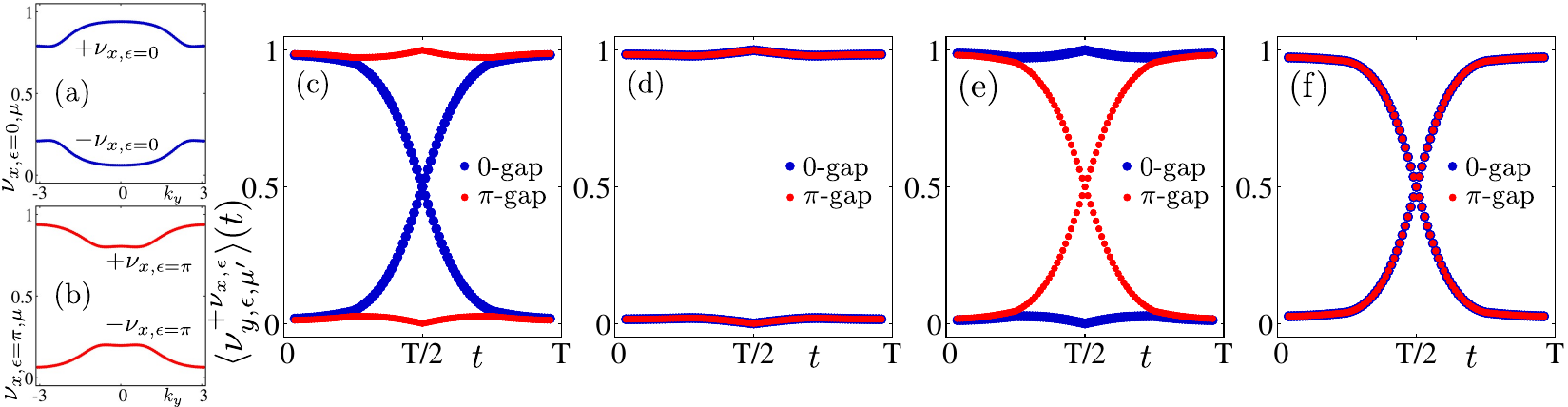}}
\caption{Dynamical polarization branches $\nu_{x,\epsilon=0,\mu}$ and $\nu_{x,\epsilon=\pi,\mu}$, associated with the $0$- and $\pi$-gaps, are shown as functions of $k_y$ in panels (a) and (b) at $t=T/2$, with the system in phase R4. Panels (c)–(f) display the average quadrupolar motion $\langle \nu_{y,\epsilon,\mu'}^{+\nu_{x,\epsilon}} \rangle (t)$ for R1, R2, R3, and R4 (see Fig.~\ref{2DSOTSC}), depicted as functions of time $t$. The results reveal gapless crossings between opposite branches. Blue and red markers indicate contributions from the $0$- and $\pi$-gaps, respectively. Further details are discussed in the main text. This figure is adapted from~\cite{GhoshDynamical2022} with permission, Copyright (2022) by the American Physical Society.}
\label{quadrupole}
\end{figure}

In the presence of the FHOTSC phase, the eigenvalues $\nu_{x,\epsilon,\mu}(k_{j \neq x},t)$ remain gapped throughout the full driving cycle $t \in [0,T]$, and can be separated into two distinct groups labeled as $\pm \nu_{x,\epsilon}$~\cite{GhoshDynamical2022}. The projection onto these folded subspaces using the projector $P_{\pm \nu_{x,\epsilon}}(t)$, enables us to obtain 
the dynamical second-order polarization, where the relative motion along the $y$-direction is investigated. The corresponding first-order nested Wilson loop, $W_{y,\epsilon,\vect{k}}^{\pm \nu_{x,\epsilon}}(t)$, is constructed as a path-ordered product of the link elements
$\left[Q_{y,\epsilon,\vect{k}}^{\pm \nu_{x,\epsilon}}(t)\right]_{\mu_1 \mu_2} = 
\sum_{mn} \left[\nu_{x,\epsilon,\mu_1}(\vect{k}+\Delta_y \vect{e}_y,t)\right]^*_m 
\left[Q_{y,\epsilon,\vect{k}}(t)\right]_{mn} 
\left[\nu_{x,\epsilon,\mu_2}(\vect{k},t)\right]_n$. The eigenvalue problem for this nested Wilson loop yields the dynamical second-order (quadrupolar) branches, $W_{y,\epsilon,\vect{k}}^{\pm \nu_{x,\epsilon}}(t) 
\ket{\nu_{y,\epsilon,\mu'}^{\pm \nu_{x,\epsilon}}(\vect{k},t)} = 
e^{-2\pi i \nu_{y,\epsilon,\mu'}^{\pm \nu_{x,\epsilon}}(k_{j\neq y},t)}
\ket{\nu_{y,\epsilon,\mu'}^{\pm \nu_{x,\epsilon}}(\vect{k},t)} $. These quadrupolar branches serve as the topological markers of the anomalous 2D FSOTSC phase, as illustrated in Fig.~\ref{quadrupole}.

\subsubsection{Dynamical quadrupole moment} 
We illustrate the features in Fig.~\ref{quadrupole}(a) and (b), corresponding to the $0$- and $\pi$-gaps at $t=T/2$, with the system is in phase R4~\cite{Huang2020}. The mirror symmetry $\mathcal{M}_x$ enforces the quadrupolar branches obtained from opposite first-order sectors to coincide, i.e., $\nu_{y,\epsilon,\mu'}^{+\nu_{x,\epsilon}}(k_x,t)=\nu_{y,\epsilon,\mu'}^{-\nu_{x,\epsilon}}(k_x,t)$, while $\mathcal{M}_y$ ensures that the quadrupolar branches occur in pairs, such that $\nu_{y,\epsilon,\mu'_1}^{+\nu_{x,\epsilon}}(k_x,t)=-\nu_{y,\epsilon,\mu'_2}^{+\nu_{x,\epsilon}}(k_x,t)$~\cite{Huang2020}.  The averaged quadrupolar motion, $\langle \nu_{y,\epsilon,\mu'}^{+\nu_{x,\epsilon}} \rangle (t)=\frac{1}{L_x}\sum_{k_x}\nu_{y,\epsilon,\mu'}^{+\nu_{x,\epsilon}}(k_x,t)$,  
is central to understanding the topology of the Floquet modes. Since the periodized evolution operator satisfies $U_{2\rm D,\epsilon}(0)=U_{2\rm D,\epsilon}(T)=\mathbb{I}$, the dynamics correspond to a closed trajectory over $t \in [0,T]$, enforcing the fixed-point conditions $\langle \nu_{y,\epsilon,\mu'}^{+\nu_{x,\epsilon}} \rangle (0)=\langle \nu_{y,\epsilon,\mu'}^{+\nu_{x,\epsilon}} \rangle (T)=0 \ (\mathrm{mod}\ 1)$.  In a topologically trivial phase, these fixed points are connected adiabatically throughout the cycle without any branch crossing, as illustrated in Fig.~\ref{quadrupole}(c) ($\pi$-gap), (d) (both gaps), and (e) ($0$-gap). By contrast, when MCMs are present (see Fig.~\ref{2DSOTSC}), the motion of $\langle \nu_{y,\epsilon,\mu'}^{+\nu_{x,\epsilon}} \rangle (t)$ is obstructed, resulting in a branch crossing at $\tfrac{1}{2}\ (\mathrm{mod}\ 1)$ precisely at $t=T/2$; this is shown in Fig.~\ref{quadrupole}(c) ($0$-gap), (e) ($\pi$-gap), and (f) (both $0$- and $\pi$-gaps).  This behavior leads to the quantization of the dynamical quadrupole moment, $Q_{\epsilon}^{+\nu_{x,\epsilon}}=\frac{1}{T}\int_0^T dt \, \langle \nu_{y,\epsilon,\mu'}^{+\nu_{x,\epsilon}} \rangle (t) = [0]\ 1 \ (\mathrm{mod}\ 2)$, corresponding to the trivial and topological cases, respectively, thereby establishing a $\mathbb{Z}_2$ classification.

\subsection{Brief Summary}
In this section, we study the Floquet  generation of  SOTSC phases while the initial phase of the matter is trivial.
In Sec. \ref{subsec:2D-TSC1}, we demonstrate that the $\delta$-kick in magnetic field or HOT mass term dynamically break the $C_4$ symmetry while particle-hole and chiral symmetry are still respected. Note that this mass term directly facilitates the SOTSC gap while the FOTSC gap is always maintained. On the other hand,   we adopt the on-site  and off-site step drive  
in Sec. \ref{Subsec:DynamicalConstruction}
 to illustrate the second-order MCMs where one can systematically  control the FOT and HOT terms while dynamically preserving mirror and particle-hole symmetries. 
\section{Floquet Third-order TSC in  3D systems}~\label{Sec:3D}
In this section, we discuss the emergence of Majorana hinge modes~(MHMs) and MCMs in 3D SOTSC and TOTSC, respectively. In Sec.~\ref{subsec:HierarchyHOTSC}, we discuss a model of 3D TI in proximity to a bulk $s$-wave superconductor, where two Wilson Dirac mass terms are used to generate a hierarchy of TSC such as SOTSC and TOTSC while they remain extrinsic. In Sec.~\ref{subsec:DynamicalTOTSC}, we discuss the dynamical generation of  intrinsic Floquet TOTSC  harboring both $0$-and $\pi$-modes, with proximity-induced $d$-wave superconductivity. A dynamical octupole moment is utilized for the complete topological characterization.

\subsection{Hierarchy of HOTSCs in three dimensions} \label{subsec:HierarchyHOTSC}

\subsubsection{$s$-wave topological superconductor model}\label{Sec:II}
We start from a generic lattice Hamiltonian defined on a cubic geometry, representing a 3D HOTI with proximity-induced $s$-wave pairing~\cite{Ghosh2021PRB}:
\begin{eqnarray}
	H(k)&=&2 \lambda\sum_{j=1}^{3} \sin k_{j} \Gamma_{j} + \left[(m_0 - 6t) +2t \sum_{j=1}^{3} \cos k_{j}\right] \Gamma_{4} \nonumber\\
	&& +\Delta_0 \Gamma_{5} + V_1 \Gamma_{6} + V_2 \Gamma_{7} + h_{x} \Gamma_{8} \nonumber \\
	&=& \vect{N(k)}. \vect{\Gamma}~\label{Ham},
\end{eqnarray}
where $\lambda, m_0, t$, and $\Delta_0$ represent SOC, crystal field splitting strength, nearest neighbour hopping and proximity-induced superconducting gap, respectively. Additional perturbations $V_1$, $V_2$, and $h_x$ drive the system into higher-order topological superconducting phases~\cite{Ghosh2021PRB,benalcazar2017,Nag2020}: see Eq.~(\ref{Ham}). Here we use the notation $k_{1,2,3}=k_{x,y,z}$, ${\bm N}(\bm k)=( N_1(\bm k),\dots,N_8(\bm k))$, and ${\bm \Gamma}=(\Gamma_1,\dots,\Gamma_8)$, with the $\Gamma$ matrices being $16\times16$ operators. Here,  $N_6=V_1=\sqrt{3} \Lambda_1 \left(\cos k_x-\cos k_y \right)$ and $N_7=V_2=\Lambda_2 \left(2\cos k_z-\cos k_x-\cos k_y \right)$.
Their explicit forms are given by $\Gamma_1=\mu_x\sigma_x s_x \tau_z$, $\Gamma_2=\mu_x \sigma_x s_y\tau_z$,  $\Gamma_3=\mu_x\sigma_x s_z \tau_z$, $ \Gamma_4=\mu_x\sigma_z \tau_z$, $\Gamma_5=  \tau_x$, 
$\Gamma_6=\mu_x \sigma_y$, $\Gamma_7=\mu_z \tau_z$ and $\Gamma_8=s_x$~\cite{Ghosh2021PRB}, where  ${\bm \mu}$, ${\bm \sigma}$, ${\bm s}$, and ${\bm \tau}$  act on sub-lattice $(A,B)$, orbital $(\alpha,\beta)$, spin $(\uparrow, \downarrow)$, and particle–hole (e-h)  degrees of freedom, respectively. For $V_1 \neq 0$ and $V_2=0$, one-dimensional dispersive Majorana hinge modes (MHMs) appear along the $z$ direction at the intersections of the $xz$ and $yz$ surfaces. Including both $V_1 \neq 0$ and $V_2 \neq 0$ in Eq.~(\ref{Ham}) gaps out these hinge channels along the body diagonals, as $V_2$ vanishes there~\cite{Ghosh2021PRB}, leaving only corner-localized zero-energy states characteristic of the TOTSC phase.   The above model breaks $C_4$ and $\mathcal {T}$ while preserving $C_4 \mathcal {T}$, $\mathcal {C}$ and $\mathcal {P}$ symmetries.

The existence of MHMs can be analyzed through the effective surface Hamiltonians $H^S_{ij}$, obtained from Eq.~(\ref{Ham}) by imposing open boundary conditions along the perpendicular direction~\cite{ghosh2020floquet2}. Here $M_{\Delta}=\Delta_0$, $M_{\Lambda_1}=\tfrac{\sqrt{3} m_0\Lambda_1}{2t}$, and $M_{\Lambda_2}=\tfrac{m_0 \Lambda_2}{2t}$. Since the Wilson–Dirac mass $M_{\Lambda_1}$ switches its sign between surfaces, one finds 16 gapless hinge channels when $M_{\Lambda_1}>M_{\Delta}$~\cite{ghosh2020floquet2}. This follows from the fact that all eight gap factors, labeled by $\tau_x,\sigma_x,s_z=\pm1$, switch sign accordingly. In contrast to HOTI, which appears for arbitrarily small $M_{\Lambda_1}$, the superconducting case requires $M_{\Lambda_1}$ to exceed a threshold value $M_{\Lambda_1}^T=M_{\Delta}$ for the SOTSC phase to be realized.  

To capture the MCMs, we further examine hinge Hamiltonians $H^H_{i,ij}$ obtained from the corresponding surface models under additional open boundaries. This set of hinge Hamiltonians predicts the number of MCMs when $M_{\Lambda_2}\neq 0$, which gaps out the hinge channels. The relative sign structure of the gap factors (for $s_x,\tau_x=\pm1$) changes only within the $s_x=+1$ block, indicating that each corner can, in principle, accommodate two localized Majorana zero modes.


\subsubsection{Floquet Generation of HOTSC}\label{Sec:VI}
\begin{figure}[]
	\centering
	\subfigure{\includegraphics[width=0.49\textwidth]{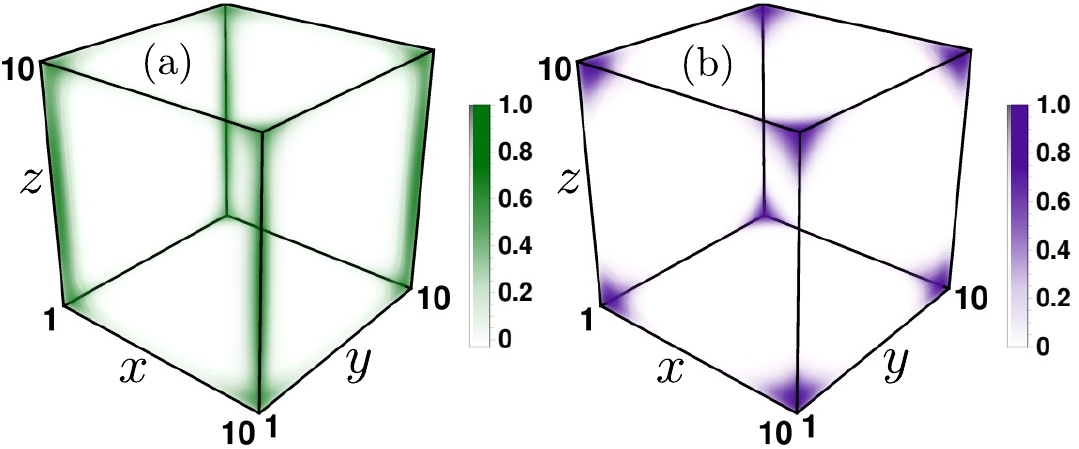}}
	\caption{ (a) LDOS corresponding to quasi-energy $\epsilon_m=0$ is shown under OBC along all directions, using the driving protocol mentioned in Eq.~(\ref{kick}) with $\Lambda_1=0.4$ and $\Lambda_2=0.0$. In this case, Floquet SOTSC hosts MHMs propagating along the $z$ axis, localized at the intersections of the $xz$ and $yz$ surfaces. (b)~The LDOS profile for $\Lambda_1=0.4$ and $\Lambda_2=0.4$ is presented, where MCMs emerge, signaling the realization of the FTOTSC phase. The color bars represent the weight of the normalized LDOS. The remaining model parameters are fixed at $m=2.0$, $t=\lambda=1.0$, $\Delta=0.4$, and $T=0.628$.  This figure is adapted from~\cite{Ghosh2021PRB} with permission, Copyright (2021) by the American Physical Society. }
	\label{FloquetLDOS}
\end{figure}


\begin{figure}[]
	\centering
	\subfigure{\includegraphics[width=0.49\textwidth]{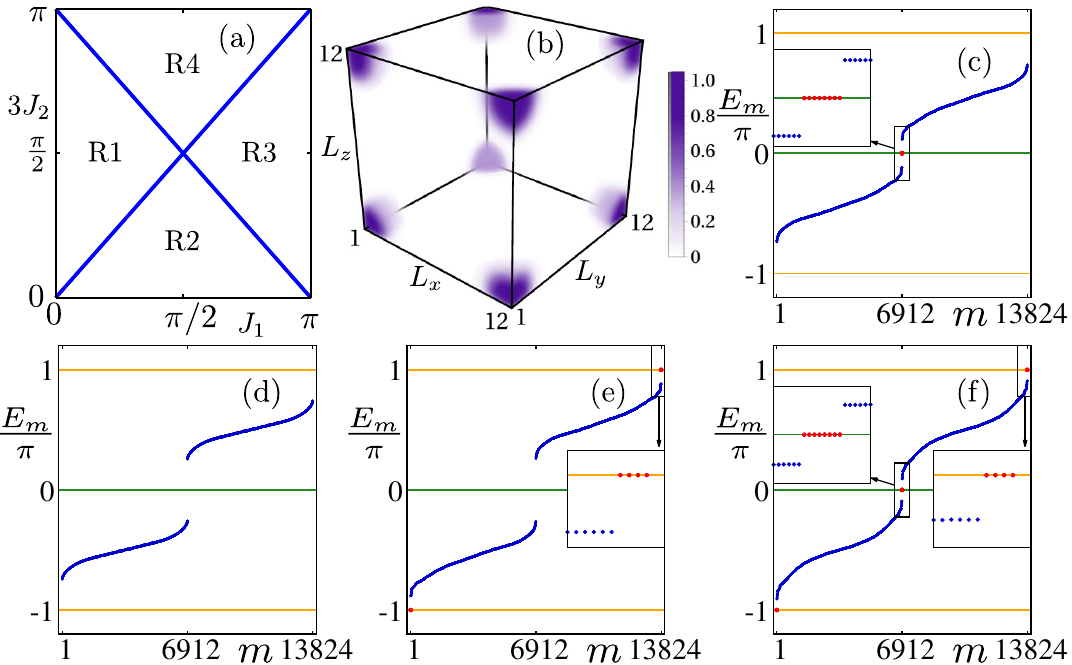}}
\caption{Results analogous to Fig.~\ref{2DSOTSC} are shown, now obtained for a 3D cubic lattice of size $L_x \times L_y \times L_z$ with $\Delta_1=\Delta_2=1.0$, while all other model parameters are considered to be identical to those used in Fig.~\ref{2DSOTSC}. The color bar represents the weight of the normalized LDOS. This figure is adapted from~\cite{GhoshDynamical2022} with permission, Copyright (2022) by the American Physical Society.}
\label{3DTOTSC}
\end{figure}

Having analyzed the hierarchy of static HOTSC phases, we now address their Floquet realization starting from a 3D topological insulator placed in close proximity to an $s$-wave superconductor. To this end, we consider a periodic $\delta$-kick driving protocol~\cite{Nag19,ghosh2020floquet,ghosh2020floquet2,Nag2020}, 
given by  
\begin{eqnarray}
	V(t)&=& {\tilde {\bm M}}({\bm k}) \cdot{\tilde {\bm \Gamma}}   \sum_{r=1}^{\infty} \delta(\tilde{t}-rT)\ ,
	\label{kick}
\end{eqnarray}
where, ${\tilde{\bm M}}({\bm k})=(V_1,V_2)$ and ${\tilde{\bm \Gamma}}=(\Gamma_6,\Gamma_7)$. Here $T$ is the driving period and $\tilde{t}$ denotes time. As in the static case, choosing $V_1\neq 0$, $V_2=0$ ($V_1\neq 0$, $V_2\neq 0$) realizes FSOTSC (FTOTSC) phases.  

Using the static Hamiltonian [Eq.~(\ref{Ham})] with $V_1=V_2=h_x=0$, the Floquet operator takes the form  
$U(T)=\exp(-iH(\bm k)T)\exp(-iV)$.  
From this, the effective Floquet Hamiltonian, valid at arbitrary frequency, can be written as  
\small
\begin{eqnarray}
	H_{\rm eff}&=&\frac{\xi_{\bm k}}{\sin\xi_{\bm k}T}\Bigg[\sin(|{\bm N}({\bm k})| T) \cos(|{\bm M}({\bm k})|) \sum_{j=1}^{5}n_j \Gamma_j \non \\ &&+\cos(|{\bm N}({\bm k})| T) \sin(|{\bm M}({\bm k})|) \sum_{j=1,2} m_j \Gamma_{j+5}\non \\
	&&+\sin(|{\bm N}({\bm k})| T) \sin(|{\bm M}({\bm k})|) \sum_{j=1}^{5} \left(n_j m_1 \Gamma_{j6} + n_j m_2 \Gamma_{j7}\right) \Bigg]\ , \non \\
\end{eqnarray}
\vskip -0.2cm
\normalsize
with $\xi_{\bm k}=\tfrac{1}{T}\cos^{-1}\!\left[\cos(|{\bm N}({\bm k})|T)\cos(|{\bm M}({\bm k})|)\right]$,  
$n_j=N_j({\bm k})/|{\bm N}({\bm k})|$, $m_j=M_j({\bm k})/|{\bm M}({\bm k})|$,  
$|{\bm N}({\bm k})|=\sqrt{\sum_{j=1}^{5}N_j^2({\bm k})}$, and $|{\bm M}({\bm k})|=\sqrt{\sum_{j=1}^{2}M_j^2({\bm k})}$.  The above Hamiltonian preserves the chiral and particle-hole symmetry which are essential to obtain the HOT phases. 

The Floquet eigenvalue equation $U(T)\ket{\phi_m}=e^{-i\epsilon_m T}\ket{\phi_m}$ is solved numerically to obtain the quasienergy spectrum, where $\ket{\phi_m}$ and $\epsilon_m$ denote the quasienergy states and their eigenvalues. In Figs.~\ref{FloquetLDOS}(a) and (b) we show the LDOS associated with states near zero quasienergy ($\epsilon_m\simeq 0$), revealing Floquet MHMs and MCMs for suitable driving parameters. These Floquet excitations are stabilized by particle–hole symmetry $\mathcal{C}$. The topological character of FSOTSC and FTOTSC phases can further be captured through Floquet Wannier spectra, which exhibit mid-gap eigenvalues pinned at $0.5$, in direct analogy with the static case~\cite{ghosh2020floquet,ghosh2020floquet2}.  

For clarity, we focus on the case without the need for a magnetic field ($h_x=0$). Notably, in the high-frequency regime, the driven HOTSC phases generated by this driving protocol [Eq.~(\ref{kick})] do not support anomalous quasienergy modes at $\epsilon_m=\pi/T$~\cite{ghosh2020floquet,ghosh2020floquet2}. The possibility of engineering anomalous HOTSC phases (Floquet MHMs or MCMs at $\pi/T$) via alternative driving schemes~\cite{Huang2020,WuPRBL2021} is discussed
in the next sub-section. 


\subsection{Dynamical construction of Octupolar topological superconductors} \label{subsec:DynamicalTOTSC}



We adopt the step drive protocol akin to Eq.~(\ref{drive}) to realize a 3D TOTSC phase that hosts MCMs~\cite{GhoshDynamical2022}. 
In this framework, we set $h_{1,3\rm D}(\vect{k})=h_{3,3\rm D}(\vect{k})=\tau_z \sigma_z$, while 
$h_{2,3\rm D}(\vect{k})=\epsilon_{3\rm D}(\vect{k})\tau_z \sigma_z+\Lambda_{3\rm D}(\vect{k})+\Delta_{3\rm D}(\vect{k})$, 
with $\epsilon_{3\rm D}(\vect{k})=\cos k_x+\cos k_y+\cos k_z$, 
$\Lambda_{3\rm D}(\vect{k})=\sin k_x \tau_z\sigma_x s_x+\sin k_y \tau_z\sigma_x s_y+\sin k_z \tau_z\sigma_x s_z$, 
and $\Delta_{3\rm D}(\vect{k})=\Delta_1(\cos k_x-\cos k_y)\tau_x+\Delta_2(2\cos k_z-\cos k_x-\cos k_y)\tau_y$. 
This form incorporates the mixed $d_{x^2-y^2}+i d_{3z^2-r^2}$ pairing symmetry in three dimensions~\cite{RoyPRBL2021,Roydwave2019}. See Fig. \ref{3DTOTSC} for the pictorial illustration of 3D FTOTSC phases along with the energy spectrum.

Extending the analysis of second-order polarization, the octupolar phase ensures that quadrupolar branches remain gapped during the entire cycle $t\in[0,T]$, 
and can therefore be organized into two distinct sets $\pm\nu_{y,\epsilon}^{\pm\nu_{x,\epsilon}}$~\cite{GhoshDynamical2022}. The projection onto these folded subspaces using the projector $P_{\pm\nu_{y,\epsilon}^{\pm\nu_{x,\epsilon}}}(t)$ enables us to compute the dynamical third-order polarization, which measures the relative motion along the $z$ direction. In this context, the second-order nested Wilson loop $W_{z,\epsilon,\vect{k}}^{\pm\nu_{y,\epsilon}^{\pm\nu_{x,\epsilon}}}(t)$ is obtained from the path-ordered product of the line element 
$\left[Q_{z,\epsilon,\vect{k}}^{\pm\nu_{y,\epsilon}^{\pm\nu_{x,\epsilon}}}(t)\right]_{\mu'_1\mu'_2}=\sum_{\substack{mn\mu_1\mu_2}} \left[\nu_{y,\epsilon,\mu'_1}^{\pm\nu_{x,\epsilon}}(\vect{k}+\Delta_z\vect{e}_z,t)\right]_{\mu_1}^* \left[\nu_{x,\epsilon,\mu_1}(\vect{k}+\Delta_z\vect{e}_z,t)\right]_m^*$ ~
 $\left[Q_{z,\epsilon,\vect{k}}(t)\right]_{mn} \left[\nu_{x,\epsilon,\mu_2}(\vect{k},t)\right]_n \left[\nu_{y,\epsilon,\mu'_2}^{\pm\nu_{x,\epsilon}}(\vect{k},t)\right]_{\mu_2}$. 
The eigenvalues of $W_{z,\epsilon,\vect{k}}^{\pm\nu_{y,\epsilon}^{\pm\nu_{x,\epsilon}}}(t)$ yield the third-order octupolar branches $\nu_{z,\epsilon,\mu''}^{\pm\nu_{y,\epsilon}^{\pm\nu_{x,\epsilon}}}(k_{j\neq z},t)$, which serve as the dynamical topological indicator of the 3D FTOTSC phase, as illustrated in Fig.~\ref{octupole}. 

\subsubsection{Dynamical octupole moment} 
\begin{figure}[]
	\centering
	\subfigure{\includegraphics[width=0.49\textwidth]{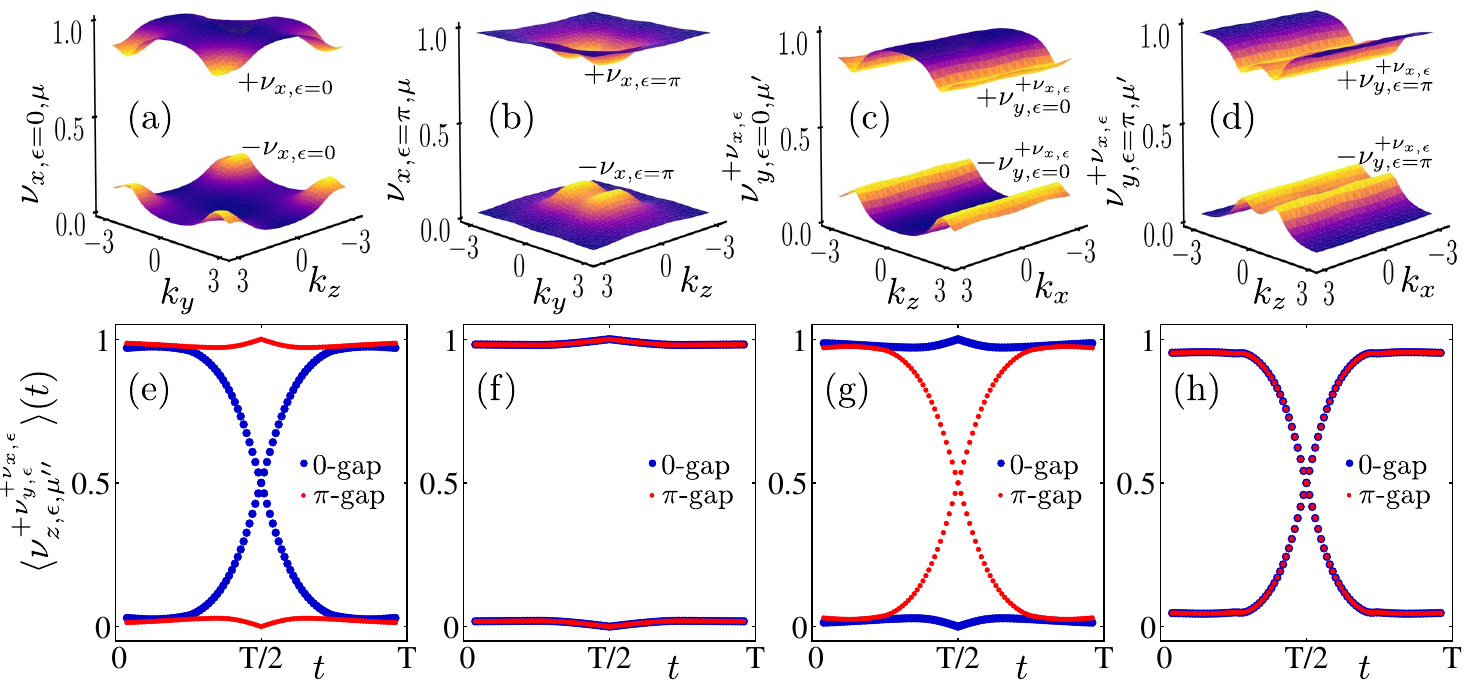}}
	\caption{Panels (a) and (b) reproduce the results of Figs.~\ref{quadrupole}(a) and (b) for a 3D cubic lattice, now shown as a function of $k_z$ and $k_x$. 
    The dynamical quadrupolar branches $\nu_{y,\epsilon=0,\mu'}^{+\nu_{x,\epsilon}}$ and $\nu_{y,\epsilon=\pi,\mu'}^{+\nu_{x,\epsilon}}$, both gapped, are presented in panels (c) and (d) at $t=\tfrac{T}{2}$ in region R4. Panels (e)–(h) display the average octupolar motion $\langle \nu_{z,\epsilon,\mu''}^{+ \nu_{y,\epsilon}^{+ \nu_{x,\epsilon}}}\rangle(t)$ for regime R1, R2, R3, and R4, respectively, uncovering gapless crossings between opposite branches. Here, blue and red markers denote contributions arising from the $0$- and $\pi$-gaps, respectively, as further discussed in the text. This figure is adapted from~\cite{GhoshDynamical2022} with permission, Copyright (2022) by the American Physical Society.}
	\label{octupole}
\end{figure}

We illustrate the dynamical first-order branches in Figs.~\ref{octupole}(a) and (b), corresponding to the $0$- and $\pi$-gaps, respectively, at $t=\tfrac{T}{2}$ when the system is in phase R4. The mirror symmetry $\mathcal{M}_x$ ensures that the second-order branches obtained from opposite first-order branches $\pm \nu_{x,\epsilon,\mu}(k_y,k_z,t)$ coincide, $\mathcal{M}_y$ enforces the condition $\nu_{y,\epsilon,\mu'_1}^{+ \nu_{x,\epsilon}}(k_z,k_x,t)=-\nu_{y,\epsilon,\mu'_2}^{+ \nu_{x,\epsilon}}(k_z,k_x,t)$, 
and $\mathcal{M}_z$ requires $\nu_{y,\epsilon,\mu'}^{+ \nu_{x,\epsilon}}(k_z,k_x,t)=\nu_{y,\epsilon,\mu'}^{+ \nu_{x,\epsilon}}(-k_z,k_x,t)$. 
The corresponding dynamical second-order branches are shown in Figs.~\ref{octupole}(c) and (d) for $0$- and $\pi$-gaps at $t=\tfrac{T}{2}$, also in region R4. 
Similar to the first-order case, the quadrupolar branches remain gapped, paving the way for the construction of the dynamical third-order (octupolar) branch. 
Here, $\mathcal{M}_x$ and $\mathcal{M}_y$ ensure that the octupolar branches originating from different quadrupolar branches coincide, while $\mathcal{M}_z$ dictates that they appear in pairs, 
$\nu_{z,\epsilon,\mu''_1}^{+ \nu_{y,\epsilon}^{+ \nu_{x,\epsilon}}}(k_x,k_y,t)=-\nu_{z,\epsilon,\mu''_2}^{+ \nu_{y,\epsilon}^{+ \nu_{x,\epsilon}}}(k_x,k_y,t)$.  

Following the 2D construction, we define the average octupolar motion as 
$\langle \nu_{z,\epsilon,\mu''}^{+ \nu_{y,\epsilon}^{+ \nu_{x,\epsilon}}} \rangle (t)=\frac{1}{L_x L_y}\sum_{k_x,k_y}\nu_{z,\epsilon,\mu''}^{+ \nu_{y,\epsilon}^{+ \nu_{x,\epsilon}}}(k_x,k_y,t)$. 
In the trivial case, $\langle \nu_{z,\epsilon,\mu''}^{+ \nu_{y,\epsilon}^{+ \nu_{x,\epsilon}}} \rangle (t:0\to T)$ returns to its initial value (mod $1$) without any gap closings between branches [see Figs.~\ref{octupole}(e) $\pi$-gap, (f) $0$- and $\pi$-gaps, and (g) $0$-gap]. 
In contrast, the non-trivial case is characterized by a gap closing between distinct octupolar branches at $\tfrac{1}{2}~(\mathrm{mod}~1)$, where 
$\langle \nu_{z,\epsilon,\mu''}^{+ \nu_{y,\epsilon}^{+ \nu_{x,\epsilon}}} \rangle (t\to 0)=0~(1)$ evolves to 
$\langle \nu_{z,\epsilon,\mu''}^{+ \nu_{y,\epsilon}^{+ \nu_{x,\epsilon}}} \rangle (t\to T)=1~(0)$ [see Figs.~\ref{octupole}(e) $0$-gap, (g) $\pi$-gap, and (h) $0$- and $\pi$-gaps]. 
Therefore, the 3D octupole moment admits a $\mathbb{Z}_2$ invariant analogous to the quadrupole moment in 2D: 
$O_{\epsilon}^{+ \nu_{y,\epsilon}^{+ \nu_{x,\epsilon}}}=\tfrac{1}{T}\int_0^T dt \,\langle \nu_{z,\epsilon,\mu''}^{+ \nu_{y,\epsilon}^{+ \nu_{x,\epsilon}}}\rangle(t)=[0]~1~(\mathrm{mod}~2)$ for the trivial [topological] phase~\cite{GhoshDynamical2022}. 

\begin{widetext}

\begin{table}[]

\begin{tabular}{|c|c|c|c|c|c|}
\hline
Driving Protocol &
  Topological order &
  Dimension &
  Boundary modes & Symmetry &
  Topological indices \\ \hline
\multirow{3}{*}{Step drive} &
  \begin{tabular}[c]{@{}c@{}}
  Floquet FOTSC: Sec.~\ref{subsec:FloquetMEMRashba}
  \end{tabular} &
  1D &
  $0$-, $\pi$-MEMs  & $\mathcal {P}$, $\mathcal{C}$ &
  Winding numbers \\ \cline{2-5} 
 &
  \begin{tabular}[c]{@{}c@{}}
  Floquet SOTSC: Sec.~\ref{Subsec:DynamicalConstruction}
  \end{tabular} &
  2D &
  $0$-, $\pi$-MCMs & $\mathcal {M}_{x,y}$, $\mathcal {P}$ &
  Dynamical quadrupolar moments \\ \cline{2-6} 
 &
  \begin{tabular}[c]{@{}c@{}}
  Floquet TOTSC: Sec.~\ref{subsec:DynamicalTOTSC}
  \end{tabular} &
  3D &
  $0$-, $\pi$-MCMs & $\mathcal {M}_{x,y}$, $\mathcal {P}$ &
  Dynamical octupolar moments \\ \hline
Harmonic drive &
  \begin{tabular}[c]{@{}c@{}}
  Floquet FOTSC: Sec.~\ref{subsec:HelicalShiba}
  \end{tabular} &
  1D &
  $0$-, $\pi$-MEMs &  $\mathcal {P}$, $\mathcal{C}$ &
  Winding numbers  \\ \hline
\multirow{4}{*}{Periodic kick} &
\begin{tabular}[c]{@{}c@{}}
  Floquet FOTSC: Sec.~\ref{subsec:KitaevDriven}
  \end{tabular} &
  1D &
  $0$-, $\pi$-MEMs & $\mathcal {P}$, $\mathcal{C}$ &
  Winding numbers \\ 
  \cline{2-5} 
&
  \begin{tabular}[c]{@{}c@{}}
  Floquet SOTSC: Sec.~\ref{subsec:2D-TSC1}
  \end{tabular} &
  2D &
  $0$-MCMs &  $\mathcal {P}$, $\mathcal{C}$  &
  Quandrupole moment, Wannier spectra \\ \cline{2-5} 
 &
  \begin{tabular}[c]{@{}c@{}}
  Floquet SOTSC: Sec.~\ref{subsec:HierarchyHOTSC}
  \end{tabular} &
  3D &
  $0$-MHMs &  $\mathcal {P}$, $\mathcal{C}$  &
  Quandrupole moment, Wannier spectra \\ \cline{2-5} 
 &
  \begin{tabular}[c]{@{}c@{}}
  Floquet TOTSC: Sec.~\ref{subsec:HierarchyHOTSC}
  \end{tabular} &
  3D &
  $0$-MCMs &  $\mathcal {P}$, $\mathcal{C}$  &
  Octupole moment, Wannier spectra \\ \hline
\end{tabular}

\caption{Summary of all the Floquet topological superconductor models discussed, including the driving protocol, topological order, dimension, boundary modes, and topological indices.}
\label{tab:Floquet}
\end{table}

\end{widetext}

\subsection{Brief Summary}
In this section, we study the Floquet generation of
HOTSC phases while the initial phase of the matter is
trivial. In  
Sec. \ref{subsec:HierarchyHOTSC}, we show that $\delta$-kick of HOT mass terms indeed cause the $C_4$ symmetry to break over six different surface  of a cube while the underlying FOT gap is always maintained. On the other hand, 
in Sec. \ref{subsec:DynamicalTOTSC}, we demonstrate the on-site and off-site step drive to engineer HOT mass terms where the underlying FOT mass term can be tuned. 

\section{Discussions and Outlook}~\label{Sec:discussion_outlook} 
The emergence of topological boundary, corner, or hinge modes can be understood from the Jackiw–Rebbi mechanism: a change in the sign of an effective mass term across a boundary or domain wall gives rise to localized topological states. This principle applies to both static and periodically driven systems. In Floquet systems, the external periodic drive renormalizes or generates effective mass terms in the Floquet Hamiltonian. As the drive parameters are varied, the effective mass gap can close and reopen with an opposite sign, signaling a topological phase transition and the appearance of new boundary, corner, or hinge Majorana modes.

We summarize all the driven TSC models discussed in the manuscript in Table~\ref{tab:Floquet}. Our review focuses on the systematic realization of HOTSC phases in both equilibrium and non-equilibrium settings, along with the topological characterization of $\pi$-Majorana modes. While existing studies largely treat static and Floquet FOTSC, SOTSC, and TOTSC phases separately, the emergence of $\pi$-Majorana modes in Floquet HOTSCs remains largely unexplored. We therefore, in this review,  provide a unified framework  to engineer diverse topological superconducting phases by systematically tuning symmetry-breaking terms in the BdG Hamiltonian, both in equilibrium and under periodic driving.

The generation and emergence of the regular $0$- and $\pi$-Majorana modes in different dimensions, as discussed in the manuscript, offers a versatile opportunity in quantum technology. 
In order to achieve fault-tolerant quantum computation, the unitary operations, executed by the quantum gates, are essentially carried out by braiding of the topological quantum objects, namely anyons. Therefore, the braiding-based topological quantum computation will become an efficient platform for the future information 
processing~\cite{NayakRMP2008,collins2006computing,venema2008dreamweaver,das2006topological}.  It has been shown that two successive counterclockwise exchanges fail to return the system to its original state, signifying the nontrivial braiding of anyons. In this context, the manifestation of the non-Abelian braiding statistics is best mimicked by the well-separated MZMs~\cite{sarma2015majorana}. The fractional quantum Hall systems and Abrikosov vortices in a topological superconductor are known for hosting the non-Abelian anyons where interaction and disorder play important roles, 
respectively~\cite{nayak08,PhysRevB.100.115147}. These anyonic modes are protected by an energy gap that yields topological immunity. This robustness is further supported by the fact that the temperature has to be much lower than the energy gap, and the inter-anyonic separation has to be larger than the correlation length. These conditions, therefore, ensure that the system lies in the regime of exponentially small Majorana energy splitting.

On the other hand, the field of topological computation has gained substantial momentum following the observation of a zero-bias tunneling conductance peak at a finite external magnetic field in semiconductor nanowires (InSb or InAs) proximitized by a conventional metallic superconductor (Al or Nb)~\cite{Mourik2012Science}.  However, these experiments neither shed light on the exponential localization with system length nor the anyonic braiding behavior.  Note that the $2e^ {2}/h$ quantization of {Majorana} peak is delicately thermal sensitive, and the quantization gets destroyed when thermal energy becomes larger than the width of the peak. Briefly added, topological quantum computation in these systems relies on the exponential localization of isolated Majorana modes and their non-Abelian braiding properties. For such an application, the detection of these modes can be performed via differential conductance~\cite{DM_trans_2025} and Josephson current~\cite{pal2025_AM} measurements. One can check for all the above studies, the emergence of Majorana modes and their signatures are stable against weak disorder. 
The fractional  
$4 \pi$ periodic Josephson effect is considered to be another deterministic signature of MZMs when two topological superconductors are in 
contact~\cite{Kitaev_2001,wiedenmann20164} with a tunnel barrier. This is markedly different from the regular $2\pi$ periodic Josephson effect when two normal superconductors are in contact. Interestingly, when a topological Josephson junction is exposed to electromagnetic radiation, the DC voltage experiences missing intermediate steps due to the $4\pi$ periodicity~\cite{rokhinson2012fractional}. Thus, exploring these transport behaviors for the Floquet Majorana modes in these first- and higher-order TSC remains an attractive future direction. While it would be also instructive to study their shot noise properties to ascertain the true Majorana nature, even if they exist in multiple numbers and at finite temperature as well. At present, a comprehensive theoretical and experimental understanding of other detection schemes, such as noise-correlation measurements for periodically driven $\pi$-Majorana modes, remains limited, thus allowing for the future research  directions.

{Given that Floquet dynamics can give rise to anomalous $\pi$-Majorana modes, which require a distinct characterization from conventional $0$-Majorana modes, one can further investigate their real-time emergence in other non-equilibrium settings.
One such scenario could be the aperiodic drive where the discrete time translation symmetry is broken. 
In particular}, the random aperiodic drives have grabbed attention in the community for their intriguing outcomes in problems of non-equilibrium statistical physics as well as condensed matter physics ~\cite{Bhattacharya18,Maity18,Nag21,kundu2024statistical,prg5-3srn,Albert16}. The emerging ensemble distribution for an aperiodically driven system is significantly different from that of a periodically driven system. 
{In a temporal model for anomalous quantum Hall insulator, 
adopting a periodic drive with two incommensurate frequencies, it has been shown that pumping of energy from one drive to the other is created by the non-zero Chern number~\cite{Crowley19}}. However, the aperiodicity-mediated topological properties are less explored to date, thereby offering a lot of future scope.  
In the field of topological superconductivity, the role of aperiodic drive is not explored yet.  {The present review, being focused on the periodic Floquet dynamics, does not offer scope to discuss the aperiodic/quasiperiodic dynamics. However, 
the quasi-periodic nature of the drive may also induce significant modification to the emergence of Majorana modes for FOTSC and HOTSC models that we leave for future studies. }



The effect of dissipation is another interesting avenue, which is always omnipresent in a physical system. The effect of the dissipative superconductor has been studied for a 1D Rashba nanowire employing Keldysh Green function’s formalism~\cite{YangPRL2021}. It has been shown that both the $0$- and $\pi$-Majorana modes become dissipative under some conditions, and a Floquet Majorana poisoning model has been employed to understand the dissipative modes. While dissipative has also been utilized to show the generation of Majorana modes in both static~\cite{GhoshSP2024} and driven systems~\cite{GogoiPRB2026}. Thus, it is natural to ask the fate of $0$- and $\pi$-Majorana that appear in the FHOTSC phase, with a particular interest in engineering the Majorana modes using dissipation~\cite{GhoshSP2024, GogoiPRB2026}.

\section{Summary and Conclusions}~\label{Sec:summary_conclusion} 
In this article, we demonstrate various periodic driving schemes for Floquet engineering of TSC in one, two, and three dimensions. We illustrate the emergence of regular $0$- and dynamical anomalous $\pi$-Majorana modes along with their appropriate topological characterizations in 1D realistic model systems. We find discrete-time drives, such as the delta kick and step drive, to be more efficient in obtaining the desired outcomes given the richness of phase diagrams. We establish 
the dynamical winding of the Floquet higher-order MCMs in two and three dimensions, which is a novel phenomenon for a driven system only. The true Majorana nature could only be guaranteed by the non-abelian braiding of these quasi-particles, irrespective of their static and dynamic nature, which we leave for future studies. However, the topological characterization suffices for the dynamical Majorana nature of $0$- and $\pi$- energy states using the BdG basis.

\section{Acknowledgements}
T.N. acknowledges the NFSG ``NFSG/HYD/2023/H0911” from Birla Institute of Technology and Science (BITS) Pilani. T.N. and A.S. thank ANRF, Govt. of India ``ANRF/ARG/2025/003163/PS". 


%

\end{document}